%% file: tesi.tex
\documentclass[a4paper,notitlepage,twoside,openany]{report}  
\newcommand{\PUBNUM}{hep-ph/9906517} 
\usepackage{jaume_de} 

\begin{document} 
 
\include{davant}

\include{inici}

\begin{fmffile}{fmftesi}     
\input{feyn_def.tex}       

\setcounter{chapter}{-1} 
\include{abstract} 
\include{intro}

\include{mssm} 
\include{renorm} 
\include{tbh}                   
\include{fcnctop}                   
\include{sbdecay}

\include{conclu}            

 
\addcontentsline{toc}{chapter}{\bibname} 
\include{biblio}

\addcontentsline{toc}{chapter}{\listfigurename} 
\listoffigures

\newpage
\addcontentsline{toc}{chapter}{\listtablename} 
\listoftables 
 
\appendix 
 
\include{tbh_c}

\end{fmffile} 
 
\end{document}

%% file: davant.tex

\pagestyle{empty}

\input{portada}
\input{certif_fitxa}
\input{agraim}


%% file: portada.tex
\begin{titlepage}
\vspace*{\fill}

\hfill  \PUBNUM
\vspace{2cm}

\hfill\resizebox{!}{3cm}{\includegraphics{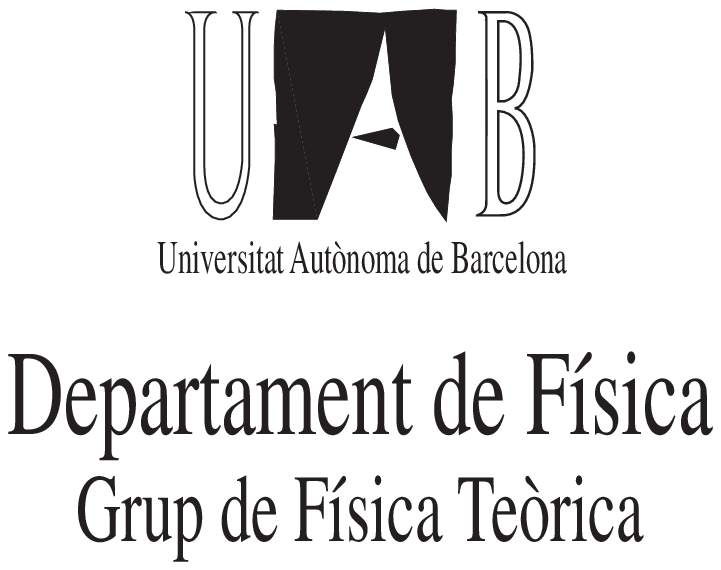}}

\vspace{2cm}

{\huge\bf\begin{center}
Supersymmetric radiative corrections to top quark and Higgs boson physics
\end{center}}
\vspace{1cm}
{\large
\begin{center}
Jaume Guasch Inglada
\vspace{1cm}

Universitat Aut{\`o}noma de Barcelona

Grup de F{\'\i}sica Te{\`o}rica 

Institut de F{\'\i}sica d'Altes Energies  
\end{center}
\vspace{3cm}
\begin{center}
\footnotesize
Mem{\`o}ria presentada com a Tesi Doctoral per a
optar al t{\'\i}tol de Doctor en F{\'\i}sica per la 
Universitat Aut{\`o}noma de Barcelona.
\end{center}

}

\vspace*{\fill}
\end{titlepage}


%% file: certif_fitxa.tex
\begin{titlepage}
\vspace*{\fill}

\noindent
\begin{center}

Aquesta mem{\`o}ria {\'e}s la tesi doctoral 

\vspace{1cm}

{\huge \bf ``Supersymmetric radiative corrections to top quark and Higgs boson
  physics''}

\vspace{1cm}

Fou realitzada per en 

\vspace{1cm}

{\large Jaume Guasch Inglada }

\vspace{1cm}

sota la direcci{\'o} del 

\vspace{1cm}

{\large Dr. Joan Sol{\`a} i Peracaula}

\vspace{1cm}

professor titular de F{\'\i}sica Te{\`o}rica de la Facultat de Ci{\`e}ncies de la
Universitat Aut{\`o}noma de Barcelona.

\vspace{1cm}

Va ser llegida
el dia 18 de gener de 1999 a la sala de seminaris de l'IFAE de la Universitat
Aut{\`o}noma de Barcelona.
\end{center}
\vspace*{\fill}
Tesi publicada per la Universitat Aut{\`o}noma de Barcelona amb ISBN 84-490-1544-8
\vfill
\end{titlepage}


%% file: agraim.tex
\begin{titlepage}
\vspace*{\fill}
{\small
Podriem dir que aquesta tesi {\'e}s {\em b{\`a}sicament} filla meva, per{\`o} les criatures
tamb{\'e} t{\'e}nen un avi, sense el qual mai haurien pogut arribar a n{\`e}ixer, en aquest
cas l'{\em avi} {\'e}s en Joan Sol{\`a}. Des d'un primer moment (i fins a l'ultim minut!)
hem treballat dur.  
Quan sorgeix algun problema imprevist, sempre hi sol haver alguna carpeta
d'apunts que ens simplifica la vida. A partir de la teva tesi hem pogut anar
desgranant els camins cap a nous mons. He tingut la oportunitat de treballar, ben
encarrilat, per{\`o} amb for{\c c}a llibertat. Gr{\`a}cies per haver tingut la oportunitat de
treballar amb tu.

Les criatures mai van soles, si no estan acompanyades de canalla de la mateixa
edat (any m{\'e}s, any menys) els falta alguna cosa, i als pares tamb{\'e}! 
aquesta tesi ha anat creixent al costat d'altres, amb co{\lgem}aboracions, caf{\`e}s,
subrutines FORTRAN agafades, deixades, robades \ldots, 
codi {\em Mathematica} i \LaTeX\  amunt i avall, 
discussions (de f{\'\i}sica i,
sobretot, d'altres coses), acudits, \ldots, m{\'e}s d'un ensurt ({\em on c\ldots\
  defineixes la massa del bottom!!!, H\ldots\ no ho s{\'e}!!}), 
 que, afortunadament, gaireb{\'e} sempre s'acaben en res ({\em ufff! {\'e}s al common} {\tt
     Other\_Standard\_Model\_Masses\_new\_2}), 
 tot aix{\`o} gr{\`a}cies al com\-panys
meravellosos del  
{\bf SUSY Team} de l'IFAE, en David, en Ricard i en Toni, si els hagu{\'e}s d'agrair
tot el que m'han ajudat (comen{\c c}ant pels inicis dif{\'\i}cils amb els ordinadors, i
acabant per nombrosos suggeriments i correccions) no cabrien en aquesta plana. 

A la f{\'\i}sica no tot es SUSY, hi ha tamb{\'e} reticles, bombolletes a l'univers, etc., i
gr{\`a}cies a aquestes coses hi ha companys de doctorat que entre caf{\`e}s, sopars, i
costellades, ens ajuden a pujar la moral. 

Gracies tamb{\'e} als membres del Grup de F{\'\i}sica Te{\`o}rica de l'U.A.B. per haver-me ofert la
possibilitat de realitzar-hi el doctorat. Voldria agrair adem{\'e}s la
co{\lgem}aboraci{\'o} del Prof. Wolfgang Hollik en alguns dels treballs presentats
en aquesta tesi.

A la vida no tot es feina, tot que de vegades es barregin les coses. Gr{\`a}cies
Siannah per tot el que has fet, pel teu suport, i tamb{\'e} per les nombroses
correccions i suggeriments a aquest manuscrit. Perdona'm el que durant la seva
preparaci{\'o} no t'hagi tractat com et mereixes.

Retornant al tema familiar, voldria agrair el suport que sempre he rebut per
part dels meus pares, Rosa i Ramon M., els quals sempre m'han ajudat i animat en
tot all{\`o} que he volgut fer.

Aquest treball ha estat possible gr{\`a}cies a la beca de la Generalitat de Catalunya
1995FI-02125PG.
}
\vspace*{\fill}

{
\noindent
\includegraphics{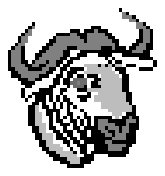}
{\small

\noindent
This Thesis has been written using Free Software.

\noindent
The \LaTeX$2_\varepsilon$\  Typesetting system.

\noindent
Feynman graphs using \texttt{feyn}\textsf{MF}\ --T. Ohl, Comput. Phys. Commun. {\bf 90}
(1995) 340, {\tt hep-ph/9505351}. 

\noindent
Plots using Xmgr plotting tool ({\tt http://plasma-gate.weizmann.ac.il/Xmgr/}).

\noindent
GNU Emacs. 

\noindent
Running in a GNU Linux system -- the Free Software Foundation ({\tt
  http://www.gnu.org}).

}
}
\vspace*{\fill}

\end{titlepage}

%% file: inici.tex

\setcounter{page}{1}
\pagenumbering{Roman}

\input{head_i}

\tableofcontents                 
\cleardoublepage
\input{headings}
\pagenumbering{arabic}
\setcounter{page}{1}


%% file: head_i.tex

\addtolength{\headheight}{0.5cm}
\addtolength{\textheight}{-0.5cm}

\lhead[\fancyplain{}{}]{ \fancyplain{}{}   }
\rhead[ \fancyplain{}{} ]{ \fancyplain{}{}   }
\chead[ \fancyplain{}{}  ]{\fancyplain{}{}}
\cfoot{\fancyplain{\thepage}{\thepage}}

\pagestyle{fancyplain}

%% file: headings.tex

\pagestyle{fancyplain}         


\renewcommand{\chaptermark}[1]{\markboth{#1}{#1}}           
\renewcommand{\sectionmark}[1]{\markright{\thesection\ #1}} 


%
%

\lhead[ \fancyplain{}{\bf\thepage}  ]{ \fancyplain{}{\it\rightmark} }
\rhead[ \fancyplain{}{\it\leftmark} ]{ \fancyplain{}{\bf\thepage}   }
\chead[ \fancyplain{}{}]{\fancyplain{}{} }

%
%
\cfoot{\fancyplain{\thepage}{}} 


%% file: feyn_def.tex
\fmfcmd{%
  style_def mass_boson expr p=
  cdraw (wiggly p);
  shrink (1);
  cfill (arrow p);
  endshrink;
  enddef;
}

\fmfcmd{%
  style_def gluino expr p=
    cdraw (curly p);
  shrink (2);
 draw_plain p;
  cfill (arrow p);
  endshrink;
  enddef;
}




%% file: abstract.tex
\chapter{Abstract}

In this Thesis we have investigated some effects appearing in top quark
observables, in the framework of the Minimal Supersymmetric Standard Model
(MSSM).

The Standard Model (SM) of Strong (QCD) and Electroweak (EW) interactions has
had a great success in describing the nature at the Electroweak scale, and its
validity has been tested up the the quantum level in past and present
accelerators, such as the LEP at CERN or the Tevatron at Fermilab. The last great
success of the SM was the discovery in 1994 of its last matter building block, namely the
top quark, with a mass of $m_t=173.8\pm 3.2 \pm 3.9 \GeV/c^2$. However the
mechanism by which all the SM particles get their mass is still unconfirmed,
since no Higgs scalar has been found yet. The fermions couple to Higgs
particles with a coupling proportional to its mass, so one expects that the
large interactions between top and Higgs particles give rise to large quantum
effects.

We have focused our work in the MSSM. This is an extension of the SM that
incorporates Supersymmetry (SUSY). Supersymmetry is an additional transformation that
can be added in the action of Quantum Field Theory, leaving this action
unchanged. The main phenomenological consequence of it is that to any SM
particle ($p$) there should exist a partner of it, which we call {\em sparticle}
($\tilde p$). This extension of the SM provides elegant solutions to some
theoretical problems of the SM, such as the {\em hierarchy problem}.

We have computed the radiative corrections to some top quark observables, using
the on-shell renormalization scheme, and with a physically motivated definition
of the $\tb$ parameter. $\tb$ is the main parameter of the MSSM, and it governs
the strength of the couplings between the Higgs bosons and the fermion fields
(and its {\em superpartners}).

We have computed the SUSY-EW corrections to the non-standard top decay partial
width into a charged Higgs particle and a bottom quark $\Gamma(t\to H^+b)$. We
have found that these corrections are large  in the moderate and specially in the high regime
of \tb, where they can easily reach values of
$  \delta_{EW} (t\rightarrow H^+\,b) \simeq +30\%$
for negative $\mu$ and a ``light'' sparticle spectrum, and $\delta_{EW} (t\rightarrow H^+\,b) \simeq +20\%$
for positive $\mu$  and heavy sparticle spectrum. In both cases we have singled out
the domain $\mu A_t<0$ of the parameter space, which is the one  preferred by
the experimental data on  radiative B-meson decays ($b\rightarrow s \gamma$). We
have singled out the leading contribution to this corrections, which is the
supersymmetric contribution to the bottom quark mass renormalization constant
$\delta\mb/\mb$. It is
proportional to $-\mu\,A_t$ and shows a possible non-decoupling effect. 
The contributions from Higgs particles is tiny, and can be neglected.
We have
added this corrections to the known Strong corrections
($\delta_{QCD}\simeq-60\%$,  $\delta_{SUSY-QCD} \simeq +80\%$ and
$\delta_{SUSY-QCD} \simeq -40\%$) and we have look at its impact on the
interpretation of the Tevatron data. The standard analysis (using only
$\delta_{QCD}$) implies that for a charged Higgs mass of $\mHp=110\GeV$ the values
of $\tb\geq 50$ are excluded. If this charged Higgs boson belongs to the MSSM 
 the excluded values are $\tb\geq 35$ or $\tb\geq 75$ for the two
scenarios presented above. So no model independent bound on the charged Higgs
mass can be put from present experimental data.

We have looked at the possibility that the top quark could decay via Flavour
Changing Neutral Current (FCNC) into a neutral Higgs particle and a charm
quark. We have computed the EW contributions and the QCD contributions, using
a mass model motivated by Grand Unification Theories (GUT), but not restricted to any
specific GUT. We have included the full interaction lagrangian between all the
particles. The upper theoretical bounds are found to be 
$BR^{\rm SUSY-EW}(\tch) \lsim \mbox{ several }\times10^{-6}$ and 
$ BR^{\rm SUSY-QCD}(\tch)\lsim \mbox{ several }\times10^{-4}$
and the typical values for this ratio are $10^{-8}$ and $10^{-5}-10^{-4}$ for
the \SUSY-\EW\ and \SUSY-\QCD\  induced \FCNC\  decays respectively.
The Higgs and the purely SUSY contributions to the \EW\  induced process are of
the same size, and can be of the same or opposite sign.
We have found that the \SUSY-\QCD\  induced \FCNC\
decay widths are at least two orders of magnitude larger than the \SUSY-\EW\  ones in most
of the parameter space, thus making unnecessary the computation of the
interference terms. The value of this branching ratio is too small to be
measured either at the Tevatron or at the Next Linear Collider (LC)\footnote{\textbf{Note Added}: See however note on section~\ref{sec:con} (pg.\,\pageref{pagenoteLC}).}, but there is
chance that it could be measured at the Large Hadron Collider (LHC).

If bottom-like squarks (the {\em superpartners} of bottom quarks) are heavy
enough they could decay into a top quark and a 
chargino (the {\em superpartners} of gauge bosons and Higgs bosons): 
$\sbottom\to t\cmin$. This could serve as an unexpected source of top quarks at the
Tevatron, at the LHC, or at the LC. We have computed the QCD radiative
corrections and the leading EW corrections to this partial decay width. The QCD
corrections are dominant, they are negative in most of the parameter space, and
are of the order of  $ \delta_{QCD} (\sbottom_1\rightarrow t\,\cmin_1) \simeq -60 \%$,
  $\delta_{QCD} (\sbottom_2\rightarrow t\,\cmin_1) \simeq -20 \% $ for a wide range of the parameter space. 
\EW\  corrections can be of both signs. These corrections have been computed in
the {\em higgsino approximation}, which gives the leading behaviour of the \EW\
corrections.
Our renormalization prescription forces the physical region to a narrow range. Within this restricted region the typical corrections
vary in the range $\delta_{EW} (\sbottom_1\rightarrow t\,\cmin_1) \simeq +25 \% \mbox{ to } -15 \%$
  $\delta_{EW} (\sbottom_2\rightarrow t\,\cmin_1) \simeq +5 \% \mbox{ to } -5 \%$.
However we must recall  that these limits are
qualitative. In the edge of such regions  we find the largest \EW\
contributions. 
We stress that in this case
it is not possible to narrow down the bulk of the corrections to just the
renormalization of the bottom quark Yukawa coupling.

Our general conclusion is that the supersymmetric strong and electroweak radiative
corrections can be very important in the top/bottom-Higgs super-sector of the
\MSSM. Therefore, it is necessary
to account for these corrections in the theoretical computation of the high energy physics
observables, otherwise highly significant information on the potentially
underlying \SUSY\  dynamics could be missed. 
This is true, not only for the future experiments at the LHC and
the LC, but also for the present Run I data (and the Run II data around
the corner) at the Fermilab Tevatron collider.


%% file: intro.tex

\chapter{Introduction}
\label{cap:Intro}

Recently, the Standard Model (\SM) of the strong (\QCD) and electroweak (\EW) 
interactions\,\cite{WeinbergSM,GlashowSM,SalamSM} has been crowned with the discovery of the penultimate
building block of its theoretical structure: the top 
quark, $t$\,\cite{Abe:1995hr,Abachi:1995iq}.
At present the best determination of the
top--quark mass at the Tevatron reads as follows\,\cite{RADCORtuts}:
\begin{equation}
m_t=173.8\pm 3.2 \mbox{ (stat.) } \pm 3.9 \mbox{ (sist.)} \GeV\,.
\label{eq:CDFD0}
\end{equation} 

While the \SM\  has been a most successful framework to describe the
phenomenology of the strong and electroweak interactions
for the last thirty years,
the top quark itself stood, at a purely theoretical level --namely, on
the grounds of requiring internal consistency, such as gauge invariance 
and renormalizability-- as a firm prediction of the \SM\  
since the very confirmation of the existence of the bottom quark and
the measurement of its weak isospin quantum numbers\,\cite{Schaile:1992tm}.
With the finding of the top quark, the matter content of the \SM\  has been fully
accounted for by experiment. Still, the last building block of the \SM,
viz. the fundamental Higgs scalar, has not been found yet, which means that
in spite of the great significance of the
top quark discovery the theoretical mechanism by which all particles 
acquire their masses in the \SM\  remains experimentally unconfirmed. 
Thus, it is not clear at present whether the \SM\  will remain as the last word
in the phenomenology of the strong and electroweak interactions around
the Fermi's scale or whether it will be eventually subsumed within a larger 
and more fundamental theory. The search for physics beyond the \SM, therefore,
far from been accomplished, must continue with redoubled efforts. 
Fortunately, the peculiar nature of the top quark (in particular its large
mass--in fact, perhaps the heaviest particle 
in the \SM!-- and its characteristic interactions with the scalar particles)  
may help decisively to unearth any vestige of physics beyond the \SM. 

We envisage at least four wide avenues of interesting new 
physics potentially conveyed by top quark 
dynamics and which could offer us the clue to solving the nature of
the spontaneous symmetry breaking (\SSB) mechanism, to wit:
\begin{enumerate}
\item The ``Top Mode'' realization(s) of the SSB mechanism, 
i.e. SSB without fundamental Higgs scalars, but rather
through the existence of $t\bar{t}$ condensates\,\cite{Bardeen:1990ds};
\item The extended Technicolour Models; also without Higgs particles, and
giving rise to residual non-oblique interactions of the top quark with
the weak gauge bosons\,\cite{Chivukula:1994mn};
\item The non-linear (chiral Lagrangian)
realization of the $SU(2)_L\times U(1)_Y$
gauge symmetry\,\cite{Georgi84,Dobado:1991zh,Espriu:1992vm}, 
which may either accommodate or dispense with
the Higgs scalars. It can also generate additional (i.e. non-standard)
non-oblique interactions of the top quark with the weak gauge
bosons\,\cite{Peccei:1990kr,Hill:1995di};  
and 
\item The supersymmetric (\SUSY)
realization of the \SM, such as the Minimal Supersymmetric Standard Model
(\MSSM)\,\cite{Nilles:1984ex,Haber:1985rc,Lahanas:1987uc,Ferrara87} (see
also\cite{Martin:1997ns} for a comprehensive review), 
where also a lot of potential new
phenomenology spurred by top and Higgs physics might
be creeping in here and there. 
Hints of this new phenomenology may show up either in the form
of direct or virtual effects from supersymmetric Higgs particles or from 
the ``sparticles'' themselves
(i.e. the $R$-odd\,\cite{Nilles:1984ex,Haber:1985rc,Lahanas:1987uc,Ferrara87}
partners of the \SM\  particles).
\end{enumerate}

In this Thesis, we shall focus our attention on the fourth large avenue of
hypothetical physics beyond the \SM, namely on the (minimal) \SUSY\  extension 
of the \SM, the \MSSM, which is at present the most predictive framework
for physics beyond
the \SM\  and, in contradistinction to all other approaches, it has the virtue of 
being a fully-fledged Quantum Field Theory. Most important, 
on the experimental
side the global fit analyses to all indirect precision
data within the \MSSM\  are comparable to those in the \SM;
in particular,
the \MSSM\  analysis implies that
$\mt=172\pm 5 \GeV$\,\cite{deBoer:1996vq,deBoer:1996hd}, 
a result which is
compatible with the above mentioned experimental
determinations of $\mt$. 

Moreover the \SUSY\  theories are a step forward in the search for an unified
theory. On the ``light energy'' point of view a simple supersymmetrized version
of the \SM\  yields to a unification of the three gauge couplings of the \SM, at
a scale of $\Lambda\simeq10^{16} \GeV$\,\cite{Ellis:1992ri} .
On the ``high energy'' point of view these theories can be embedded in a more
general framework, Superstring Theories, 
these theories
provide a unification of ``Classical'' Quantum Field Theory with  Einstein's
Gravitational Theory. Some \SUSY\  models of unification have been constructed,
that provide unification at large scales,
the most hard constrain that they must fulfill is to reproduce the \SM\  at the
\EW\  scale within the present experimental constrains. Maybe the most popular
of such model is the so called ``Minimal Supergravity''
(mSUGRA)\cite{Nilles:1984ex,Haber:1985rc,Lahanas:1987uc,Ferrara87}. 

If $R$-parity is conserved \SUSY\  (i.e. $R$-odd) particles cannot decay into \SM\  ones (see
section~\ref{sec:MSSMintro}), thus the Lightest Supersymmetric Particle (LSP) is
stable. The LSP is a good candidate for cold dark matter, which
is a necessity of the present cosmological models. Cold dark matter is
necessary  to explain the
flatness of the present universe, and the structure formation mechanism.

\SUSY\  theories also provide answers to some \SM\  questions, they provide a
natural solution to the ``hierarchy problem''\,\cite{Witten:1981nf}, that is,
the impossibility, in the  
\SM, of having two scales (the \EW\  scale, and the unification scale) with a
large gap between them. This is due to the presence of 
quadratic divergences in the one-loop correction of the boson masses. These
divergences appear because of bosonic loops in the mass correction. In \SUSY\
theories each bosonic (fermionic) particle has associated a fermionic (bosonic)
partner, with the same quantum numbers and couplings, thus the fermionic loops
cancel the quadratic divergence of the bosonic loops, and the two scales remain
stable. 

Aside from these facts \SUSY\  theories can be useful also in other
subjects, for example they give us hints about quark
confinement\cite{Seiberg:1994rs}. The excitement is so great that a sole event
at the Tevatron, not expected in the SM, has produced a full analysis of its
expectation as a \SUSY\  event\cite{Ambrosanio:1996zr}.

All these in a hand, it seems that \SUSY\  could be the solution of all our
theoretical problems (or 
our theoretical prejudices) in particle physics, however no supersymmetric
particles have been found in the high energy physics experiments yet, or, to put it
in other words, only ``half'' of the \MSSM\  spectrum has been found (aside from the
Higgs sector). Thus \SUSY\  can not be an exact symmetry of nature at the \EW\
scale, and we would seem forced to abandon this nice framework. However there exist a
mechanism of breaking \SUSY\  without losing its most important properties,
it is called ``Soft-\SUSY-Breaking''\,\cite{Girardello:1982wz}. At scales lower
than the 
Soft-\SUSY-Breaking scale the model can be described by a set of parameters
which 
determine the spectrum of the \SUSY\  partners of known particles. One thinks
that when the masses of the superpartners are large enough the 
supersymmetric particles eventually decouple\cite{Dobado:1997up}, though it has
not been demonstrated yet. 

Soft-\SUSY-Breaking can be realized by means of different mechanisms. Each of
these mechanisms 
provides us with a different set of Soft-\SUSY-Breaking parameters at the \EW\
scale, determined by a small set of parameters at high energies.
 
In this Thesis we will take the point of view that the \MSSM\  is the effective
theory at the \EW\  scale of a more fundamental theory, which we do not know
about, thus we will treat the Soft-\SUSY-Breaking parameters as being arbitrary,
within the allowed experimental range. 

Radiative corrections\cite{Bohm:1986rj,Hollik:1990ii} have shown to be a powerful
tool in particle physics for the last half century.
Recall the first theoretical and experimental determination of the
electron anomalous coupling ($g-2$)\cite{gmenysdos} 
as one of its earliers 
applications, and the measurement of the radiative corrections to precision \EW\
observables (such as the relation $\mz/\mw$) at present high
energy colliders as the most recent one (see e.g.\,\cite{HollikQEMSSMReview}). Radiative
corrections are useful also 
to determine (indirectly) the existence, and the parameters, of particles yet to
be discovered.
As a matter of fact the mass of the top quark was estimated, before its direct
observation at the Tevatron\cite{Abe:1995hr,Abachi:1995iq}, 
with the help of its radiative
corrections to the $\mw-\ssw$ correlation\cite{Blondel:1993jq}. One could think of estimating also the
Higgs mass by this method, unfortunately the one-loop Higgs radiative corrections enter
this observable as the logarithm of its mass\,\cite{Veltman:1977fy}, whereas the
effect of the top 
quark grows quadratically with its mass.

It is a wonderful idea, spread all over the theoretical particle physicist community, 
the use of radiative corrections to determine if there is any physics beyond the
\SM. One can look at the precision observables, taken out of present high energy
colliders, and search for deviations of the \SM. We must note that present
precision data does not present significant deviations from the \SM\
expectations, but this has not been always the case. Some time ago there was a
a large discrepancy between the \SM\  prediction and the experimental
measurement of the hadronic fraction of $Z$ decays into $b\bar{b}$ pairs. This
discrepancy could be
cured by introducing in the theoretical estimate the \SUSY\  radiative
corrections\,\cite{Garcia:1995wv,Garcia:1995ub} (see also a complete study of the $Z$ boson in
the \MSSM\  in\,\cite{Garcia:1995wu,TesiDavid}). Nowadays this mismatch has been brought down to
non-significant deviation (less than 1  standard deviation), however we
learned that using these precision measurements a precise prediction on the
\MSSM\  spectrum could be found trough global fits to electroweak precision data\,\cite{deBoer:1996vq,deBoer:1996hd,RADCORdeBoer}.

In this Thesis we will address the important issue of the \EW\  \SUSY\  effects
on top quark and Higgs boson physics.
The top quark presents a privileged laboratory for \EW\  physics, due
to its large mass~(\ref{eq:CDFD0}), as the Higgs
particle couples to fermions proportionally to its mass. In the case of \SUSY\
theories this privilege is enhanced for several reasons. First of all the Higgs
sector is extended into a so called ``Type II Two-Higgs-Doublet
Model'' (2HDM)\,\cite{Hunter}, and the Yukawa couplings of the top--bottom weak 
doublet become (normalized with respect to the $SU(2)$ gauge coupling)
\begin{equation}
\lambda_t\equiv {h_t\over g}={m_t\over \sqrt{2}\,M_W\,\sin{\beta}}\;\;\;\;\;,
\;\;\;\;\; \lambda_b\equiv {h_b\over g}={m_b\over \sqrt{2}
\,M_W\,\cos{\beta}}\,.
\label{eq:Yukawas} 
\end{equation}
where $\tb$ is the ratio between the vacuum expectation values of the two
neutral scalar Higgs bosons (see chapter~\ref{cap:MSSM}).
Notice that in this extension of th \SM\  it is not only the top quark that can
have large Yukawa interactions with Higgs bosons.
From~(\ref{eq:Yukawas}) we see that at large
$\tb$ ($\gsim 30$) the bottom 
quark Yukawa coupling also becomes important. Second, the presence of the
superpartners of the top and bottom  quarks  (``stop'' and ``sbottom'') and those
of the Higgs bosons (``higgsinos'') raise up a very interesting top-stop-Higgs-higgsino
phenomenology. 

The \SUSY\  radiative corrections to the top quark standard decay
mode into a charged gauge boson and a bottom quark have been known since some
time ago\,\cite{Garcia:1994rq,Dabelstein:1995jt} (see also~\cite{TesiRicci} for 
an exhaustive analysis). Also the conventional strong (QCD) corrections regarding the phenomenology of top and the charged  Higgs are well
known\cite{Czarnecki:1993ig,Czarnecki:1993zm,Mendez:1990jr,Li:1991ag,Djouadi:1995gf},
and its strong \SUSY\ 
radiative corrections have
been studied too\,\cite{TesiRicci,Guasch:1995rn,Jimenez:1996wf}. Thus the following step is to
determine the importance 
of the Yukawa couplings to the top--Higgs sector phenomenology\cite{Coarasa:1996qa}.

The aim of this Thesis is to study the effects of the radiative corrections to
the top--Higgs sector in the \MSSM, by looking at unconventional decay and
production modes. We will show that \EW\  radiative
corrections are important, and this has an effect both in the interpretation
of the present experimental data (Tevatron Run I)\cite{Guasch:1998jc} and on
the prospects of 
measurements in future colliders (Tevatron Run II, Large Hadron Collider -LHC-,
and next Super Linear Collider -LC-)\cite{Guasch:1997kc}.

Moreover one expects that, if \SUSY\  particle exists, they could be an
unexpected source of top quarks at high energy colliders. The observed top quark
production cross section at the Tevatron is equal to the Drell-Yan production
cross-section convoluted over the parton distribution times the squared
branching ratio. Schematically
\begin{equation}
\sigma_{\rm obs.}(p\bar{p}\rightarrow t\bar{t})=\int dq\,d\bar{q}\,\,\sigma (q\,\bar{q}\rightarrow t\,\bar{t})\,
\times |BR (t\rightarrow W^+\,b)|^2\,.
\label{eqn:productionSM}
\end{equation}
However, in the framework of the \MSSM, we rather expect a generalization of this
formula in the following way:
\begin{eqnarray}
\sigma_{\rm obs.}&=&\int dq\,d\bar{q}\,\,\sigma (q\,\bar{q}\rightarrow t\,\bar{t})\,
\times |BR (t\rightarrow W^+\,b)|^2\nonumber\\
&+& \int dq\,d\bar{q}\,\,\sigma (q\,\bar{q}\rightarrow \tilde{g}\,\bar{\tilde{g}})\,
\times |BR (\tilde{g}\rightarrow t\,\bar{\tilde{t}}_1)|^2\times
|BR (t\rightarrow W^+\,b)|^2\nonumber\\
&+& \int dq\,d\bar{q}\,\,\sigma (q\,\bar{q}\rightarrow 
\tilde{b}_a\,\bar{\tilde{b}}_a)\,
\times |BR (\tilde{b}_a\rightarrow t\,\chi^-_1)|^2\times
|BR (t\rightarrow W^+\,b)|^2+\ldots\,,
\label{eq:productionMSSM}
\end{eqnarray}
where $\tilde{g}$ stand for the gluinos,
$\tilde{t}_1$ for the lightest stop and $\tilde{b}_a (a=1,2)$ for the sbottom quarks.
One should also include electroweak and \QCD\  radiative corrections to all these
production cross-sections within the \MSSM\@. 
For some of these processes
calculations already exist in the literature showing that one-loop effects
can be important on sparticle
production\,\cite{Kim:1996nz,Yang:1996dm,Beenakker:1996nd,Hopker:1996kx} as well
as on sparticle decays, both the \QCD\,\cite{Djouadi:1997wt,Kraml:1996kz} and
the \EW\,\cite{Guasch:1998as} \MSSM\  corrections.

Thought we have been mainly interested in a scenario where the charged Higgs
particle is lighter than the top quark, an obvious question is what would happen if this charged Higgs is heavier than
the top. We have considered this issue in Ref.\cite{Coarasa:1997ky} (see also an exhaustive analysis in\,\cite{TesiToni}).
The radiative corrections in the top-Higgs sector in the \MSSM\  should be
compared with those from the generic 2HDM's. We have been interested in these
extensions of the \SM\  in Ref.\,\cite{Coarasa:1998xy} and more work is
currently in progress.
The main result is that if a charged Higgs boson is
found, one could discriminate to what kind of model it belongs by using radiative
corrections. These calculations of radiative corrections are in the line of
completing (within the same order of perturbation theory) our previous studies
of the full set of three-body decays of the top quark in the
\MSSM\,\cite{Guasch:1997dk}.

The structure of this Thesis is as follows: in chapter~\ref{cap:MSSM} we give the
basic notation of the \MSSM\  used throughout this Thesis; in
chapter~\ref{cap:Renorm} we explore the renormalization of the \MSSM, extending
the well known formalism used in the \SM\cite{Bohm:1986rj,Hollik:1990ii}, and using a physically
motivated renormalization prescription; chapters~\ref{cap:tbh}
to~\ref{cap:sbdecay} deal with explicit effects of the one-loop corrections on
some physical processes of top quark production and decay; and finally in
chapter~\ref{conclusions} we present the general conclusions. 
At the end we include 
an appendix with some technical details and notation.


%% file: mssm.tex
\chapter{The \mssm\  (\MSSM)}
\label{cap:MSSM}

\section{Introduction}
\label{sec:MSSMintro}

It goes beyond the scope of this Thesis to study the
formal theory of Supersymmetry\,\cite{WessANDBagger,1001lessons}, 
however we would like, at least, to give a feeling
on what is it. 

Supersymmetry (\SUSY) 
can be introduced in
many manners, maybe the most straightforward one is adding to the space-time
coordinates $(t,\vec{x})$ another set of coordinates 
$\theta_\alpha$ ($\alpha=1,\ldots,n$ the space-time dimension) that are Grassmann
variables, i.e.\  they anti-commute. Now the 
general ``rotations'' in this space are a superset of the Poincar{\'e}
transformations of space-time. It is clear that being $\theta_\alpha$ Grassmann
variables the
generators of the rotations that involve these coordinates will behave in a
special way, and indeed they do. These generators (usually called $Q_\alpha$)
anti-commute with themselves, so they do not form an Algebra, but a 
{\em  Super-Algebra}, and the \SUSY\  transformations do not form a Lie Group.
However it turns out that it is the only external transformation
that can be added to the Poincar{\'e} Group, and leave the Scattering (${\cal S}$)
matrix untransformed. One can add as many ``supersymmetries'' (i.e.\  sets of
$\theta$ variables) as the dimension of the space-time, thus if we introduce a
single set of $\theta$ it is said that we have a $N=1$ supersymmetry, and so
on. The structure of the full set of coordinates $(t,\vec{x},\theta_\alpha)$ is
called {\em Superspace}.

The functions defined in the Superspace are polinomic functions of the $\theta$
variables (since $\theta_\alpha^2=0$). Thus we can decompose the functions
(superfields) of this 
Superspace in components of $\theta^0$, $\theta_\alpha$, 
$\theta_\alpha \theta_\beta$, \ldots each of these 
components will be a function of the space-time coordinates. Analogously to the
space-time, we can define in the Superspace scalar superfields, vector
superfields, \ldots For example in a 4-dimensional space 
time with $N=1$ 
supersymmetry a scalar superfield has 10 components. 

There can be defined fields with specific properties with respect to the
$\theta$ variables. We are interested in the {\it chiral} fields. A scalar
chiral field in a $4D$ $N=1$ Superspace has 4 components, two of them (the
components of $\theta_\alpha$) can be associated to be the components a Weyl spinor,
the component of $\theta^0$ is a scalar field, and the $\theta_\alpha\theta_\beta$ component is
the so called ``auxiliary'' field. This auxiliary field is not a dynamical field
since its equations of motion do not involve time derivatives. To this end we
are left with a superfield, whose components represent an ordinary scalar field
and an ordinary chiral spinor. So if nature is described by the dynamics of this
field we would find a chiral fermion and a scalar with identical quantum
numbers. That is {\em Supersymmetry relates particles which differ by spin
  1/2}. Had we started with a $N=2$ \SUSY\  we would end with a set of
particles of spin $0$, $1/2$ and  $1$ as a part of the same scalar superfield, this is
called a {\em Supermultiplet}. When a \SUSY\  transformation ($Q$) acts on a
superfield it transform spin $s$ particles into spin $s\pm1/2$ particles.

Thus, for a $N=1$ \SUSY, we find that to any chiral fermion there should be a
scalar particle with exactly the same properties. This fact is on the basis of
the absence of quadratic divergences in boson mass renormalization, since for
any loop diagram involving a scalar particle there should be a fermionic loop
diagram, which will cancel quadratic divergences between each other, though
logarithmic divergences remain.

Supersymmetric interactions can be introduced by means of generalized
gauge transformations, and by means of a generalized potential function, the
Superpotential, which give rise to masses, Yukawa-type interactions, and a scalar
potential. 

As no scalar particles have been found at the electroweak scale we may infer
that, if \SUSY\  exists, it is broken. We can allow \SUSY\  to be broken
maintaining the property that no quadratic divergences are allowed: this is the
so called Soft-\SUSY-Breaking mechanism\,\cite{Girardello:1982wz}. We can achieve this by only introducing
a small set of \SUSY-Breaking terms in the Lagrangian, to wit: masses for the
components of lowest spin of a supermultiplet; and triple scalar
interactions. However, other terms like explicit fermion masses for the matter
fields would 
violate the Soft-\SUSY-Breaking condition.

The \MSSM\  is the minimal Supersymmetric extension of the Standard Model. It is
introduced by means of a $N=1$ \SUSY, with the minimum number of new
particles. Thus for each fermion $f$ of the \SM\  there are two scalars related
to its chiral components called ``sfermions'' ($\tilde{f}_{L,R}$), for each
gauge vector $V$ there is also a chiral fermion: ``gaugino'' ($\tilde{v}$), and
for each Higgs scalar $H$ another chiral fermion: ``higgsino'' ($\tilde{h}$). 
In the \MSSM\  it turns out that, in order to be able of giving masses to up-type and down-type
fermions, we must introduce two Higgs doublets with opposite hypercharge, and so
the \MSSM\  Higgs sector is of the so called Type II (see section~\ref{sec:hmas}
and Ref.\,\cite{Hunter}).

We can define the following quantum number
$$
R=(-1)^{2S+L+3B}\,\,,S\equiv {\rm spin}\,\, ,L\equiv\mbox{lepton
  number}\,\,,B\equiv\mbox{barion number}\,\, ,
$$
called $R$-parity, which is $1$ for the \SM\  fields and $-1$ for its
supersymmetric partners. 
In the way the \MSSM\  is implemented $R$-parity is conserved, this means that
$R$-odd particles (the superpartners of \SM\  particles) can only be created in
couples, also that in the final product decay of an $R$-odd particle at least
one \SUSY\  particle exists, and that the Lightest Supersymmetric Particle (LSP)
is stable.

In this chapter we review the \MSSM\  at the tree-level: its field content~(in
sec.~\ref{sec:MSSMfield}); its Lagrangian in the Electroweak
basis~(sec.~\ref{sec:MSSMlagrangian}); its mass
spectrum~(sec. \ref{sec:MSSMspectrum}); in 
section~\ref{sec:MSSMinteractions} the interactions in the mass eigenstate
basis; and finally we make a short
revision of the experimental constraints on the parameters in
section~\ref{sec:limits}.

\section{Field content}
\label{sec:MSSMfield}
The field content of the \MSSM\  consist of the fields of the \SM\  plus all their
supersymmetric partners, and an additional Higgs doublet, so the superfield
content of the model will be:
\begin{itemize}
\item the matter fields:
  \begin{equation}
    \label{eq:smatter}
    \ba{ccccc}
    L=\left(\ba{c}\nu\\ l^-\ea\right)
    &R=l^+_L&Q=\left(\ba{c}u\\d\ea\right)_L&D=d_L^c&U=u_L^c\,\,,\\
    \tilde L=\left(\ba{c}\tilde \nu\\ \tilde l^-\ea\right)_L 
    &\tilde R= \tilde l^+_R&\tilde Q=\left(\ba{c}\tilde u\\ \tilde d\ea\right)_L
    & \tilde D= \tilde d_R^*
    & \tilde U= \tilde u_R^* \,\,,\\
    Y=-1&Y=2&Y=\frac{1}{3}&Y=\frac{2}{3}&Y=-\frac{4}{3}\,\,,
    \ea
  \end{equation}
for each generation of fermions
\item the gauge superfields, which in the Wess-Zumino gauge consist of:
  \begin{equation}
    \label{eq:sgauge}
    \ba{cc}
    W_1^\mu W_2^\mu W_3^\mu & \tilde w_1 \tilde w_2 \tilde w_3 \,\,,\\
    B^0_\mu & \tilde B^0 \,\,,\\
    g_\mu & \sg \,\,,
    \ea
  \end{equation}
\item and the two Higgs/higgsino doublets:
  \begin{equation}
    \label{eq:shiggs}
    \label{eq:H1H2}
    \ba{cc}
    H_1=\left(\ba{c}H_1^0 \\ H_1^- \ea \right) 
    & H_2=\left(\ba{c}H_2^+ \\ H_2^0 \ea \right)\,\,, \\
    \tilde H_1=\left(\ba{c}\tilde H_1^0 \\ \tilde H_1^- \ea \right) 
    & \tilde H_2=\left(\ba{c}\tilde H_2^+ \\ \tilde H_2^0 \ea \right) \,\,,\\
    Y=-1 & Y=1\,\,.
    \ea
  \end{equation}
\end{itemize}

All these fields suffer some mixing, so the physical (mass eigenstates) fields
look much different from these ones. The gauge fields mix up to give the well
known gauge bosons of the \SM, 
$W^\pm_\mu$, $Z^0_\mu$, $A_\mu$, 
the gauginos and higgsinos mix up to give the
chargino and neutralino fields, and finally the Left- and Right-chiral sfermions
mix 
among themselves in sfermions of indefinite chirality. Let aside the
intergenerational mixing between fermions and sfermions that give rise to the
well known Cabibbo-Kobayashi-Maskawa (\ckm) matrix. For the sake of simplicity
in most of our work we will take no intergenerational mixing, except in
chapter~\ref{cha:fcnc}, where we make an analysis of some \FCNC\  effects.

\section{Lagrangian}
\label{sec:MSSMlagrangian}
The \MSSM\  interactions come from three different kinds of sources:

\begin{itemize}
\item{Superpotential: 
\begin{equation}
W=\epsilon_{ij}\left[ f \hat H_1^i \hat L^j \hat R
+h_d \hat H_1^i \hat Q^j \hat D
+h_u \hat H_2^j \hat Q^i \hat U-\mu \hat H_1^i \hat H_2^j
\right]\,\,.\\
\label{eq:superp}\end{equation}
The superpotential contributes to the interaction Lagrangian~(\ref{eq:ltotal}) 
with two different kind of interactions. The first one is the Yukawa
interaction, which is obtained from~(\ref{eq:superp}) just replacing two of the
superfields by its fermionic field content, whereas the third superfield is
replaced by its scalar field content:
\begin{equation}
\ba{lcl}
V_Y&=&\epsilon_{ij}\left[ f  H_1^i L^j R
+h_d H_1^i Q^j D
+h_u H_2^j Q^i U-\mu \tilde H_1^i \tilde H_2^j
\right]\\
~&~&+\epsilon_{ij}\left[ f \tilde H_1^i L^j \tilde  R
+h_d \tilde  H_1^i Q^j \tilde D
+h_u \tilde H_2^j Q^i \tilde U
\right]\\
&~&+\epsilon_{ij}\left[ f \tilde H_1^i \tilde L^j R
+h_d \tilde  H_1^i \tilde Q^j D
+h_u \tilde H_2^j \tilde Q^i U
\right]\\
~&~&+\mbox{ h.c.}\,\,.
\ea\label{eq:vyukawa}\end{equation}
The second kind of interactions are obtained by means of taking the
derivative of the superpotential:
\begin{equation}
V_W=\sum_i \left|\frac{\partial W\left(\varphi\right)}{\partial
    \varphi_i}\right|^2\,\,,
\label{eq:superder}\end{equation}}
$\varphi_i$ being the scalar components of superfields.
\item{Interactions related to the gauge symmetry, which contain:
\begin{itemize}
\item the usual gauge interactions 
\item the gaugino interactions:
\begin{equation}
V_{\tilde G \psi \tilde \psi}=
i \sqrt{2} g_a \varphi_k \bar \lambda ^a \left( T^a \right)_{kl}
\bar\psi_l+\mbox{ h.c.}
\label{eq:vgfsf}\end{equation}
where $(\varphi,\psi)$ are the spin $0$ and spin $1/2$ components of a chiral
superfield respectively, $T^a$ is a
generator of the gauge symmetry, $\lambda_a$ is the gaugino field and $g^a$ 
its coupling constant.
\item and the $D$-terms, related to the gauge structure of the theory, but that
  do not contain neither gauge bosons nor gauginos:
\begin{equation}
V_D=\frac{1}{2}\sum D^a D^a\,\,,
\label{eq:vd}\end{equation}
with
\begin{equation}
D^a= g^a \varphi_i^* \left(T^a\right)_{ij} \varphi_j \,\,,
\end{equation}
$\varphi_i$ being the scalar components of the superfields.
\end{itemize} }
\item{Soft-\SUSY-Breaking interaction terms:
\begin{equation}
V_{\rm soft}^{\rm I}=\frac{g}{\sqrt{2} M_W \cos{\beta}} \epsilon_{ij}\left[
m_l A_l H_1^i \tilde L^j \tilde R+
m_d A_d H_1^i \tilde Q^j \tilde D-m_u A_u H_2^i \tilde Q \tilde U
\right]+\mbox{ h.c. } \,\,.
\label{eq:softsusy}\end{equation}
The trilinear Soft-\SUSY-Breaking couplings $A_f$ can play an important role,
specially for the third generation interactions and masses, and they are in the
source of the large value of the bottom quark mass renormalization effects (see
section~\ref{sec:tbhnumeric}).}
\end{itemize}

The full \MSSM\  Lagrangian is then:
\begin{eqnarray}
{\cal L}_{\rm MSSM}&=&
{\cal L}_{\rm Kinetic}+
{\cal L}_{\rm Gauge}
-V_{\tilde G \psi \tilde \psi}-V_D-
V_Y-\sum_i \left|
\frac{\partial W\left(\varphi\right)}{\partial \varphi_i}\right|^2\nonumber\\
~&~&
-V_{\rm soft}^{\rm I}
-m_1^2\, H_1^\dagger H_1
-m_2^2\, H_2^\dagger H_2
-m_{12}^2\, \left(H_1 H_2+H_1^\dagger H_2^\dagger\right)\nonumber\\
~&~&
-\frac{1}{2}m_{\sg}\, \psi^a_{\sg} \psi^a_{\sg}
-\frac{1}{2}M \,\tilde w_i \tilde w_i
-\frac{1}{2}M^\prime\, \tilde B^0 \tilde B^0\nonumber\\
~&~&-m_{\tilde L}^2\, \tilde L^* \tilde L
-m_{\tilde R}^2\, \tilde R^* \tilde R
-m_{\tilde Q}^2\, \tilde Q^* \tilde Q
-m_{\tilde U}^2\, \tilde U^* \tilde U
-m_{\tilde D}^2\, \tilde D^* \tilde D\,\,,
\label{eq:ltotal}\end{eqnarray}
where we have also included the Soft-\SUSY-breaking masses.

From the Lagrangian~(\ref{eq:ltotal}) we can obtain the full \MSSM\  spectrum,
as well as the interactions, which contain the usual \SM\  gauge interactions,
the fermion-Higgs interactions that correspond to a Type II Two-Higgs-Doublet
Model~\cite{Hunter}, and the pure \SUSY\  interactions. A very detailed
treatment of this Lagrangian, and the process of derivation of the forthcoming
results can be found in\,\cite{TesiJefe}.

\section{\MSSM\  spectrum}
\label{sec:MSSMspectrum}
\subsection{Higgs sector}
\label{sec:hmas}
When a Higgs doublet is added to the \SM\  there exist two possibilities for
incorporating it, avoiding Flavour Changing Neutral Currents (\FCNC) at tree
level\,\cite{Hunter}. The first possibility is not to allow a coupling between
the second doublet and the fermion fields, this is the so called Type I
\thdm. The second possibility is to allow both Higgs doublets to couple with
fermions, the first doublet only coupling to the Right-handed down-type
fermions, and the second one to Right-handed up-type fermions, this is the so
called Type II \thdm. 

The Higgs sector of the \MSSM\  is that of a Type II \thdm\,\cite{Hunter}, with some
\SUSY\  restrictions. After expanding~(\ref{eq:ltotal})
the Higgs potential reads 
\begin{eqnarray}
V &=& m_1^2\,|H_1|^2+m_2^2\,|H_2|^2-m_{12}^2\,\left(
\epsilon_{ij}\,H_1^i\,H_2^j+{\rm h.c.}\nonumber\right)\\
&+&{1\over 8}(g^2+g'^2)\,\left(|H_1|^2-|H_2|^2\right)^2
+{1\over 2}\,g^2\,|H_1^{\dagger}\,H_2|^2\,.
\label{eq:potential}
\end{eqnarray}
The neutral Higgs bosons fields acquire a vacuum expectation
value (\vev),
\begin{equation}
\langle H_1 \rangle_0=\pmatrix{v_1\cr0}
 ~~ \langle H_2 \rangle_0 =\pmatrix{0\cr v_2}\,\,.
\label{eq:VeVfields}\label{eq:vv1v2}
\end{equation}
We need two physical parameters in order to know their value, which are usually
taken to be
\begin{equation}
M_W^2=\frac{1}{2} g^2 (v_1^2+v_2^2)\equiv g^2\frac{v^2}{2}~~,~~\tan\beta=\frac{v_2}{v_1}\,\,.
\label{eq:VEV}\end{equation}

These \vev's make the Higgs fields to mix up. There are five physical Higgs
fields: 
a couple of charged Higgs bosons ($H^\pm$); a pseudoscalar Higgs ($CP=-1$) 
$A^0$; and two scalar Higgs bosons ($CP=1$) $H^0$ (the heaviest) and  $h^0$ (the
lightest). There are also the Goldstone bosons $G^0$ and $G^\pm$. The relation
between the physical Higgs fields and that fields of~(\ref{eq:shiggs}) is
\begin{equation}
\ba{lcl}
H_1^-&=&-(\cos\beta\, G^--\sin\beta\, H^-) \,\,,\\
H_2^+&=&\sin\beta\,G^++\cos\beta\, H^+ \,\,,\\
H_1^0&=&v_1+\frac{1}{\sqrt{2}} \left(
\cos \alpha\, H^0-\sin\alpha\, h^0-i(\cos\beta\, G^0- \sin \beta \, A^0)\right)\,\,,\\
H_2^0&=&v_2+\frac{1}{\sqrt{2}} \left(
\sin \alpha \, H^0+\cos\alpha\, h^0+i (\sin\beta \, G^0+\cos \beta\, A^0)\right)\,\,,
\ea
\label{eq:HiggsDefin}
\end{equation}
where $\alpha$ is given in~(\ref{eq:alfa})\cite{Hunter}.

All the masses of the Higgs sector of the \MSSM\  can be obtained with only two
parameters, the first one is $\tb$~(\ref{eq:VEV}), and the second one is a
mass; usually this second parameter is taken to be either the pseudoscalar Higgs
mass $\mA$ or the charged Higgs mass $\mHp$. We will take the last option, as
the charged Higgs plays an important role in most of our
studies. From~(\ref{eq:potential}) one can obtain the
tree-level mass relations between the different Higgs particles,
\begin{eqnarray}
\mAs&=&\mHps-\mws \nonumber\,\,,\\ 
M_{H^0,h^0}^2&=&\frac{1}{2} \left(
\mAs+\mzs\pm\sqrt{\left(\mAs+\mzs\right)^2-4\,\mAs\,\mzs
\cos^2 2\beta}
\right)\,\,,
  \label{eq:mh}
\end{eqnarray}
and the mixing angle between the two scalar Higgs is obtained by means of:
\begin{equation}
\cos 2\alpha= - \cos 2 \beta
\left(\frac{\mAs-\mzs}{\mHs-\mhs}\right)\,\,,\,\,
\sin 2 \alpha=- \sin 2 \beta \left(\frac{\mHs+\mhs}{\mHs-\mhs}\right)\,\,.
\label{eq:alfa}
\end{equation}

\subsection{The \sm\  sector}
\label{sub:massasm}
In this section we give some expressions to obtain some \MSSM\  parameters as a
function of the \sm\  parametrization.

As stated above~(sec. \ref{sec:hmas}) the \vev's can be obtained by means
of~(\ref{eq:VEV}), and the $Z$ mass can be obtained at tree-level by the relation:
$$
\sin^2 \theta_W=1-\frac{M_W^2}{M_Z^2}\,\,.
$$

Fermion masses are obtained from the Yukawa potential~(\ref{eq:vyukawa}) letting
the neutral Higgs fields acquire their \vev~(\ref{eq:VeVfields}). The up-type
fermions get their masses from the $H_2^0$ whereas $H_1^0$ gives masses to
down-type fermions, so
$$
  m_u=h_u v_2 = \frac{h_u \sqrt{2} M_W \sin\beta}{g}\,\,,\,\, 
  m_d=h_d v_1 = \frac{h_d \sqrt{2} M_W \cos\beta}{g}\,\,,
$$
and the Yukawa coupling can be obtained as
\begin{equation}
\lambda_u=\displaystyle\frac{h_u}{g}=  \frac{ m_u}{\sqrt{2} M_W \sin\beta}\,\,,\,\, 
\lambda_d=\displaystyle\frac{h_d}{g}=\frac{ m_d}{\sqrt{2} M_W \cos\beta}\,\,.
\label{eq:Yukawasgeneric}
\end{equation}
\subsection{Sfermion sector}
\label{sec:sfmas}
The sfermion mass term is obtained from the derivative of the
superpotential~(\ref{eq:superder}), the $D$-terms~(\ref{eq:vd}) and the 
Soft-\SUSY-Breaking terms~(\ref{eq:ltotal}) letting the
neutral Higgs 
fields get their \vev~(\ref{eq:VeVfields}), and one obtain the following mass
matrices:
\begin{equation}
  {\cal M}_{\tilde{q}}^2 =\pmatrix{
    M_{\tilde{q}_L}^2+m_q^2+\cos{2\beta}(\TqL-\Qq s_W^2) M_Z^2 & 
    m_q M_{LR}^q\cr 
    m_q M_{LR}^q &
    M_{\tilde{q}_R}^2+m_q^2+ \cos{2\beta}\,\Qq\,s_W^2\, M_Z^2
    }\,,
  \label{eq:stopmatrix}
  \label{eq:sbottommatrix}
  \label{eq:mmsu2}
  \label{eq:mmsd2}
\end{equation}
being $Q$ the corresponding fermion electric charge, $\TqL$ the third component
of weak isospin,  $M_{{\tilde{q}}_{L,R}}$ the Soft-\SUSY-Breaking squark
masses~\cite{Nilles:1984ex,Haber:1985rc,Lahanas:1987uc,Ferrara87} (by
$SU(2)_L$-gauge invariance, we must have
$M_{\tilde{t}_L}=M_{\tilde{b}_L}$, whereas $M_{{\tilde{t}}_R}$,
$M_{{\tilde{b}}_R}$ are in general independent parameters),
$s_\theta=\sin\theta_W$, and 
\begin{equation}
M_{LR}^u=A_u-\mu \cot{\beta}\,\,,\,\,\nonumber\\
M_{LR}^d=A_d-\mu \tan{\beta}\,\,.
\label{eq:MLRdefinition}
\end{equation}
We define the sfermion mixing matrix as 
($\tilde{q'}_a=\{\tilde{q'}_1\equiv \tilde{q}_L,\,\, \tilde{q'}_2\equiv
\tilde{q}_R\}$ are the weak-eigenstate squarks, and $\sq_a=\{\sq_1,\sq_2\}$
are the mass-eigenstate squark fields)
\begin{eqnarray}
 \tilde{q'}_a&=&\sum_{b}
     R_{ab}^{(q)}\tilde{q}_b,\nonumber\\ R^{(q)}&
     =&\pmatrix{ \cos{\theta_q} & -\sin{\theta_q} \cr 
         \sin{\theta_q} & \cos{\theta_q}}\,.
\label{eq:rotation}
\end{eqnarray}
\begin{equation}
R^{(q)\dagger} {\cal M}_{\tilde{q}}^2 R^{(q)}=
      {\rm diag}\{m_{\tilde{q}_2}^2,m_{\tilde{q}_1}^2\}
\ \ \ \ \ (m_{\tilde{q}_2}\geq m_{\tilde{q}_1})\,, 
\end{equation}    
\begin{equation}
 \tan2{\theta_q} = {2\,m_q\,M_{LR}^q\over
M_{\tilde{q}_L}^2-M_{\tilde{q}_R}^2+\cos{2\beta}(\TqL-2\Qq s_W^2) M_Z^2}\,.
       \label{eq:thetarot}
\end{equation}

From eq.~(\ref{eq:mmsu2})  we can see that the sfermion
mass is dominated by the Soft-\SUSY-Breaking parameters  ($M_{\tilde f}
\gg m_f \mbox{ for } f \not= \mbox{top}$), and that the non-diagonal terms could
be neglected, except in the case of the top squark (and bottom squark at large \tb), however we will maintain
those terms, the reason is that, although the $A$ parameters do not play any
role when computing the sfermion masses, they do play a role in the
Higgs-sfermion-sfermion coupling -see eq.~(\ref{eq:Matrixhsfsf})-, and thus it
has an effect on the Higgs self-energies. Moreover these $A$ parameters are
constrained by the approximate (necessary) condition of absence of
colour-breaking minima,
\begin{equation}
A_q^2<3\,(m_{\tilde{t}}^2+m_{\tilde{b}}^2+M_H^2+\mu^2)\,,
\label{eq:necessary}
\end{equation}
where $m_{\tilde{q}}$
is of the order of the average squark masses for
$\tilde{q}=\tilde{t},\tilde{b}$~\cite{Frere:1983ag,Claudson:1983et,Kounnas:1984td,Gunion:1988qv}.

All the Soft-\SUSY-Breaking parameters are free in the strict \MSSM, however some
simplifications must be done to be able of making a comprehensive numerical
analysis. As the main subject of study are the third generation squarks we 
make a separation between that and the rest of sfermions. This separation is
justified by the evolution of the squark masses from the (supposed) unification
scale down to the electroweak scale~\cite{Martin:1997ns} (see also
section~\ref{sec:parameters} for a more detailed discussion).

So we will use the following approximations:
\begin{itemize}
\item equality of the diagonal elements of eq.~(\ref{eq:stopmatrix})  
  \begin{equation}
    \label{eq:msqDiag}
    {\cal M}_{\tilde{q}}^2{}_D\equiv{\cal M}_{\tilde{q}}^2{}_{11}={\cal M}_{\tilde{q}}^2{}_{22}\,\,,
  \end{equation}
  for each charged slepton and each squark of the
  the first and second generation.
\item the up and charm type sfermions share the same value of the
  parameter~(\ref{eq:msqDiag}).
\item the first and second generation squarks share the same value of the $A$
  parameter~(\ref{eq:MLRdefinition}).
\item sleptons also share the same value for~(\ref{eq:msqDiag}) and $A$ parameters.
\end{itemize}

\subsection{Charginos and neutralinos}
\label{sec:cnmas}
Gauginos and higgsinos develop mixing due to the breaking of the gauge
symmetry. To find the mass eigenstates we construct the following sets of
two-component Weyl spinors
\begin{equation}
\ba{lcl}
\Gamma^+&\equiv&(-i \tilde W^+,\tilde H_2^+) \,\,,\\
\Gamma^-&\equiv&(-i \tilde W^-,\tilde H_1^-) \,\,,\\
\Gamma^0&\equiv&(-i \tilde B^0,-i \tilde W_3^0,\tilde H_2^0,\tilde H_1^0) \,\,.
\ea
\end{equation}
Then from~(\ref{eq:vyukawa}) (higgsino mass parameter $\mu$), the
Soft-\SUSY-Breaking masses~(\ref{eq:ltotal}) (gaugino mass terms $M$,
$M^\prime$), and 
replacing the Higgs fields by its \vev's in~(\ref{eq:vgfsf}), we obtain the
following chargino and neutralino mass Lagrangian
\begin{equation}
{\cal L}_M=-\frac{1}{2}\pmatrix{\Gamma^+,\Gamma^-}
\pmatrix{
0&{\cal M}\cr
{\cal M}^T&0
}
\pmatrix{ \Gamma^+\cr \Gamma^-}
-\frac{1}{2} \pmatrix{\Gamma_1,\Gamma_2,\Gamma_3,\Gamma_4}
{\cal M}^0 \pmatrix{\Gamma_1\cr \Gamma_2 \cr \Gamma_3 \cr \Gamma_4}
+\mbox{ h.c.}\,\,,
\end{equation}
where we have defined
\begin{equation}
{\cal M}=\pmatrix{
M&\sqrt{2} M_W \cos{\beta} \cr
\sqrt{2} M_W \sin{\beta}&\mu
}\,\,,
\label{eq:mmassacplus}\end{equation}
\begin{equation}
{\cal M}^0=
\pmatrix{
M^\prime&0&M_Z\sin{\beta}s_\theta&-M_z\cos{\beta}s_\theta \cr
0&M&-M_Z\sin{\beta}c_\theta&M_Z\cos{\beta}c_\theta\cr
M_Z\sin{\beta}s_\theta&-M_Z\sin{\beta}c_\theta&0&-\mu \cr
-M_Z\cos{\beta}s_\theta&M_Z\cos{\beta}c_\theta&-\mu&0
}\,\,\,.
\label{eq:mmassaneut}\end{equation}
We shall assume a grand unification relationship between the gaugino parameters
\begin{equation}
\frac{M^\prime}{M}=\frac{5}{3} \tan^2 \theta_W\,\,\,.
\label{eq:GauginoMassRelation}
\end{equation}

The mass matrices~(\ref{eq:mmassacplus}) and~(\ref{eq:mmassaneut}) are
diagonalized by
\begin{equation}
\ba{lcccl}
U^* {\cal M} V^\dagger&=&{\cal M}_D&=&{\rm diag}\left(M_1,M_2\right)\,\,,\\
 N^*{\cal M}^0 N^\dagger &=&{\cal M}^0_D&=&
{\rm diag}\left(M_1^0,M_2^0,M_3^0,M_4^0\right)\,\,,
\ea
\end{equation}
where $U$, $V$ and $N$ are in general complex matrices that define the mass
eigenstates
\begin{equation}
  \Psi_i^{+}= \pmatrix{
    U_{ij}\Gamma_j^{+} \cr V_{ij}^{*}\bar{\Gamma}_j^{-}
    }
  \; \;\; \;\;,\;\;\;\;\;
  \Psi_i^{-}= C\bar{\Psi_i}^{-T} =\pmatrix{
    V_{ij}\Gamma^{-}_j \cr U_{ij}^{*}\bar{\Gamma}_j^{+} 
    }\ ,  
  \label{eq:cinos}
  \nonumber
\end{equation}
and
\begin{equation}
  \Psi_{\alpha}^0= \pmatrix{
    N_{\alpha\beta}\Gamma_{\beta}^0 \cr 
    N_{\alpha\beta}^{*}\bar{\Gamma}_{\beta}^0
    } =  
  C\bar{\Psi}_{\alpha}^{0T}\ .
  \label{eq:ninos} 
  \nonumber
\end{equation}
In practice, we have performed the calculation with real matrices
$U$, $V$ and $N$, so we have been using unphysical mass-eigenstates
(associated to non-positively definite chargino-neutralino masses).
The transition from our unphysical mass-eigenstate basis
$\{\Psi\}\equiv\{\Psi_i^{\pm}, \Psi_\alpha^0\}$ into the physical
mass-eigenstate basis 
$\{\chi\}\equiv\{\chi_i^{\pm}, \chi_\alpha^0\}$ can be done by
introducing a set of
$\epsilon$ parameters as follows: for every chargino-neutralino $\Psi$ 
whose mass matrix eigenvalue are $M_i, M_{\alpha}$,
the proper physical state, $\chi$, is given by
\begin{equation}
\chi=\left\{\ba{ll}
\Psi&\,\ {\rm if}\ \ \epsilon=1\\
\pm\gamma_5\,\Psi &\,\ {\rm if}\ \ \epsilon=-1\,,
\ea\right.
\end{equation}
and the physical masses for charginos and neutralinos are 
$m_{\chi^{\pm}_i}=\epsilon\,M_i$ and $m_{\chi_{\alpha}^0}=\epsilon\,M_{\alpha}^0$,
respectively. Needless to say, 
in this real formalism one is supposed to
propagate the $\epsilon$ parameters accordingly 
in all the relevant couplings, as shown 
in detail in Ref.\,\cite{Guasch:1997dk,Tesina}. 
This procedure is
entirely equivalent\,\cite{Gunion:1986yn} 
to use complex diagonalization matrices insuring that physical
states are characterized by a set of positive-definite mass eigenvalues; 
and for this reason we have
maintained the complex notation in all our formulae.
Whereas for computations with real sparticles the
distinction matters\,\cite{Guasch:1997dk,Tesina}, 
for virtual sparticles 
the $\epsilon$ parameters cancel out, and so one could use either basis
$\{\Psi\}$ or $\{\chi\}$ without the inclusion of the
$\epsilon$ coefficients.
We have stressed here the differences between the two bases just to make clear
what are the physical 
chargino-neutralino states, when they are referred to in the text.   

\section{Interactions in the mass-eigenstate basis}
\label{sec:MSSMinteractions}

We need to convert the interaction Lagrangian presented in
section~\ref{sec:MSSMlagrangian} to a Lagrangian in the mass-eigenstate basis,
which is the one used in the computation of the physical quantities. As the
expression for the full interaction Lagrangian in the \MSSM\  is rather lengthy we
quote only the interactions that we will need in our studies. 
Explicit Feynman rules derived from these Lagrangians can be found in~\cite{Tesina}.

\begin{itemize}
\item fermion--sfermion--(chargino or neutralino): this interaction is obtained
    from the potential~(\ref{eq:vgfsf}) -gauginos-, and form the Yukawa coupling
    term~(\ref{eq:vyukawa}) -higgsinos-, in the mass-eigenstate basis:
    \begin{eqnarray}
      \label{LcqsqLR}
      \label{eq:Lqsqcn}
      {\cal L}_{\Psi q\tilde{q}}&=&
      g\,\sum_{a=1,2}\sum_{i=1,2}\left[ 
        -\sta^*\cbim\left(\Apit\epsilon_i\PL+\Amit\PR\right)\,b
       \right.\nonumber\\
       &&\left.-\sba^*\cbip\left(\Apib\PL+\Amib\epsilon_i\PR\right)\,t 
      \right] \nonumber\\
      &+&\displaystyle{\frac{g}{\sqrt{2}}}
      \sum_{a=1,2}\sum_{\alpha=1,\ldots ,4}\left[
        -\sta^*\nba\left(\Apat\PL+\Amat\epsilon_\alpha\PR\right)\,t
        \right.\nonumber\\
       &&\left.+\sba^*\nba\left(\Apab\PL+\Amab\epsilon_\alpha\PR\right)\,b
      \right] \nonumber\\
      &+&\mbox{\rm h.c.}\ ,
    \end{eqnarray}
    where we have introduced the usual chirality projection operators
    $P_{L,R}=\frac{1}{2}\left(1\mp\gamma^5\right)$  and the matrices
    \begin{eqnarray}
        \label{V1Apm}
        \Apit &=& \Rotc\Uo^*-\lt\Rttc\Ut^*\, ,\nonumber\\
        \Amit &=& -\lb\Rotc\Vt\, ,\nonumber\\
        \Apat &=& \Rotc\left(\Nt^*+\YL\tth\No^*\right)
        +\sqrt{2}\lt\Rttc\Nth^*\, ,\nonumber\\
        \Amat &=& \sqrt{2}\lt\Rotc\Nth
        -\YRt\tth\Rttc\No\, ,\nonumber\\
        \Apib &=& \Robc\Vo^*-\lb\Rtbc\Vt^*\, ,\nonumber\\
        \Amib &=& -\lt\Robc\Ut\, ,\nonumber\\
        \Apab &=& \Robc\left(\Nt^*-\YL\tth\No^*\right)
        -\sqrt{2}\lb\Rtbc\Nf^*\, ,\nonumber\\
        \Amab &=& -\sqrt{2}\lb\Robc\Nf
        +\YRb\tth\Rtbc\No\, .
     \end{eqnarray}
    with $\YL$ and $Y_R^{t,b}$ the weak hypercharges of the left-handed $SU(2)_L$
    doublet and right-handed singlet fermion, and $\lt$ and 
    $\lb$ are -- Cf.\ eq.(\ref{eq:Yukawasgeneric}) -- the potentially significant Yukawa
    couplings normalized to the $SU(2)_L$ gauge coupling constant $g$.

\item quark--squark--gluino: the supersymmetric version of the strong
  interaction is obtained from~(\ref{eq:vgfsf}):
\begin{equation}
{\cal L}_{\sg q\tilde{q}}=- \frac{g_s}{\sqrt{2}}
\tilde q_{a,i}^{*}\, \bar \psi^{\sg}_c 
\left(\lambda^c\right)_{ij} \left(R_{1a}^{(q)*} \pl - R_{2a}^{(q)*} \pr \right) q_j
+\mbox{ h.c.}\,\,,
\label{eq:Lqsqglui}\end{equation}
where $\lambda^c$ are the Gell-Mann matrices.

\item quark--quark--Higgs: this is the usual Yukawa interaction from Type II
  2HDM, in the \MSSM\  it follows after replacing in~(\ref{eq:vyukawa}) the
  mass-eigenstate Higgs fields~(\ref{eq:HiggsDefin}):
\begin{eqnarray}
{\cal L}_{H^+ u d}&=&\frac{g}{\sqrt{2}M_W} \left[\bar u \left( m_u \cot\beta\, \pl +
 m_d \tan\beta\, \pr\right) d\, H^+ + \mbox { h.c.}\right]\nonumber\\
~&~& -\frac{g m_d}{2 M_W \cos\beta}\left[\left(\cos\alpha\, H^0 -\sin\alpha\, h^0\right) \bar d d - i \sin\beta\, \bar d \gamma_5 d\, A^0 \right]\nonumber \\
~&~& -\frac{g m_u}{2 M_W \sin\beta}\left[\left(\sin\alpha\, H^0 +\cos\alpha\, h^0\right) \bar u u - i \cos\beta\, \bar u \gamma_5 u\, A^0 \right] \,\,,
\label{LqqH}\end{eqnarray}
here we have replaced the Yukawa couplings $h_i$ in favour of masses and \tb.
\item squark--squark--Higgs: the origin of this interaction is twofold, on one
  side the superpotential derivative~(\ref{eq:superder}), and on the other the
  Soft-\SUSY-Breaking trilinear interactions,
\begin{eqnarray}
{\cal L}_{H^\pm\tilde{q}\tilde{q}}&=&
-\frac{g}{\sqrt{2}M_W}\,\su_a^{*}\,\sd_b^{}\,G_{ba} \hplus+\mbox{ h.c.}
\nonumber\\
&\equiv&
-\frac{g}{\sqrt{2}M_W}\,\su_a^{*}\,\sd_b^{}\,R_{ia}^{(u)*}\,R_{jb}^{(d)}\,g_{ij} \hplus
+\mbox{ h.c.}\,\,,
\label{eq:hsfsf}  \label{LHsqsq}\end{eqnarray}
where we have introduced the matrix\footnote{Note that our convention for the
  $\mu$ parameter in~(\ref{eq:superp}) is opposite in sign to that
  of\,\cite{Hunter}.} 
\begin{equation}
g_{ij}=\left(
\ba{cc}
M_W^2\sin{2\beta}-\left(m_d^2 \tan\beta+m_u^2\cot\beta\right) & - m_d\left(\mu+A_d \tan\beta\right) \\
-m_u\left(\mu+A_u\cot\beta\right)&-m_u m_d \left(\tan\beta+\cot\beta\right)
\ea
\right)\,\,.
\label{eq:Matrixhsfsf}\label{gdef}\label{Gdef}
\end{equation}
\item chargino--neutralino--charged Higgs: this interaction is obtained
  from~(\ref{eq:vgfsf}), we note that in the electroweak basis the only
  interaction present is the Higgs--higgsino--gaugino one
\begin{equation}
{\cal L}_{H^\pm\Psi^\mp\Psi^0}=-g H^+\, \bar \psi^+_i \left(\sin\beta\, Q_{\alpha i}^{\prime R *} \epsilon_\alpha \pl
+\cos\beta\, Q_{\alpha i}^{\prime L *}\epsilon_i \pr \right) \psi^0_\alpha
\,\,,  \label{LHcn}
\end{equation}
$$
\left\{
\ba{lcl}
Q_{\alpha i}^{\prime L}&=&U_{i1}^*N_{\alpha 3}^*+
\frac{1}{\sqrt{2}}\left(N_{\alpha 2}^*+\tan\theta_W N_{\alpha 1}^*\right)U_{i2}^*\\
Q_{\alpha i}^{\prime R}&=&V_{i1}N_{\alpha 4}-
\frac{1}{\sqrt{2}}\left(N_{\alpha 2}+\tan\theta_W N_{\alpha 1}\right)V_{i2}
\,\,.\ea
\label{eq:QLQR}
\right.$$
\item gauge interactions:
in this Thesis we only need a small subset of the gauge interactions present in
the \MSSM, so we will only quote the interactions of the $W^{\pm}$ boson, and
those of \QCD. The photon interactions are simply those of QED (and scalar
QED). For a complete set of the $Z$ boson interactions see for
example\cite{TesiDavid} 
\begin{itemize}
\item quark--$W^\pm$:
\begin{equation}
{\cal L}_{W^+ud}=\frac{g}{\sqrt{2}}\, \bar u \gamma^\mu \pl d\, W_\mu^++\mbox{ h.c.}\,\,,
\label{eq:Lqqw}\end{equation}
\item squark-$W^{\pm}$:
\begin{equation}
{\cal L}_{W^+\su\sd}=i \frac{g}{\sqrt{2}}
R_{1a}^{(u)*} R_{1b}^{(d)} W^+_\mu
\su_a^{*}\stackrel{\leftrightarrow}{\partial^\mu}\sd_b
+\mbox{ h.c.} \,\,,
\label{eq:LsqsqW}\end{equation}
\item chargino--neutralino--$W^{\pm}$:
\begin{equation}
{\cal L}_{W^+\cplus\neut}=g\, \bar\psi_\alpha^0 \gamma^\mu \left(C_{\alpha i}^L \epsilon_\alpha \epsilon_i\pl + C_{\alpha i}^R \pr
\right)\psi_i^+\, W^-_\mu+\mbox{ h.c.}\,\,,
\end{equation}
$$
\left\{
\ba{lcl}
C_{\alpha i}^L&=&\frac{1}{\sqrt{2}}N_{\alpha 3}U_{i 2}^*-N_{\alpha 2}U_{i 1}^*\\
C_{\alpha i}^R&=&-\frac{1}{\sqrt{2}}N_{\alpha 4}^* V_{i 2}-N_{\alpha 3}^* V_{i 1}\,\,.
\ea\right.
$$
\item Higgs--$W^{\pm}$: after SSB there exist three different kind of gauge
  interactions for the Higgs (and Goldstone) bosons\,\cite{Hunter}, namely
  triple interactions of a gauge boson and two scalars, 
  triple interaction of two gauge bosons and a scalar, interaction between two
  gauge bosons and two scalars. We only need the first one of these interactions
  to perform the analysis presented here, that is
\begin{eqnarray}
{\cal L}_{W^+H^-H} &=  & \frac{g}{2} W_\mu^+\left(
-\sin\left(\beta-\alpha\right)H^- i\stackrel{\leftrightarrow}{\partial_\mu} H^0
+\cos\left(\beta-\alpha\right)H^- i\stackrel{\leftrightarrow}{\partial_\mu} h^0
\right. \nonumber\\
~&~&\left. -H^- \stackrel{\leftrightarrow}{\partial_\mu} A^0
\right)+\mbox{ h.c.}\,\,.
\end{eqnarray}
\item quark strong interactions: this is the usual \QCD\  Lagrangian
  \begin{equation}
    \label{eq:QCDlagrangian}
    {\cal L}_{QCD}=\frac{g_s}{2}\,
    G_{\mu}^c\,\lambda^c_{ij}\,\bar{q}_{i}\,\gamma^\mu\,q_{j}  \,\,.
  \end{equation}
\item squark strong interactions: aside from the well known scalar \QCD\
  Lagrangian, the scalar potential~(\ref{eq:vd}) introduces quartic scalar
  interactions between squarks of order $\alpha_s$, thus we have
  \begin{eqnarray}
    \label{eq:scalarQCDLagrangian}
    {\cal L}_{G\sq\sq}&=& -i \frac{g_s}{2}\, G_{\mu}^c \,\lambda_{ij}^c\, \sq^*_{a,i}\,
    \stackrel{\leftrightarrow}{\partial^\mu}\, \sq_{a,j} \,\,,\nonumber\\
    {\cal L}_{GG\sq\sq}&=& \frac{g_s^2}{4}\, G_\mu^c \,G^{\mu\,d}\, \sq_{a,i}^* \,\sq_{a,j}^*
    (\lambda^c_{ik}\lambda^d_{kj}+\lambda^d_{ik}\lambda^d_{kj})\,\,.\nonumber\\
    {\cal L}_{\sq\sq\sq\sq}&=& \frac{g_s^2}{8}\sum_{\sq,\sq'} 
    (R_{1a}\, R_{1b} - R_{2a}\,R_{2b} )\,
    (R_{1c}\, R_{1d} - R_{2c}\,R_{2d} )\times \nonumber\\
&&    \,\sq^*_{a,i}  \,\lambda^r_{ij} \,\sq_{b,j}\,
    \,\sq^{\prime\,*}_{c,k}  \,\lambda^r_{kl} \,\sq^{\prime}_{d,l}\,\,.
  \end{eqnarray}
\end{itemize}
\end{itemize}

\section{\MSSM\  parametrization}
\label{sec:limits}
\subsection{\MSSM\  parameters}
\label{sec:parameters}
If \SUSY\  were an exact symmetry then only one parameter should be added to the
\SM\  ones (\tb), but we have to deal with a plethora of Soft-\SUSY-Breaking
parameters, namely
\begin{itemize}
\item masses for Left- and Right-chiral sfermions,
\item a mass for the Higgs sector,
\item gaugino masses,
\item triple scalar couplings for squarks and Higgs.
\end{itemize}

This set of parameters is often simplified to allow a comprehensive study. Most
of these simplifications are based on some universality assumption at the
unification scale. In minimal supergravity (mSUGRA) all the parameters of the
\MSSM\  are computed from a restricted set of parameters at the Unification
scale, to wit: \tb; a common scalar mass $m_0$; a common fermion mass for
gauginos $m_{1/2}$; a common trilinear coupling for all sfermions
$A_0$; and the higgsino mass parameter $\mu$. Then one computes the running of
each one of these parameters down to the \EW\  scale, using the Renormalization
Group Equations (RGE), and the full spectrum of the \MSSM\  is found. 

We will not restrict ourselves to a such simplified model. As stated in the
introduction we treat the \MSSM\  as an effective Lagrangian, to be embedded in
a more general framework that we don't know about. This means that essentially
all the parameters quoted above are free. However for the kind of studies we
have performed there is an implicit asymmetry of the different particle
generations. We are mostly interested in the phenomenology of the third
generation, thus we will treat top and bottom supermultiplet as distinguished from the
rest. This approach is well justified by the great difference of the Yukawa
couplings of top and bottom with respect to the rest of fermions. We are mainly
interested on effects on the Higgs sector, so the smallness of the Yukawa couplings
of the first two generations will result on small effects in our final
result. We include them, though, in the numerical analysis and the numerical
dependence is 
tested. On the other hand, if we suppose that there is unification at some large
scale, at which all sfermions have the same mass, and then evolve these masses
to the \EW\  scale, then the RGE have great differences\cite{Martin:1997ns}. Slepton
RGE are dominated by 
\EW\  gauge interactions, 1st and 2nd generation squarks RGE are dominated by
\QCD, and for the 3rd generation squarks there is an interplay between \QCD\
and Yukawa couplings. Also, as a general rule, the gauge contribution to the RGE
equations of left- and right-handed squark masses are similar, so when Yukawa
couplings are not important they should be similar at the \EW\  scale.

With the statement above in mind we can simplify the \MSSM\  spectrum by taking
an unified parametrization for 1st and 2nd generation squarks (same for
sleptons). We will use: a common mass\footnote{Note that after diagonalization of the squark mass matrix the
  physical masses will differ slightly.} for $\su_L$ and $\sq_R$
($\msup$); an unified
trilinear coupling $A_u$ for 1st and 2nd generation; a common mass for all 
$\tilde \nu_L$ and $\tilde l_R$ ($\mstau$); and a common trilinear coupling
$A_\tau$\footnote{See section~\ref{sec:sfmas} for the concrete definitions of
  these parameters.}.

For the third generation we will use different trilinear couplings $A_t$ and
$A_b$, as these can play an important role in the kind of processes we are
studying (see chapter~\ref{cap:tbh}). Stop masses can present a large gap (due
to its Yukawa couplings), being the right-handed stop the lightest one. We will
use a common mass for both chiral sbottoms, which we parametrize with the
lightest sbottom mass ($\msbo$), and the lightest stop quark
mass ($\msto$), as the
rest of mass inputs in this sector. This parametrization is useful in processes
where squarks only appear as internal particles in the loops (such as the ones
studied in chapters~\ref{cap:tbh} and~\ref{cha:fcnc}), as one-loop corrections
to these parameters would appear as two-loop effects in the process subject of
study. However in chapter~\ref{cap:sbdecay} we deal with squarks as the main
subject of the process and in this case a more physical set of inputs must be
used. We have chosen to use the physical sbottom masses ($\msbo$, $\msbt$) and the
sbottom and stop mixing 
angles ($\theta_{\sbottom}$, $\theta_{\stopp}$) to be our main inputs. Again
one-loop effects on 
other parameters (such as 
$A_b$) would show up as two-loop corrections to the observables we are interested
in. 

For the same reasons \EW\  gaugino sector is also supposed to have small effects
in our studies. Thus the grand unification relation introduced in
expression~(\ref{eq:GauginoMassRelation}). Gluino mass (\mg), on the other hand, is
let free.

For the Higgs sector two choices are available, we can use the
pseudoscalar mass $\mA$, or the charged Higgs mass $\mHp$. Both choices are on
equal footing. As the charged Higgs particle is a main element for most of our
studies we shall use its mass as input parameter in most of our work. However
in chapter~\ref{cha:fcnc} it is more useful to use $\mA$.

Standard model parameters are well known, we will use present determinations
of \EW\  observables\cite{RADCORteubert,PDB,vanRitbergen:1998yd} 
\begin{eqnarray}
\mz &=& 91.1867\pm 0.0021 \GeV\nonumber\\
\mw&=&80.352\pm 0.054 \GeV\nonumber\\
G_F&=& (1.16637 \pm 0.00001)\times 10^{-5} \GeV^{-2} \nonumber\\
m_\tau&=&1777.05 \pm 0.29 \mbox{ MeV} \nonumber\\
\alpha(\mz)^{-1}&=&128.896\pm0.090 \,\,.
\label{eq:EWprecisiondata}
\end{eqnarray}
\QCD\  related observables are not so precise. On the other hand as the main
results are not affected by specific value of these observables we will use the
following ones
\begin{eqnarray}
  \label{eq:parametersQCD}
  \mt&=&175 \GeV \nonumber\\
  \mb&=&5 \GeV \nonumber\\
  \alpha_s(\mt)&=&0.11\,\,
\end{eqnarray}
(the last figure corresponds to $\alpha_s(\mz)=0.12$).

\subsection{Constraints}
\label{sec:MSSMconst}
The \MSSM\  reproduces the behaviour of the \SM\  up to energy scales probed so
far\,\cite{RADCORdeBoer}. Obviously  this is not for every point of the full
parameter space! 

There exists direct limits on sparticle masses based on direct searches
at the high energy colliders (LEP II, SLC, Tevatron). Although hadron colliders
can achieve larger center of mass energies than $e^+e^-$ ones, its samples
contain large backgrounds that make the analysis more difficult. This drawback
can be avoided if the ratio signal-to-background is improved, in fact they can
be used for precision measurements of ``known'' observables (see
e.g.\,\cite{RADCORgianotti}). $e^+e^-$ colliders samples are more clean, and
they allow to put absolute limits on particle masses in a model independent
way. 

The most stringent bound to the \MSSM\  parameter space is the LEP II bound to
the 
mass of charged particles beyond the \SM. At present\,\cite{Acciarri:1998eg,Abbiendi:1998rz,Abreu:1998jy} this limit is roughly
\begin{equation}
M_{\rm charged} \gsim 90 \GeV\,\,.
\label{eq:limmsuper}
\end{equation}

Specific searches for Supersymmetric particles are being performed at LEP II,
negative neutralino searches rise up a limit on neutralino masses of\cite{Abbiendi:1998rz}
\begin{equation}
M_{\neut_1} \gsim 30 \GeV\,\,,
\label{eq:limneutralinos}
\end{equation}
it turns out that after translating this limit to the $\mu-M$ parameters it is
less restrictive than the one obtained for the charginos
from~(\ref{eq:limmsuper}). 

Actual Higgs searches at LEP II imply that, for the \MSSM\  neutral Higgs sector\cite{Barate:1998gx}
\begin{equation}
  \label{eq:limitsLEPHiggs}
  \mh >  72.2 \GeV\,\,,\,\, \mA >  76.1 \GeV\,\, .
\end{equation}
Notice that without the \MSSM\  relations there is no model independent
bound on $\mA$ from LEP\,\cite{Krawczyk:1996fe}. Actual fits to the \MSSM\
parameter space project a preferred value for the charged Higgs mass of
$\mHp\simeq 120\GeV$\,\cite{HollikQEMSSM}.

Hadron colliders bounds are not so restrictive as those from $e^+e^-$
machines. Most bounds on squark and gluino masses are obtained by supposing
squark mass unification in simple models, such as mSUGRA. At present the limits
on squarks (1st and 2nd generation) and gluino masses are\,\cite{PDB}
\begin{equation}
  \label{eq:limitsCDFD0squarks}
  m_{\sq} > 176 \GeV \,\,,\,\, \mg > 173 \GeV \,\,.
\end{equation}

From the top quark events at the Tevatron a limit on the branching ratio
$BR(t \rightarrow H^+\,b)$ can be extracted, and thus a limit on the $\tb-\mHp$
relation. We will treat this limit in detail in chapter~\ref{cap:tbh}.

Finally indirect limits on sparticle masses are obtained from the \EW\
precision data. We apply these limits through all our computations by computing
the contribution of sparticles to these observables and requiring that they
satisfy the bounds from \EW\  measurements. We require new contributions to 
the $\rho$ parameter to be smaller than present experimental error on
it, namely
\begin{equation}
  \label{eq:deltarhoNEW}
  \delta\rho_{\rm new} < 0.003\,\,.
\end{equation}
We notice that as $\delta \rho_{\rm new}$ is also the main contribution from
sparticle contributions to $\Delta r$\,\cite{TesiDavid}, new contributions to
this parameter are also below experimental constrains. Also the corrections in
the $\alpha$- and 
$G_F$-on-shell renormalization schemes will not differ significantly (see section~\ref{sec:renormintro}).

There exist also theoretical constrains to the parameters of the \MSSM. As a
matter of fact the \MSSM\  has a definite prediction: there should exist a
light neutral scalar Higgs boson $h^0$. Tree-level analysis put this bound to
the $Z$ mass, however the existence of large radiative corrections to the
Higgs bosons mass relations grow this limit up to $\sim 130 \GeV$. Recently the
two-loop radiative corrections to Higgs mass relations in the \MSSM\  have been
performed\cite{Heinemeyer:1998jw,Hempfling:1994qq,RADCORhaber}, and the present
upper limit on $\mh$ is
\begin{equation}
  \label{eq:UPh0limit}
  \mh \leq 130-135 \GeV\,\,.
\end{equation}
The two figures in~(\ref{eq:UPh0limit}) have been computed by different
groups\,\cite{Heinemeyer:1998jw,Hempfling:1994qq,RADCORhaber} and there is a great interest in make them
match\cite{RADCORhaber}. It is very important to know as precise as possible this
limit, as by means of a possible Run III of the Tevatron collider (TEV33, at the
same energy, but higher luminosity) either a $h^0$ should be found, or on the contrary a
lower limit to its mass in the ballpark of $130\GeV$ will be put. Thus it is of
extreme importance to have both, a very precise prediction for the
bound~(\ref{eq:UPh0limit}), and a very precise analysis of the Tevatron data. Of
course if the \MSSM\  is extended in some way this limit can be evaded, though
not to values larger of $\sim 200\GeV$\,\cite{Masip:1998za,Masip:1998jc}.

Another theoretical constraint is the necessary condition~(\ref{eq:necessary}) on
squark trilinear coupling ($A$) to avoid colour-breaking minima. This constraint
is easily implemented when the $A$ parameters are taken as inputs, but if we
choose a different set of inputs (such as the mixing angle $\theta_{\sq}$, as in
chapter~\ref{cap:sbdecay}) then it constrains the parameter space in a
non-trivial way -eq.\,(\ref{eq:Abt}).

Whatever the spectrum of the \MSSM\  is, it should comply with the benefits that
\SUSY\  introduces into the \SM\   
which apply the following condition is fulfilled:
\begin{equation}
  \label{eq:MSUSYUPLIMIT}
  M_{\rm SUSY} \lsim {\cal O} (1 \mbox{ TeV})\,\,.
\end{equation}
If supersymmetric particles had masses heavier than the TeV scale then
problems with GUT's appear. 
This statement does
not mean that \SUSY\  would not exist, but that then the \SM\  would not gain
practical benefit from the inclusion of \SUSY. 

A similar upper bound is obtained
when making cosmological analyses, in these type of analyses one supposes the
neutralino to be part of the cold dark matter of the universe, and requires 
its annihilation rate to be sufficiently small to account for the maximum of cold
dark matter allowed for cosmological models, while at the same time sufficiently
large so that its presence does not becomes overwhelming. Astronomical
observations also restrict the parameters of \SUSY\  models, usually in the
lower range of the mass parameters (see e.g.\,\cite{Grifols:1997hi}).

For the various RGE analysis to hold the couplings of the \MSSM\  should be
perturbative all the way from the unification scale to the \EW\  scale. This
implies, among other restrictions, that top and bottom Yukawa couplings should
be below certain limits. 
In terms of \tb\  this amounts it to be confined in the approximate interval
\begin{equation}
  \label{eq:tanbetalim}
  .5 \lsim \tb \lsim 70\,\,.
\end{equation}

All these restrictions will apply in all our numerical computations.
Any deviation from this framework of restrictions will only be for
demonstrational purpouses, and will be
explicitly quoted in the text.


%% file: renorm.tex
 
\chapter{\MSSM\  renormalization} 
\label{cap:Renorm} 
\section{Introduction} 
\label{sec:renormintro} 
In this chapter we perform the renormalization of the \MSSM\  in the on-shell 
scheme. We do not pretend to make all the renormalization procedure,  
but just sketching what are the necessary ingredients of this renormalization 
and giving  
expressions for some non-\SM\  two-point functions. The renormalized three-point 
Green  
functions are the subject of the forthcoming chapters. We will not give the full 
expressions for the gauge bosons self-energies, or the $\delta \rho$ and  
$\Delta r$ parameters,  
since these have been subject of dedicated 
studies\cite{Garcia:1995wv,Garcia:1995wu,TesiDavid,TesiJefe,Garcia:1994sb,Grifols:1985xs,Grifols:1984gu,Sola:1991ju,Chankowski:1994eu,Bertolini:1986ia}. On the other 
hand the various counterterms and self-energies given in this chapter are 
general. We have left some expressions out of this chapter as they are  
approximations only valid in the context where they are used (see 
chapter~\ref{cap:sbdecay}).  
 
We address the renormalization of the \MSSM\  extending the \SM\  on-shell 
procedure described 
in\cite{Bohm:1986rj,Hollik:1990ii,Hollik1995Llibre,Jegerlehner:1991dq,Denner:1993kt}\footnote{Our 
  conventions differ from those  
  of~\cite{Bohm:1986rj,Hollik:1990ii}.}.  
We may use both the $\alpha$ or the $G_F$ parametrizations. 
At one-loop order, we shall call 
the former the ``$\alpha$-scheme'' and the latter the ``$G_F$-scheme''. 
In the ``$\alpha$-scheme'', the 
structure constant $\alpha\equiv \alpha_{\rm em}(q^2=0)$  
and the masses of the gauge bosons, fermions and scalars are 
the renormalized parameters: $(\alpha, M_W, M_Z, M_H, m_f, M_{SUSY},\ldots)$  
--$M_{SUSY}$ standing for the collection of renormalized sparticle masses.  
Similarly, the ``$G_F$-scheme'' is characterized by the set of inputs 
$(G_F, M_W, M_Z, M_H, m_f, M_{SUSY},\ldots)$. Beyond lowest order, 
the relation between the two on-shell schemes is given by  
\begin{equation} 
{G_F\over\sqrt{2}}={\pi\alpha\over 2 M_W^2 s_W^2} 
(1+\Delta r^{MSSM})\,, 
\label{eq:DeltaMW} 
\end{equation}  
where $\Delta r^{MSSM}$ is the prediction of the parameter 
$\Delta r$\,\cite{Bohm:1986rj,Hollik:1990ii,Hollik1995Llibre}  
in the MSSM\footnote{$\Delta r^{MSSM}$ has  
been subject of dedicated studies, see\,\cite{TesiJefe,Garcia:1994sb,Chankowski:1994eu}.}.

Let us sketch the renormalization procedure affecting the parameters and 
fields related to the various processes subject of study. 
In general, the renormalized  \MSSM\  Lagrangian 
${\cal L}\rightarrow {\cal L}+\delta{\cal L}$ 
is obtained following a similar pattern as in the \SM, i.e.\ 
by attaching multiplicative renormalization constants to each free 
parameter and field: $g_i\rightarrow (1+\delta g_i/g_i)g_i$, 
$\Phi_i\rightarrow Z^{1/2}_{\Phi_i}\Phi_i$. As a matter of fact, field  
renormalization (and so Green's functions renormalization)  
is unessential and can be either omitted or be carried out 
in many different ways without altering physical ($S$-matrix) amplitudes. 
In our case, in the line of Refs.\cite{Garcia:1994rq,Dabelstein:1995jt},    
we shall basically 
use minimal field renormalization, i.e.\  one renormalization constant per 
gauge symmetry multiplet\,\cite{Bohm:1986rj,Hollik:1990ii,Hollik1995Llibre}.  
In this way  the counterterm  
Lagrangian, $\delta{\cal L}$, as well as the various Green's functions 
are automatically gauge-invariant.  
 
The specific sign convention  of the various two-point functions used all over 
this Thesis is based on the prescription that the unrenormalized self-energy 
always add up to the bare mass parameter (or the squared mass, depending on the 
kind of particle), which is equivalent to say that,  
in the on-shell scheme, the mass parameter counterterm is 
{\em minus}  
the unrenormalized self-energy, that is 
$$ 
m^0+{\rm Re}\left(\Sigma(k^2)\right)= 
m+\delta m + {\rm Re}\left(\Sigma(k^2)=0\right)\,\,, 
\,\,\delta m=-{\rm Re}\left(\Sigma(k^2)\right)\,\,, 
$$ 
where $m$ is the physical mass parameter --the mass for fermions, the squared mass for 
bosons-- and $\delta m$ the corresponding counterterm (see next sections for the 
concrete definition in each case). The convention for each 
kind of particle can be seen in table~\ref{taula:sigmes}.  
 
\taulasigmes 
 
For the regularization of the ultraviolet divergent integrals we use the 
Dimensional Reduction (DRED)\cite{Siegel:1979wq,Capper:1980ns} prescription, as 
it respects \SUSY\@. As  
a matter of fact one-loop computations with only $R$-even external particles 
yield the same result in DRED and Dimensional Regularization, however this is 
not necessary true for higher loop computation, or for computations with $R$-odd 
external particles. 
 
\section{A note on the gauge sector renormalization} 
\label{sec:gaugerenorm} 
For the sake of fixing notation,  
in this section we review some well known features of the renormalization of 
the electroweak gauge sector, which is identical to the \SM\  one. 
We refer 
to~\cite{Bohm:1986rj,Hollik:1990ii,Hollik1995Llibre} for a  
comprehensive exposition of the subject, and 
to~\cite{Garcia:1995wv,Garcia:1995wu,TesiDavid,TesiJefe,Grifols:1985xs,Grifols:1984gu,Sola:1991ju,Bertolini:1986ia} for the \MSSM\  expressions of 
the various self-energies. 
 
For the $SU(2)_L$ gauge field we have 
\begin{equation} 
W^{\pm}_{\mu}\rightarrow {(Z_2^W)}^{1/2} W^{\pm}_{\mu}\pm 
i\,{\delta Z_{HW}\over M_W}\,\partial_{\mu} H^{\pm}\,, 
\label{eq:ZHW} 
\end{equation} 
$Z_2^W=1+\delta Z_2^W$ is the usual 
$SU(2)_L$ gauge triplet renormalization 
constant given by the formula 
\begin{equation} 
\delta Z_2^W  =  
\left.  {\Sigma_{\gamma}(k^2)\over k^2}\right|_{k^2=0}-2{c_W 
\over s_W}\,{\Sigma_{\gamma Z}(0)\over M_Z^2} 
+{c_W^2\over s_W^2} 
\left({\delta M_Z^2\over M_Z^2}-{\delta M_W^2\over M_W^2}\right)\,, 
\label{eq:gtriplet} 
\end{equation} 
and 
\begin{equation} 
\delta M_W^2=-\Sigma_W(k^2=M_W^2)\,\,,\ \ \ \ \ 
\delta M_Z^2=-\Sigma_Z(k^2=M_Z^2)\,, 
\label{eq:MCT} 
\end{equation} 
are the gauge boson mass counterterms 
enforced by the usual on-shell mass renormalization conditions. The $\Sigma$ 
functions denote 
the (real part of the) unrenormalized two-point Green functions. 
$\delta Z_{HW}$ on eq.(\ref{eq:ZHW}) 
is a dimensionless constant associated to the 
wave-function renormalization mixing among the bare $H^{\pm}$ 
and $W^{\pm}$ fields; its meaning and value is discussed together with the Higgs 
renormalization procedure (section~\ref{sec:renormHiggs}).

For the $SU(2)_L$ gauge coupling constant, we have  
\begin{equation} 
g \rightarrow  (1+\frac{\delta g}{g}) g={(Z_1^W)}\,{(Z_2^W)}^{-3/2}\, g\,,  
\end{equation} 
where $Z_1^W$ refers to the renormalization constant associated to the 
triple vector boson vertex. Therefore, from charge renormalization, 
\begin{equation} 
\frac{\delta\alpha}{\alpha} =     
-\left.  {\Sigma_{\gamma}(k^2)\over k^2}\right|_{k^2=0}-2{s_W 
\over c_W}\,{\Sigma_{\gamma Z}(0)\over M_Z^2}\,, 
\end{equation} 
and the bare relation $\alpha=g^2\,s^2_W/4\pi\rightarrow 
\alpha+\delta\alpha=(g^2+\delta g^2)\,(s^2_W+\delta s^2_W)/4\pi$,   
one gets for the counterterm to $g$:   
\begin{equation} 
\frac{\delta g^2}{g^2} =\frac{\delta\alpha}{\alpha}- 
{c_W^2\over s_W^2} 
\left({\delta M_Z^2\over M_Z^2}-{\delta M_W^2\over M_W^2}\right)\,, 
\label{eq:dg} 
\end{equation}  
and as a by-product 
\begin{equation} 
\delta Z_1^W =  \frac{1}{2}\frac{\delta g^2}{g^2} 
+\frac{3}{2} \delta Z_2^W\,. 
\end{equation}

\section{Fermion renormalization} 
\label{sec:renormfermion} 
Following the directives from section~\ref{sec:renormintro} and 
references~\cite{Garcia:1994rq,Dabelstein:1995jt} we introduce the fermion wave 
function renormalization constants 
\begin{eqnarray} 
& &\pmatrix{ 
t_L \cr b_L 
} 
\rightarrow  Z_L^{1/2}\, 
\pmatrix{ 
t_L \cr b_L 
} 
\rightarrow \pmatrix{ 
{(Z_L^t)}^{1/2}t_L \cr {(Z_L^b)}^{1/2}b_L 
}\,,    
\nonumber\\ 
& & b_R \rightarrow {(Z_R^b)}^{1/2}b_R\,, \ \ \ 
t_R \rightarrow{(Z_R^t)}^{1/2}t_R\,. 
\label{eq:FR} 
\end{eqnarray} 
Here $Z_i=1+\delta Z_i$ are the doublet ($Z_L$) and singlet ($Z_R^{t,b}$) 
field renormalization constants for the top and bottom quarks. 
Although in the minimal field renormalization scheme there is only one 
fundamental constant, $Z_L$, per matter doublet, it is useful 
to work with $Z_L^b=Z_L$ and $Z_L^t$, where the latter differs from the 
former by a {\it finite} 
renormalization effect\,\cite{Bohm:1986rj,Hollik:1990ii,Hollik1995Llibre}.  
To fix all these constants one starts from the usual on-shell  
mass renormalization condition for fermions, $f$, together with the 
``${\rm residue}=1$'' condition on the renormalized propagator. These are 
completely standard procedures, and  
in this way one obtains\footnote{We understand 
  that in all formulae defining counterterms we are taking the real part of the 
  corresponding functions.} 
\begin{equation} 
  {\delta m_f\over m_f}=-\left[{\Sigma^f_L(m_f^2)+\Sigma^f_R(m_f^2)\over 2} 
    +\Sigma^f_S(m_f^2)\right]\,, 
  \label{eq:deltamf} 
\end{equation}  
and 
\begin{eqnarray} 
  \delta Z_{L,R}^f &=& \Sigma^f_{L,R}(m_f^2)+m_f^2[\Sigma^{f\,\prime}_L 
  (m_f^2)+\Sigma^{f\,\prime}_R(m_f^2) 
  +2\Sigma^{f\,\prime}_S(m_f^2)]\,, 
  \label{eq:DSRC} 
\end{eqnarray} 
where we have  
decomposed the fermion self-energy according to 
\begin{equation} 
\Sigma^f(p)=\Sigma^f_L(p^2)\slas{p}\,\pl+\Sigma^f_R(p^2)\slas{p}\,\pr 
+m_f\,\Sigma^f_S(p^2)\,, 
\label{eq:Sigma} 
\end{equation} 
and used the notation $\Sigma'(p)\equiv \partial\Sigma(p)/\partial p^2$. 
 
\diagautobottom 
 
\diagautotop 
 
The one-loop Feynman diagrams contributing to these various self-energies can be 
seen in figures~\ref{diag:autobottom} and~\ref{diag:autotop} for the bottom and 
top quarks respectively. To express the various self-energies and vertex 
functions we use the standard one-, two- and three-point one-loop functions from 
Refs.\,\cite{'tHooft:1979xw,Passarino:1979jh,Consoli:1979xw,Axelrod:1982yc} which we have 
collected in Appendix~\ref{ap:pointfun}. 
Using this notation the bottom quarks self-energies read as 
\begin{eqnarray} 
  \label{SUSYSelfb} 
  \Sigma_{\{L,R\}}^b(p^2)&=&\left.\Sigma_{\{L,R\}}^b(p^2) 
   \right|_{(a)+(b)}\nonumber\\ 
 &=&-ig^2\left[\abs{\Apmit}^2\Bo\left(p,\mi,\msta\right) 
  +\frac{1}{2}\abs{\Apmab}^2\Bo\left(p,\ma,\msba\right)\right]\,,\nonumber\\ 
  \mb\Sigma_S^b(p^2)&=&\left.\mb\Sigma_S^b(p^2) 
   \right|_{(a)+(b)}\nonumber\\ 
 &=&ig^2\left[\mi {\rm Re}\left(\Apitc\Amit\right) 
    \Bz\left(p,\mi,\msta\right)\right.\nonumber\\ 
 &&+\left.\frac{1}{2}\ma {\rm Re}\left(\Amabc\Apab\right) 
    \Bz\left(p,\ma,\msba\right)\right]\,, 
\end{eqnarray} 
from \SUSY-\EW\  particles, and 
\begin{eqnarray} 
  \label{HiggsSelfb} 
  \Sigma_{\{L,R\}}^b(p^2)&=&\left.\Sigma_{\{L,R\}}^b(p^2) 
   \right|_{(c)+(d)}\nonumber\\ 
  &=&\frac{g^2}{2i\mws}\left\{m_{\{t,b\}}^2\left[ 
   \{\ctbs,\tbs\} \Bo(p,\mt,\mHp)+\Bo(p,\mt,\mw) 
   \right]\right. \nonumber\\ 
  &+&\frac{\mbs}{2\cbts}\left[\cas\,\Bo(p,\mb,\mH) 
   +\sas\,\Bo(p,\mb,\mh)\right. \nonumber\\ 
  &&\ \ \ +\left.\left. \sbts\,\Bo(p,\mb,\mA) 
   +\cbts\,\Bo(p,\mb,\mz)\right]\right\}\,,\nonumber\\ 
  \Sigma_S^b(p^2)&=&\left.\Sigma_S^b(p^2) 
   \right|_{(c)+(d)}\nonumber\\ 
 &=&-\frac{g^2}{2i\mws}\left\{\mts\left[ 
   \Bz(p,\mt,\mHp)-\Bz(p,\mt,\mw)\right]\right.\nonumber\\ 
  &+&\frac{\mbs}{2\cbts}\left[\cas\,\Bz(p,\mb,\mH)+ 
   \sas\,\Bz(p,\mb,\mh)\right.\nonumber\\ 
  &&\ \ \ -\left.\left. \sbts\,\Bz(p,\mb,\mA) 
   -\cbts\,\Bz(p,\mb,\mz)\right]\right\}\,, 
\end{eqnarray} 
from Higgs and Goldstone bosons in the Feynman gauge. 
To obtain the corresponding expressions for an up-like fermion, $t$, just 
perform the label substitutions $b\leftrightarrow t$ 
on eqs.\,(\ref{SUSYSelfb})-(\ref{HiggsSelfb}); and  
on eq.\,(\ref{HiggsSelfb})  
replace $\sa\leftrightarrow \ca$ and 
$\sbt\leftrightarrow \cbt$ (which also implies replacing 
$\tb\leftrightarrow \ctb$).

The ``strong'' (\QCD) self-energies from Figs.\,\ref{diag:autobottom}~(e) and (f) are 
\begin{eqnarray} 
  \label{eq:QCDselfb} 
  \Sigma_{\{L,R\}}^{b}(p^2)&=& \left.\Sigma_{\{L,R\}}^b(p^2) 
  \right|_{(e)+(f)}\nonumber\\ 
  &=& -i\,8\,\pi\,\alpha_s\,C_F\, \left[ B_1(p,\mb,\lambda)- 
  \left|R_{\{1,2\}a}^{(b)}\right|^2 (B_0-B_1) (p,\msba,\mg) \right] \nonumber\\ 
  \Sigma_{S}^b(p^2)&=& \left.\Sigma_{S}^b(p^2) \right|_{(e)+(f)}\nonumber\\ 
  &=& -i\,8\,\pi\,\alpha_s\,C_F\,\left[2\,\Bz(p,\mb,\lambda) 
  +\frac{\mg}{\mb} {\rm Re}( R_{1a}^{(b)} R_{2a}^{(b)*}) B_0(p,\msba,\mg)\right] 
\end{eqnarray} 
where $C_F=(N_C^2-1)/2\,N_C=4/3$ is a colour factor and we have introduced a 
small gluon mass $\lambda$ to regularize the infrared divergences. The top quark 
ones from Figs.\,\ref{diag:autotop}~(e) and (f) are easily obtained by 
substituting in~(\ref{eq:QCDselfb})  
the particle indexes $b\rightarrow t$ and $\sbottom\rightarrow\stopp$. 
 
\section{Higgs sector} 
\label{sec:renormHiggs} 
  
One also assigns doublet renormalization constants to the two Higgs doublets~(\ref{eq:H1H2}) of the \MSSM: 
\begin{equation} 
\pmatrix{ 
H_1^0 \cr H_1^{-} 
} \rightarrow Z_{H_1}^{1/2} 
\pmatrix{ 
H_1^0 \cr H_1^{-} 
}\,,\ \ \ \ \  
\pmatrix{ 
H_2^{+} \cr H_2^0  
}\rightarrow Z_{H_2}^{1/2} 
\pmatrix{ 
H_2^{+} \cr H_2^0  
}\,. 
\label{eq:ZH1H2}   
\end{equation} 
Following the on-shell procedure we prefer to fix the counterterms 
using the physical fields. In this approach we have decided to take the charged 
Higgs as the renormalized particle, for the charged Higgs mass will be a natural 
input in most of our  
computations. This  
will induce finite renormalization constants in the other fields of the sector 
($A^0$, $h^0$ and $H^0$). Of course other equivalent choices can be made. For 
example we could have taken $A^0$ to be the renormalized Higgs as 
in\cite{Dabelstein:1995hb,Dabelstein:1995js,Chankowski:1994er,Chankowski:1995ri} 
and in this case the charged Higgs sector would have received finite renormalization 
effects. To this effect we introduce  wave function renormalization constants 
for the Higgs particles in the mass-eigenstate basis 
$Z_{H^\pm},\,Z_{G^\pm},\,\ldots$, which are only shortcuts for certain 
combinations of $Z_{H_1}$ and $Z_{H_2}$, and fix these constants as usual in the 
on-shell scheme by using as input particle the charged Higgs. We will discuss in 
detail the charged Higgs sector renormalization whereas the neutral sector has been 
extensively discussed 
in\cite{Dabelstein:1995hb,Dabelstein:1995js,Chankowski:1994er,Chankowski:1995ri}. We will expose 
two  
equivalent approaches: the renormalization in the Feynman gauge and in the 
Unitary gauge.

\subsection{Feynman gauge} 
 
\diagautomixhwren 
 
As we carry out our calculations in the Feynman gauge, we would also 
like to perform 
the renormalization of the Higgs sector in that gauge. The Lagrangian 
is sketched as follows:  
\begin{equation} 
L=L_{C}+L_{GF}+L_{FP}\,\,, 
\end{equation} 
where $L_{C}$ is the classical Lagrangian, $L_{GF}$ stands for the gauge-fixing 
term in that gauge,  
\begin{equation} 
L_{GF}=-F^{+}\,F^{-}+\ldots\,\ \ \ \ \ (F^{\pm}\equiv \partial^{\mu}W^+_{\mu}\mp iM_W\,G^+)\,\,, 
\label{eq:GF} 
\end{equation} 
and $L_{FP}=\bar{\eta}^a\,\left(\partial F^a/\partial\theta^b\right)\,\eta^b$  
is the Faddeev-Popov ghost Lagrangian constructed from FP and  
anti-FP Grassmann scalar fields 
$\eta, \bar{\eta}$.  Since we are interested in the 
charged gauge-Higgs ($W^{\pm}-H^{\pm}$) 
and charged Goldstone-Higgs ($G^{\pm}-H^{\pm}$) mixing terms in that gauge,  
we have singled out just the relevant term on eq.(\ref{eq:GF}).  
 
As is well-known, although the classical Lagrangian, $L_{C}$, also contains 
a nonvanishing mixing among the weak gauge boson 
fields, $W^{\pm}$, and the Goldstone boson fields, $G^{\pm}$, namely 
\begin{equation}  
{\cal L}_{GW}=i M_W\,W_\mu^- \partial^\mu G^+ +{\rm h.c.}\,, 
\end{equation}   
the latter is canceled (in the action) by a piece  
contained in $L_{GF}$. 
Now, after substituting the renormalization transformation  
for the Higgs doublets, 
eq.(\ref{eq:ZH1H2}), on the Higgs boson kinetic term with 
$SU(2)_L\times U(1)_Y$ gauge covariant 
derivative, one projects out the following relevant counterterms 
\begin{eqnarray} 
\delta {\cal L}&= & \delta Z_{H^\pm}\,\partial_\mu H^+ \partial^\mu H^{-} 
+ \delta Z_{G^\pm}\,\partial_\mu G^+ \partial^\mu G^-\nonumber\\ 
&+& \delta Z_{HG}\,\left(\partial_\mu H^- \partial^\mu G^+ 
+ {\rm h.c.}\right)  
+\delta Z_{HW}\,\left(i M_W\,W_\mu^-\partial^\mu H^+ 
+ {\rm h.c.} \right)+\ldots  
\label{eq:lclas} 
\end{eqnarray} 
where 
\begin{equation}  
\pmatrix{\delta Z_{H^{\pm}}\cr \delta Z_{G^{\pm}} } 
   =\pmatrix{ c_{\beta}^2 & s_{\beta}^2 \cr  
         s_{\beta}^2 & c_{\beta}^2} 
   \pmatrix{\delta Z_{H_1}\cr \delta Z_{H_2} }\,, 
\label{eq:deltaZs1} 
\end{equation}  
and 
\begin{eqnarray}    
\delta Z_{HG} &=& s_{\beta}\,c_{\beta}\,\left(\delta Z_{H_2} 
-\delta Z_{H_1}\right)\,,\nonumber\\ 
\delta Z_{HW} &=& s_{\beta}\,c_{\beta}\,\left[{1\over 2}\left(\delta Z_{H_2} 
-\delta Z_{H_1}\right)+{\delta\tan\beta\over\tan\beta}\right]\,\,, 
\label{eq:deltaZs2} 
\end{eqnarray} 
behave as if they were renormalization constants introduced as 
\begin{equation} 
  \label{eq:defdzhg} 
 \pmatrix{H^{\pm} \cr G^{\pm}}  
\rightarrow \pmatrix{ 
    Z_{H^{\pm}} & \delta Z_{HG} \cr 
     \delta Z_{HG} & Z_{G^{\pm}}} 
 \pmatrix{ H^{\pm} \cr G^{\pm}}\,\,, 
\end{equation} 
and 
\begin{equation} 
  \label{eq:defdzhw} 
  W^\pm_\mu \rightarrow {\left(Z_2^W\right)}^{1/2} W^\pm_\mu \pm i \frac{\delta Z_{HW}}{M_W}\partial_\mu H^\pm\,\,\,, 
\end{equation} 
with $Z_2^W$ being the usual $SU(2)_L$ gauge triplet renormalization 
constant~(\ref{eq:gtriplet}). A note regarding neutral Higgs particles is worth 
here. As the pseudoscalar Higgs and neutral Goldstone boson undergo the same 
mixing procedure as their charged partners, the same procedure above can be used 
for the $A^0-G^0-Z^0_\mu$ sector, with the only proviso that a factor $1/2$ must 
be put in front of the definition 
of the mixing terms $\delta Z_{AG}$ and $\delta Z_{AZ}$ in~(\ref{eq:defdzhg}) 
and~(\ref{eq:defdzhw}) to take into account the neutral nature of the particles.

The renormalization transformation for the \vev's of the Higgs potential~(\ref{eq:potential}), 
\begin{eqnarray} 
v_i\rightarrow Z_{H_i}^{1/2} (v_i+\delta v_i) 
=\left(1+{\delta v_i\over v_i}+{1\over 2}\,\delta Z_{H_i}\right)\,v_i\,, 
\label{eq:renvevs} 
\end{eqnarray} 
implies that the counterterm to $\tan\beta$ 
is related to the fundamental 
counterterms in the Higgs potential by\footnote{For a more detailed 
discussion on the $\delta\tb$ counterterm, see Sec.\,\ref{sec:tbren}.} 
\begin{equation}  
{\delta\tan\beta\over\tan\beta}={\delta v_2\over v_2}-{\delta v_1\over v_1} 
+{1\over 2}\,\left(\delta Z_{H_2}-\delta Z_{H_1}\right)\,. 
\label{eq:cdeltatan} 
\end{equation}  
If one imposes the usual on-shell renormalization conditions for the 
$A^0$-boson, one has 
\begin{equation} 
\delta Z_{H_2}-\delta Z_{H_1}= 
-{\tan\beta+\cot\beta\over \,M_Z^2}\,\Sigma_{AZ} (M_{A^0}^2)\,. 
\end{equation}  
There exists also another mixing term between $H^{\pm}$ and $G^{\pm}$  
originating from the mass matrix of the Higgs 
sector~\cite{Dabelstein:1995hb,Dabelstein:1995js,Chankowski:1994er,Yamada:1994kj}.  
This one-loop mixture is contained in: 
\begin{equation} 
V^b= \pmatrix{ {H^+}^b & {G^+}^b \cr } 
\pmatrix{ M_{H^\pm}^{b\,2} & \frac{t_{0}^b}{\sqrt{2} v^b} \cr 
\frac{t_{0}^b}{\sqrt{2} v^b} 
& \frac{t_{1}^b}{\sqrt{2} v^b} 
} 
\pmatrix{ {H^-}^b \cr {G^-}^b 
}\,, 
\end{equation} 
where we have attached a superscript ${\rm b}$ to bare quantities, 
and $t_i$ are the tadpole counterterms 
\begin{equation} 
\begin{array}{lcl} 
t_{0}&=&-\sin\left(\beta-\alpha\right)~t_{H^0} 
+\cos\left(\beta-\alpha\right)~t_{h^0}\,\,,\\ 
t_{1}&=&\sin\left(\beta-\alpha\right)~t_{h^0} 
+\cos\left(\beta-\alpha\right)~t_{H^0}\,. 
\end{array} 
\end{equation} 
We are now ready to find an expression for the mixed 2-point Green functions 
(Figs.\,\ref{diag:automixhwren}(a) and \ref{diag:automixhwren}(b)).  
For the $W^{\pm}-H^{\pm}$ mixing (Fig.\,\ref{diag:automixhwren}(a)) we can write 
the renormalized 2-point Green function as 
\begin{equation} 
\Delta^{HW}_{\mu}\equiv \frac{i}{k^2-M_{H^\pm}^2} \left[k^\nu\frac{-i\Sigma_{HW} (k^2)}{M_W}  
+ ik^\nu~M_W^2~\frac{\delta Z_{HW}}{M_W}\right]  
\frac{-i g_{\mu\nu} }{k^2-M_W^2}\,. 
\label{eq:greenHWFG} 
\end{equation} 
which allows to define a renormalized self-energy as follows 
\begin{equation} 
\hat\Sigma_{HW}(k^2)=\Sigma_{HW}(k^2)- M_W^2 \delta Z_{HW}\,. 
\label{eq:sigmahat}\end{equation} 
Now we must impose a renormalization condition on $\hat\Sigma ^{HW}(k^2)$;  
and we 
choose it in a way that the physical Higgs does not mix with the physical 
$W^\pm$: 
\begin{equation} 
\hat \Sigma_{HW} (M_{H^\pm}^2) =0 \Longrightarrow 
\delta Z_{HW}=\frac{\Sigma_{HW}(M_{H^\pm}^2)}{M_W^2}\,. 
\label{eq:renorm} 
\end{equation} 
Notice also that with this renormalization procedure on-shell $W^{\pm}$ do not 
mix also with $H^{\pm}$ since the renormalized 2-point Green function~(\ref{eq:greenHWFG}) is 
proportional to the external momentum $k^{\nu}$. 
 
We still have another ingredient, the  
mixed $H^{\pm}-G^{\pm}$ 2-point Green function: 
\begin{equation} 
\Delta^{HG}\equiv\frac{i}{k^2-M_{H^\pm}^2}\left(-i \Sigma_{HG}(k^2)+i k^2 \delta Z_{HG} - i 
\frac{t_0^b}{\sqrt{2} v^b}\right) \frac{i}{k^2-M_W^2}\,. 
\end{equation} 
This allows to define renormalized self-energy 
\begin{equation} 
\hat \Sigma_{HG}(k^2)=\Sigma_{HG}(k^2) - k^2 \delta Z_{HG}  
+ \frac{t_0^b}{\sqrt{2} v^b}\,. 
\end{equation} 
The mixed self-energies $\hat\Sigma_{HW}(k^2)$ and $\hat\Sigma_{HG}(k^2)$   
obey the following  Slavnov-Taylor identity: 
\begin{equation} 
 k^2 \hat\Sigma_{HW}(k^2)-M_W^2 \hat\Sigma_{HG}(k^2) = 0\,. 
\label{eq:STI} 
\end{equation} 
This identity is derived from a BRS   
transformation involving the Green function constructed with an anti-FP field and 
the charged Higgs field: $<0|\delta_{BRS}\,(\bar{\eta}^+\,H^+)|0>=0$. 
Following the standard procedure\,\cite{Collins84}  
one 
immediately gets: 
\begin{equation} 
<0|F^{+}\,H^+|0>=<0|\partial^{\mu}W^-_{\mu} H^+ - iM_W\,G^-H^+|0>=0\,, 
\end{equation} 
which in momentum space reads 
\begin{equation} 
k^\mu\,\Delta^{HW}_{\mu}+M_W\,\Delta^{HG}=0\,\,, 
\label{eq:MSSTI} 
\end{equation} 
with 
\begin{equation} 
\Delta^{HG}\equiv \frac{i}{k^2-M_{H^\pm}^2}\,\, 
\left[-i\,\hat{\Sigma}_{HG}(k^2) 
\right]\,\,\frac{i}{k^2-M_W^2}  
\,. 
\end{equation} 
Clearly, eq.(\ref{eq:MSSTI}) implies eq.(\ref{eq:STI}). The latter identity guarantees 
that the contribution from diagrams with external charged Higgs particles  in 
Figs.\,\ref{diag:automixhwren}(a)  and \ref{diag:automixhwren}(b) vanishes since no 
mixing is generated among the physical boson $H^{\pm}$ and the 
renormalized fields  $G^{\pm}$ and  $W^{\pm}$: 
 $\hat{\Sigma}_{HG}(M_{H^{\pm}}^2)=\hat{\Sigma}_{HW}(M_{H^{\pm}}^2)=0$. 
There is of course another Slavnov-Taylor identity, derived 
in a similar manner, which insures that the renormalized 
mixing between $G^{\pm}$ and $W^{\pm}$ also vanishes.

\subsection{Unitary gauge} 
 
It is useful also to have a look at the renormalization procedure in the Unitary 
gauge as well, where it is straightforward. Introducing the 
counterterm~(\ref{eq:defdzhw}) into the Unitary gauge Lagrangian one obtains: 
\begin{equation} 
L_{UG}=-\frac{1}{4}F_{\mu\nu}F^{\mu\nu}+M_W^2 W^+_\mu W{^-}^\mu \rightarrow 
L_{ct}=i M_W \delta Z_{H W} (W^-_\mu 
 \partial^\mu H^+ - W^+_\mu \partial^\mu H^-)\,. 
\label{eq:lct} 
\end{equation} 
In this gauge the corresponding renormalized 2-point Green function reads 
(Fig.\,\ref{diag:automixhwren}(a)): 
\begin{equation} 
\frac{i}{k^2-M_{H^\pm}^2} \left[k^\nu\frac{-i\Sigma_{HW} (k^2)}{M_W} 
+ ik^\nu~M_W^2~\frac{\delta Z_{HW}}{M_W}\right] 
\frac{-i\left(g_{\mu\nu}-\frac{k_\mu k_\nu}{M_W^2}\right) }{k^2-M_W^2}\,.  
\label{eq:green}\end{equation} 
which is identical to~(\ref{eq:greenHWFG}) but with the $W^{\pm}$ propagator in 
the Unitary gauge. Thus a renormalized self-energy can be defined with the same 
formal expression as~(\ref{eq:sigmahat}) and so it  applies the same renormalization 
condition to obtain~(\ref{eq:renorm}). 
 
Thus we have proven that the expression for $\delta Z_{HW}$ is  
formally the 
same in both Unitary and Feynman gauges, but that in the latter gauge one 
must take into account the additional renormalization of 
the mixed self-energy $\Sigma_{HG}$. Moreover, it is possible to use different 
gauge fixing for particles inside and outside the loops\cite{Papavassiliou:1995fw,Philippides:1995qi}, so we can use 
the Unitary gauge renormalization, but still maintain Goldstone bosons inside loops. 
 
\subsection{Higgs masses and wave functions} 
Whether in the Feynman or in the Unitary gauge, the charged Higgs counterterms can 
be introduced as 
\begin{equation} 
  \label{eq:countercharged} 
  H^{\pm}\rightarrow (Z_{H^{\pm}})^{1/2} H^{\pm} \ \ , \ \  \mHp\rightarrow \mHp+\delta\mHp\,\,, 
\end{equation} 
from which a renormalized self-energy can be defined as follows: 
\begin{equation} 
\hat{\Sigma}_{H^{\pm}}(k^2)= \Sigma_{H^{\pm}}(k^2)+\delta M_{H^{\pm}}^2 
-(k^2-M_{H^{\pm}}^2)\,\delta Z_{H^{\pm}}\,, 
\end{equation} 
where $\Sigma_{H^{\pm}}(k^2)$ is the corresponding unrenormalized self-energy.  
 
In order to determine the counterterms, 
we impose the following renormalization conditions:  
 
i) On-shell mass renormalization condition: 
\begin{equation} 
\hat{\Sigma}_{H^{\pm}}(M_{H^{\pm}}^2)=0\,, 
\end{equation}                                                                                                                                             
 
ii) ``Residue $=1$'' condition for the renormalized propagator 
at the pole mass: 
\begin{equation} 
\left. {\partial\hat{\Sigma}_{H^{\pm}}( k^2) 
\over \partial k^2}\right|_{k^2=M_{H^{\pm}}^2} 
\equiv\, \hat{\Sigma}_{H^{\pm}}^{\prime}(M_{H^{\pm}}^2) =0\,. 
\end{equation}                                                                                                                                             
From these conditions one derives 
\begin{eqnarray}   
\delta M_{H^{\pm}}^2 &=& -\Sigma_{H^{\pm}}(M_{H^{\pm}}^2)\,,\nonumber\\ 
\delta Z_{H^{\pm}} &=& +\Sigma_{H^{\pm}}^{\prime}(M_{H^{\pm}}^2)\,. 
\end{eqnarray} 
 
Having fixed $\delta \mHp$, $\delta Z_{H^{\pm}}$, $\delta Z_{HW}$, and  
$\delta\tb/\tb$ (see section~\ref{sec:tbren} below) all the renormalization 
constants of the Higgs sector are fixed, and one can find the value of the original 
counterterms~(\ref{eq:ZH1H2}) by inverting the set of 
equations~(\ref{eq:deltaZs1}) and~(\ref{eq:deltaZs2}).  
 
On the other hand, the renormalization of the neutral Higgs sector has been studied in the series 
of works~\cite{Dabelstein:1995hb,Dabelstein:1995js,Chankowski:1994er,Chankowski:1995ri}. We can use the expressions for the 
one-loop neutral Higgs masses given in these references by translating the 
corrections for the charged Higgs mass to corrections to the pseudoscalar Higgs 
mass, that is, from the relations  
\begin{eqnarray} 
  \label{eq:neutralhiggsMA} 
  \mHps &=& \left.\mHps\right|_{\rm Tree}+\Delta\mHps(\mA,M_{\rm SUSY})\,\,, \nonumber\\ 
  \mhs &=& \mhs(\mA,M_{\rm SUSY})\,\,, \nonumber\\ 
  \mHs &=& \mHs(\mA,M_{\rm SUSY})\,\,, \nonumber 
\end{eqnarray} 
we invert the first equation above, and use the computed value of \mAs\  as input 
for the other ones.

\subsection{\tb\  renormalization} 
\label{sec:tbren} 
At this stage a prescription to renormalize 
$\tb=v_2/v_1$,  
\begin{equation} 
\tb\rightarrow\tb+\delta\tb\,, 
\end{equation} 
is still called for. Indeed, eq.(\ref{eq:cdeltatan}) given in the 
previous section was just a formal expression which was unrelated to any 
physical input. 
There are many possible strategies. The ambiguity is  
related to the fact that \tb\  is just a Lagrangian parameter  
and as such it is not a physical observable. 
Its value beyond the tree-level is 
renormalization scheme dependent. (The situation is similar to the definition 
of the weak mixing angle $\theta_W$, or equivalently of 
$\sin^2\theta_W$.) 
However, even within a given scheme, e.g.\  the 
on-shell renormalization scheme, there are some ambiguities that must be 
fixed.  
For example, we may wish to define \tb\  in a  
process-independent (``universal'') way 
as the ratio $v_2/v_1$ between the true \vev's after renormalization of the 
Higgs 
potential\,\cite{Dabelstein:1995hb,Dabelstein:1995js,Chankowski:1994er,Yamada:1994kj,Ellis:1991zd,Brignole:1991pq,Haber:1991aw,Haber:1993an,Haber93}.    
In this case a consistent choice (i.e.\  a choice capable of renormalizing 
away the tadpole contributions) is to simultaneously  
shift the \vev's and the mass parameters 
of the Higgs potential, eq.(\ref{eq:potential}),  
\begin{eqnarray}  
v_i &\rightarrow & Z_{H_i}^{1/2} (v_i+\delta v_i)\,,\nonumber\\ 
m_i^2 &\rightarrow & Z_{H_i}^{1\over 2}\,(m_i^2+\delta m_i^2)\,,\nonumber\\ 
m_{12}^2 &\rightarrow & Z_{H_1}^{1\over 2}\,Z_{H_2}^{1\over 2} 
\,(m_{12}^2+\delta m_{12}^2)\,, 
\end{eqnarray} 
($i=1,2$) in such a way that 
$\delta v_1/v_1=\delta v_2/v_2$.  This choice generates the following 
counterterm for \tb\  in that scheme --see eq.\,(\ref{eq:cdeltatan})--: 
\begin{equation} 
{\delta\tb\over\tb}= 
{1\over 2}\,\left(\delta Z_{H_2}-\delta Z_{H_1}\right)\,. 
\label{eq:ddtan} 
\end{equation}  
Nevertheless, this procedure looks very formal and 
one may eventually like to  
fix the on-shell renormalization condition on \tb\  in a more physical 
way, i.e.\  by relating it to some concrete physical observable, so that it is 
the measured value of this observable that is taken as an input rather  
than the \vev's of the Higgs potential.  
Following this practical attitude, we choose as a physical observable  
the decay width of the charged Higgs boson into $\tau$-lepton and associated 
neutrino: $H^{+}\rightarrow\tau^{+}\nu_{\tau}$.  
This should be a good choice, 
because: 
\begin{enumerate}  
\item  
When $\mHp < \mt-\mb$, the decay 
$H^{+}\rightarrow\tau^{+}\nu_{\tau}$ 
is the dominant decay of $H^{\pm}$ already for $\tb\gsim 2$; 
\item  
From the experimental point of view 
there is a well-defined method to separate the 
final state $\tau$'s originating from $\hplus$-decay from those coming out of 
the conventional decay $W^{+}\rightarrow\tau^{+}\nu_{\tau}$, so that  
$H^{+}\rightarrow\tau^{+}\nu_{\tau}$ should be physically accessible; 
\item  
At high \tb, the charged Higgs decay of the top quark   
can have a sizeable branching ratio, serving as a source of charged Higgs 
particles; and  
\item  
If $\mHp>\mt$ the branching ratio for $H^+\rightarrow \tau^+ 
\nu_\tau$ never becomes negligible in a wide range of Higgs masses to be 
explored at the LHC rather than at the Tevatron\,\cite{Coarasa:1997ky,TesiToni}.   
\end{enumerate} 
 
The interaction Lagrangian describing the decay  
$H^{+}\rightarrow\tau^{+}\nu_{\tau}$ is  
directly proportional to \tb   
\begin{equation} 
{\cal L}_{H\tau\nu}={g\, m_{\tau}\tb\over\sqrt{2}M_W} 
\,H^-\,\bar{\tau}\,\pl\,\nu_{\tau}+{\rm h.c.}\,, 
\label{eq:LtaunuH} 
\end{equation} 
and the relevant decay width is proportional to 
$\tan^2\beta$. Whether in the $\alpha$-scheme or in the $G_F$-scheme, it 
reads: 
\begin{equation} 
\Gamma(H^{+}\rightarrow\tau^{+}\nu_{\tau})= 
{\alpha m_{\tau^{+}}^2\,M_{H^{+}}\over 8 M_W^2 s_W^2}\,\tan^2\beta=  
{G_F m_{\tau^{+}}^2\,M_{H^{+}}\over 4\pi\sqrt{2}}\,\tan^2\beta\,  
(1-\Delta r^{MSSM})\,, 
\label{eq:tbetainput} 
\end{equation} 
where we have used the relation (\ref{eq:DeltaMW}).  
By measuring this decay width  
one obtains a physical definition of \tb\  
which can be used beyond the tree-level. 
A combined measurement 
of $M_{H^{\pm}}$ and \tb\  from charged Higgs decaying 
into $\tau$-lepton 
in a hadron collider has been described 
in the literature\,\cite{Barnett:1990rv,Godbole:1991vd,Gunion:1992er,Atlas}  
by comparing the size of the various signals for charged 
Higgs boson production, such as the multijet channels accompanied by a 
$\tau$-jet or a large missing $p_T$, and the two-$\tau$-jet 
channel. 
At the upgraded Tevatron, the 
conventional mechanisms $gg(q\bar{q})\rightarrow t\bar{t}$ followed by 
$t\rightarrow \hplus\,b$ have been studied and 
compared with the usual $t\rightarrow W^+\,b$, 
and the result is that for $M_{H^{\pm}}\simeq 100\GeV$ the charged Higgs 
production is at least as large as the $W^{\pm}$ production, apart from a gap 
around  
$\tb\simeq 6$\,\cite{Barnett:1990rv,Godbole:1991vd,Gunion:1992er} (see also chap.~\ref{cap:tbh}).  
 
 
Insofar as the determination of the counterterm $\delta\tb$ 
in our scheme, it can be 
fixed unambiguously from  
our Lagrangian definition of \tb\  on eq.(\ref{eq:LtaunuH}) and  
the renormalization procedure described above (and in chapter~\ref{cap:tbh} 
for the process-dependent terms).  
It is straightforward to find: 
\begin{equation} 
{\delta\tb\over \tb} 
={\delta v\over v}-\frac{1}{2}\delta Z_{H^\pm} 
+\cot\beta\, \delta Z_{HW}+  
\Delta_{\tau}\,. 
\label{eq:deltabeta} 
\end{equation}     
Notice the appearance of the vacuum counterterm 
\begin{equation} 
{\delta v\over v}=\frac{1}{2}{\delta v^2\over v^2}=\frac{1}{2} 
\frac{\delta M_W^2}{M_W^2}-\frac{1}{2}\frac{\delta g^2}{g^2}\,, 
\label{eq:dv2} 
\end{equation} 
which is associated to $v^2=v^2_1+v^2_2$, and whose structure is fixed from 
eq.(\ref{eq:VEV}).  
The last term on eq.(\ref{eq:deltabeta}),  
\begin{equation} 
\Delta_{\tau}=-{\delta m_{\tau}\over m_{\tau}} 
-\frac{1}{2}\delta Z_L^{\nu_{\tau}}-\frac{1}{2}\delta Z_R^{\tau}-F_{\tau}\,, 
\label{eq:deltatau} 
\end{equation}     
is the (finite) process-dependent part of the counterterm (see 
section~\ref{sec:tbhoneloop}). Here  
$\delta m_{\tau}/ m_{\tau}$, $\delta Z_L^{\nu_{\tau}}$ and 
$\delta Z_R^{\tau}$ are 
obtained from eqs.(\ref{eq:deltamf}) and (\ref{eq:DSRC}) 
(with $m_{\nu_{\tau}}=0$ );  
they represent the contribution from the 
mass and wave-function renormalization of the $(\nu_{\tau},\tau)$-doublet, 
including the finite renormalization of the neutrino  
leg. Finally, $F_{\tau}$ on eq.(\ref{eq:deltatau}) is the form 
factor describing the vertex  
corrections to the amplitude of $H^{+}\rightarrow\tau^{+}\nu_{\tau}$; its value 
can be inferred from the expressions of the vertex functions in 
chapter~\ref{cap:tbh} by substituting the bottom and top quarks (and squarks) 
masses and couplings by those of $\tau$ and $\nu_\tau$ leptons (and sleptons).

On comparing eqs.(\ref{eq:ddtan}) and (\ref{eq:deltabeta}) we see that 
the first definition of \tb\  appears as though it is free from 
process-dependent contributions. 
In practice, however,   
process-dependent terms are inevitable, irrespective of the definition 
of \tb. In fact, the  
definition of \tb\  where 
 $\delta v_1/v_1=\delta v_2/v_2$  
will 
also develop process-dependent contributions, 
as can be seen by trying to relate the ``universal'' value 
of \tb\  in that scheme with a physical quantity directly read off  
some physical observable. For instance, if $M_{A^0}$  
is heavy enough, one may define \tb\  as follows: 
\begin{eqnarray} 
{\Gamma (A^0\rightarrow b\,\bar{b})\over 
\Gamma (A^0\rightarrow t\,\bar{t})}&=& 
\tan^4\beta\,{m_b^2\over m_t^2}\,\left(1-{4\,m_t^2\over 
 M_{A^0}^2}\right)^{-1/2}\,\left[1+4\,\left({\delta v_2\over v_2}- 
{\delta v_1\over v_1}\right) 
\right.\nonumber\\ 
& &\left.+2\,\left({\delta m_b\over m_b}  
+\frac{1}{2}\delta Z_L^b +\frac{1}{2}\delta Z_R^b 
-{\delta m_t\over m_t}  
-\frac{1}{2}\delta Z_L^t-\frac{1}{2}\delta Z_R^t\right) 
 +\delta V\right]\,, 
\label{eq:tanbeta3} 
\end{eqnarray} 
where we have neglected $m_b^2\ll M_{A^0}^2$, and $\delta V$ stands for the vertex 
corrections to the decay processes $A^0\rightarrow b\,\bar{b}$ and  
$A^0\rightarrow t\,\bar{t}$. 
Since the sum of the mass and wave-function renormalization 
terms along with the vertex corrections is UV-finite, one can consistently choose 
$\delta v_1/v_1=\delta v_2/v_2$ leading to eq.(\ref{eq:ddtan}). 
Hence, deriving \tb\  from eq.(\ref{eq:tanbeta3})  
unavoidably incorporates also some 
process-dependent contributions.

Any definition of \tb\  is 
in principle as good 
as any other; and in spite of the fact that 
the corrections themselves may show some dependence 
on the choice of the particular definition, 
the physical observables should not depend at all on that choice. 
However, it can be a practical matter what definition to use 
in a given situation.    
For example, our definition of \tb\  given on eq.(\ref{eq:tbetainput}) 
should be most adequate for $M_{H^{\pm}}<m_t-m_b$ and large \tb, since then 
$\hplus\rightarrow\tau^+\,\nu_{\tau}$ is the dominant decay of $\hplus$, 
whereas the definition based on 
eq.(\ref{eq:tanbeta3}) requires also a large value of \tb\  (to avoid 
an impractical suppression of the $b\,\bar{b}$ mode); moreover, in order to  
be operative, it also requires 
a much heavier charged Higgs boson, since  
$M_{H^{\pm}} \simeq M_{A^0}>2\,m_t$ when 
the decay $A\rightarrow t\bar{t}$ is kinematically open in the MSSM\@. 
(Use of light quark final states would, of course, be extremely 
difficult from the practical 
point of view.)

\subsection{Unrenormalized self-energies}

Now we can write down the expressions for the counterterms necessary in the 
charged Higgs renormalization. 
 
 \diagautohiggs 

\diagautomixhw 
 
The various Feynman diagrams contributing to the charged Higgs self-energy can 
be seen in Fig.\,\ref{diag:autohiggs}, 
\begin{eqnarray} 
  \label{dZHSusy} 
  \delta Z_{H^\pm}&=&\left.\delta Z_{H^\pm} 
                     \right|_{(a)+(b)+(c)+(d)} 
    =\,\Sigma_{H^\pm}'(\mHps)\nonumber\\ 
  &=&-\frac{ig^2N_C}{\mws}\left[(\mbs\tbs+\mts\ctbs) 
   (\Bo+\mHps\Bo'+\mbs\Bz')\right. \nonumber\\ 
  &&\left.\mbox{\hspace{1.5cm}}+2\mbs\mts\Bz'\right](\mHp,\mb,\mt)\nonumber\\ 
  &&+\frac{ig^2}{2\mws}N_C\sum_{ab}\abs{\Gba}^2 
   \Bz'(\mHp,\msbb,\msta)  \nonumber\\ 
  &&-2ig^2\sum_{i\alpha}\left[ 
\left(\abs{Q^L_{\alpha i}}^2\cbts+\abs{Q^R_{\alpha i}}^2\sbts\right) 
  (\Bo+\mHps \Bo'+\mas\Bz')\right.                              \nonumber\\ 
  &&\left.\phantom{\abs{Q^L_i}^2}+2\mi\ma Re 
   \left(Q^L_{\alpha i}Q^{R*}_{\alpha i}\right) 
   \sbt\cbt\Bz'\right](\mHp,\ma,\mi)\,. 
\end{eqnarray} 
Notice that diagram \ref{diag:autohiggs}~(c) gives a vanishing contribution to 
$\delta Z_{H^\pm}$. The mixed self-energy diagrams are in 
Fig.\,\ref{diag:automixhw} and their contribution read 
\begin{eqnarray} 
  \label{dZHWSusy} 
  \delta Z_{HW}&=&\left.\delta Z_{HW}\right|_{(a)+(b)+(c)} 
 =\frac{\Sigma_{HW}(\mHps)}{\mws}\nonumber\\ 
 &=&-\frac{ig^2N_C}{\mws}\left[\mbs\tb(\Bz+\Bo) 
  +\mts\ctb\Bo\right](\mHp,\mb,\mt)  \nonumber\\ 
 &&-\frac{ig^2N_C}{2\mws}\sum_{ab}\Gba\Rot  
  R_{1b}^{(b)*}\left[2\Bo+\Bz\right](\mHp,\msbb,\msta) \nonumber\\ 
 &&+\frac{2ig^2}{\mw}\sum_{i\alpha} 
  \left[\ma\left(\HiaR\CL+\HiaL\CR\right)(\Bz+\Bo)\right.  \nonumber\\ 
 &&+\left.\mi\left(\HiaL\CL+\HiaR\CR\right)\Bo\right] 
  (\mHp,\ma,\mi)\,.  
\end{eqnarray} 
A sum is understood over all generations.

\section{Sfermion renormalization} 
\label{sec:renormsfermion} 
 
We follow the renormalization procedure with the scalar superpartners of the 
matter fields. In fact this sector is similar to that of the scalar Higgs, for 
it involves two scalars that mix between themselves, but is different because it 
does not involve tadpole terms, and on the other hand there exist mixing terms 
between the up-type and the down-type sfermions. In the electroweak basis, and 
for a doublet $(\stopp,\sbottom)$, the counterterms needed are 
\begin{equation} 
  \label{eq:sbottomcounters} 
  ( \delta \mb,\delta A_b, \delta m_{\sbottom_R}, \delta Z_R^{\sbottom} ) 
\end{equation} 
for  sbottom particles, 
\begin{equation} 
  \label{eq:stopcounters} 
  ( \delta \mt,\delta A_t, \delta m_{\stopp_R}, \delta Z_R^{\stopp} ) 
\end{equation} 
for  stop particles, and the common counterterms 
\begin{equation} 
  \label{eq:sfermioncounters} 
  ( \delta \mz,\delta \mw,\delta\tb, \delta \mu, \delta m_{\sq_L}, \delta Z_L^{\sq} )\,\,. 
\end{equation} 
Of course if we would like to perform a supersymmetric renormalization procedure 
we should use a single wave function renormalization constant for fermions and 
sfermions, thus we should have 
$$ 
\delta Z_L^{\sq}=\delta Z_L^{q}\,\,,\,\,\delta Z_R^{\sbottom}=\delta Z_R^{b} 
\,\,,\,\,\delta Z_R^{\stopp}=\delta Z_R^{t}\,\,. 
$$ 
However, supersymmetry is 
explicitly broken and so we may take different renormalization 
constants for fermions and sfermions.  
 
Though the counterterms~(\ref{eq:sbottomcounters}), (\ref{eq:stopcounters}) 
and~(\ref{eq:sfermioncounters}) are the fundamental blocks of the sfermion 
sector it is more convenient for the on-shell renormalization scheme to use a 
different set of counterterms. In the \EW\  basis we define 
$$ 
{\cal M}_{\tilde{q}}^2 = 
\left( 
  \begin{array}{cc} 
    M_{\sq_{11}}^2+\delta M_{\sq_{11}}^2 &  
    M_{\sq_{12}}^2+\delta M_{\sq_{12}}^2\\ 
    M_{\sq_{12}}^2+\delta M_{\sq_{12}}^2 & 
    M_{\sq_{22}}^2+\delta M_{\sq_{22}}^2  
  \end{array}     
\right)\,, 
$$ 
where the various $\delta M_{\sq_{ij}}^2$ are different combinations of the 
parameter counterterms in~(\ref{eq:sbottomcounters}), (\ref{eq:stopcounters}) 
and~(\ref{eq:sfermioncounters}), except in the case of $\delta M_{\stopp_{11}}$ 
and  $\delta M_{\sbottom_{11}}$ which are related by $SU(2)$ gauge 
invariance. Thus from~(\ref{eq:stopmatrix}) one can obtain 
$$ 
\delta M_{\stopp_{11}}^2= \delta M_{\sbottom_{11}}^2 + \delta \mt^2 -\delta 
\mb^2 +\cos{2\beta} M_W^2 \left(\frac{\delta (\cos 2 \beta)}{\cos 2\beta} 
+\frac{\delta \mw^2}{\mw^2}\right)\,\,. 
$$ 
One can also derive the relation between this new set of counterterms and the 
original from expressions~(\ref{eq:stopmatrix}) and~(\ref{eq:MLRdefinition}): 
\begin{eqnarray} 
  \label{eq:renormsfermionrel1} 
  \delta M_{\sbottom_{11}}^2&=&  
  \delta M_{\sq_L}^2+\delta\mbs+\cos{2\beta}(-\frac{1}{2}+\frac{1}{3} s_W^2) 
  \mz^2 \left(\frac{\delta (\cos 2 \beta)}{\cos 2\beta}+ 
    \frac{1}{3} \frac{\delta s_W^2}{-\frac{1}{2}+\frac{1}{3} s_W^2} 
    +\frac{\delta\mzs}{\mzs}  
  \right)\,\,,\nonumber\\  
  \frac{\delta M_{\sbottom_{12}}}{M_{\sbottom_{12}}}&=&\frac{\delta \mb}{\mb} 
  +\frac{\delta A_b-\delta \mu\, \tb -\mu\, \delta \tb}{A_b-\mu \tb}\,\,,\nonumber\\ 
  \frac{\delta M_{\stopp_{12}}}{M_{\stopp_{12}}}&=&\frac{\delta \mt}{\mt} 
  +\frac{\delta A_t-\delta \mu\, \ctb -\mu\, \delta \ctb}{A_t-\mu 
    \ctb}\,\,,\nonumber\\ 
  \delta M_{\sq_{22}}^2&=&\delta M_{\sq_R}^2+\delta m_q^2 
  + \cos{2\beta}\,\Qq\,s_W^2\, \mzs\left( 
    \frac{\delta (\cos 2 \beta)}{\cos 2\beta}+\frac{\delta s_W^2}{s_W^2}+\frac{\delta\mzs}{\mzs}  
  \right)\,\,\,, 
\end{eqnarray} 
where the last expression is valid for both 
type of sfermions, just performing the appropriate substitution 
$q\rightarrow\{t,b\}$.  
 
If we had to deal with observables in which all parameters in the RHS 
of~(\ref{eq:renormsfermionrel1}) appear in the  
tree-level expressions, as in the squark decays of Higgs particles, then we 
should invert this set of equations to obtain each counterterm corresponding to 
the appropriate variables. 
 
In the observables we will compute in this Thesis only one squark appears as an 
external particle (see chapter \ref{cap:sbdecay}), thus all the one-loop 
contributions to other squarks will be higher order contributions to these 
observables. In this situation it is better to use a different approach which 
uses the physical particles themselves. In this approach we introduce mass 
counterterms for the physical particles, the mixing angle, wave function 
renormalization constants for each squark, and mixing wave function 
renormalization constants, that is 
\begin{equation} 
  \label{eq:countersbottom} 
  (\delta \msbo,\delta \msbt,\delta \theta_{\sbottom},\delta Z^1,\delta Z^2,\delta Z^{12},\delta Z^{21}) 
\end{equation} 
for the sbottom squark. The number of parameter counterterms in this set is 
equal to the one in the electroweak basis, thus we are not introducing any new 
parameter counterterm, but 
using a new combination of the old ones. For the wave function counterterms we 
have added mixing terms between the squarks; its purpose is to make possible 
the one-particle-irreducible (1PI) renormalization procedure and  avoid the 
presence of mixing between physical squarks at one-loop. As noted above 
(section~\ref{sec:renormintro}) wave function renormalization is unnecessary, 
but it allows to renormalize the theory by renormalizing at the 
same time every Green function.

The definition of the renormalization constants~(\ref{eq:countersbottom}) is 
\begin{eqnarray} 
  \label{eq:deltazsbottom} 
  \msbas{}^0&=&\msbas+\delta \msbas\,\,,\nonumber\\ 
  \theta_{\sbottom}^{0}&=&\theta_{\sbottom}+\delta \theta_{\sbottom}\,\,, \nonumber\\ 
  \sbottom_{a}^{0}&=&(1+\frac{1}{2}\delta Z^a)\, \sbottom_a^{}+\delta 
  Z^{ab}\,\sbottom_b^{}\ \ \ (a \not= b)\,\,, 
\end{eqnarray} 
from which we write the one-loop kinetic Lagrangian 
\begin{eqnarray} 
  \label{eq:sbottomlagrangian} 
  {\cal L}^0&=&{\cal L}+\delta {\cal L}= 
  (\partial^\mu {\sbottom_a}^{*}\, \partial_\mu \sbottom_{a}^{}-\msbas\, {\sbottom_a}^{*}\, 
  \sbottom_a^{})\,(1+\delta Z^a)- 
  \delta \msbas\, {\sbottom_a}^*\, \sbottom_a^{}\nonumber\\ 
  &+& (\delta Z^{12}+\delta Z^{21})\, ( \partial^\mu \sbottom_1^{*}\, \partial_\mu 
  \sbottom_{2}^{}+\partial^\mu \sbottom_2^{*}\, \partial_\mu \sbottom_{1}^{}  ) \nonumber\\  
  &-&\delta Z^{12}\, \msbos \,(\sbottom_1^{*}\,\sbottom_{2}^{}+\sbottom_{1}^{}\,\sbottom_2^{*}) 
  -\delta Z^{21}\, \msbts\, (\sbottom_1^{*}\,\sbottom_{2}^{}+\sbottom_{1}^{}\,\sbottom_2^{*})\,\,, 
\end{eqnarray} 
that allows to obtain the one-loop inverse propagator 
\begin{equation} 
  \label{eq:sbinversepropag} 
  i\Delta^{-1}(k^2)= i 
  \left(   \begin{array}{cc} 
      \Delta^{-1}_{11}(k^2)   &    \Delta^{-1}_{12}(k^2) \\ 
      \Delta^{-1}_{21}(k^2)   &    \Delta^{-1}_{22}(k^2)   
  \end{array} 
    \right)\,\,, 
\end{equation} 
with 
\begin{eqnarray} 
\Delta^{-1}_{11}(k^2)&=&  (k^2-\msbos) (1+\delta Z^1)-\delta \msbo-\Sigma^{11}(k^2)\,\,, \nonumber\\ 
\Delta^{-1}_{12}(k^2)&=&   (\delta Z^{21}+\delta Z^{12})\,k^2-\msbts\, \delta Z^{21} 
  -\msbos\,\delta Z^{12}-\Sigma^{12}(k^2)\,\,, \nonumber\\ 
\Delta^{-1}_{21}(k^2)&=&   (\delta Z^{21}+\delta Z^{12})\,k^2-\msbos\,\delta Z^{21} 
  -\msbts\,\delta Z^{12}-\Sigma^{21}(k^2) \,\,,\nonumber\\ 
\Delta^{-1}_{22}(k^2)&=&   (k^2-\msbts) (1+\delta Z^2)-\delta \msbt-\Sigma^{22}(k^2) \,\,\,. 
\end{eqnarray} 
Next we follow the on-shell prescription requiring the mass parameters to be 
the physical masses, the ``residue=1'' condition and the absence of mixing 
between squarks on-shell 
\begin{eqnarray} 
  \label{eq:onshellsbottom} 
  \Delta^{-1}_{aa}(\msbas)&=&0\,\,,\nonumber\\ 
  \Delta^{-1}_{ab}(\msbas)&=&0\,\,,\nonumber\\ 
  (\Delta^{-1}_{aa})^{\prime}&=&1\,\,, 
\end{eqnarray} 
and from that obtain the counterterms 
\begin{eqnarray} 
  \label{eq:defsbottomcounter} 
  \delta \msbas&=&-\Sigma^{aa}(\msbas) \,\,,\nonumber\\ 
  \delta Z^a&=&\Sigma^{aa}{}^{\prime}(\msbas)\,\,,\nonumber\\ 
  \delta Z^{ab}&=&\frac{\Sigma^{ab}(\msbbs)}{\msbbs-\msbas}\,\,. 
\end{eqnarray} 
 
For fixing $\delta\theta_{\sbottom}$, we require that the 
renormalized mixing angle 
(that we use as an input data) does not 
feel a shift from  
the mixed sbottom bare self-energies $\Sigma^{ab}$ between the 
physical states $\sbottom_a$ and $\sbottom_b$ ($a\neq b$). 
This is similar to the prescription adopted in 
Refs.\cite{Djouadi:1997wt,Kraml:1996kz,Eberl:1996wa},  
though it is not identical.   
In our formalism, the 3-point Green functions explicitly incorporate  
the mixed field renormalization 
constants $\delta Z^{ab}$ ($a\neq b$) and are therefore renormalized also 
in the $\theta_{\sbottom}$ parameter. The  
UV-divergent parts of the 3-point functions are canceled against 
$\delta\theta_{\sbottom}$ by defining the latter as follows: 
\begin{equation} 
\delta\theta_{\sbottom}=\frac12\,(\delta Z^{12}-\delta Z^{21}) 
=\frac{1}{2}\,{\Sigma^{12}(m_{\sbottom_2}^2)+\Sigma^{12}(m_{\sbottom_1}^2) 
\over m_{\sbottom_2}^2-m_{\sbottom_1}^2}\,. 
\label{eq:rentheta} 
\end{equation} 
Of course another equivalent choice could just be 
$\delta\theta_{\sbottom}=\delta Z^{12}$ (or $-\delta Z^{21}$), but  
eq.(\ref{eq:rentheta}) is more symmetrical; the numerical  
differences among the finite 
parts of the two choices are negligible\cite{Djouadi:1997wt}. This renormalization prescription 
deviates somewhat from the on-shell philosophy, but, contrary to the \tb\  case, 
it is not clear by now how the squark angle will ever be measured; thus it is better 
to use a generic criteria, like eq.~(\ref{eq:rentheta}) or the ones in Refs.\cite{Djouadi:1997wt,Kraml:1996kz,Eberl:1996wa}. 
 
\diagautoenersbqcd 
 
The various \QCD\  Feynman diagrams contributing to the sbottom self-energies can 
be seen in  
Fig.\,\ref{diag:autoenersbqcd}. We have  also computed the 
\EW\  contributions in the Yukawa approximation and they are presented in 
chapter~\ref{cap:sbdecay}. From these diagrams, and with the  
interaction Lagrangians of chapter~\ref{cap:MSSM}, the unrenormalized 
self-energies can be computed. In the following we describe the contributions 
corresponding to each diagram. 
 
The gluon graph from Fig.\,\ref{diag:autoenersbqcd}~(a) only contributes to the 
diagonal self-energy 
\begin{eqnarray} 
  \label{eq:selfsbottomQCDgluon} 
\left.\Sigma^{aa}(k^2)\right|_{(a)}&=&-i\,4\,\pi\,\alpha_s\,C_F\left( 
  -A_0(\msbas)+2\,A_0(\lambda^2)\right.\nonumber\\ 
 &&\left.+ (2\,k^2-\lambda^2+2\,\msbas)B_0(k,\lambda,\msba)\right)\,\,, 
\end{eqnarray} 
where we introduced a small gluon mass $\lambda$ to regularize the infrared 
divergence, and $C_F=(N_C^2-1)/2\,N_C=4/3$ is a colour factor. The wave function 
renormalization constant derived from this expression is 
\begin{equation} 
  \label{eq:wavesbottomQCDgluon} 
  \left.\delta Z^a\right|_{(a)}=-i\,8\,\pi\,\alpha_s\,C_F\left( 
    2\,\msbas\,\Bzp+\Bz\right) (\msba,\lambda,\msba)\,\,, 
\end{equation} 
which is highly simplified if we use the appropriate limits for the 
various scalar functions $B_{*}$ of Appendix~\ref{ap:pointfun}, obtaining 
\begin{equation} 
  \label{eq:wavesbottomQCDgluonsimple} 
  \left.\delta Z^a\right|_{(a)}=-2\,\frac{\alpha_s}{3\,\pi}(\Delta+\log\frac{\lambda^2}{\mu^2})\,\,, 
\end{equation} 
where $\mu$ is the scale factor and $\Delta$ represents the UV divergence, as 
defined in equations~(\ref{eq:muUV}) and~(\ref{eq:DeltaUV}) respectively. The 
other gluonic diagram in Fig.\,\ref{diag:autoenersbqcd} (b) is zero as it is 
proportional to 
$$ 
\Az(\lambda)\,\,\, 
\raisebox{-6pt}{$\stackrel{\textstyle \longrightarrow}{\scriptscriptstyle 
      \lambda\rightarrow 0}$}\,\,\,0\,\,. 
$$ 
 
The purely gluino contribution depicted in Fig.\,\ref{diag:autoenersbqcd} (c) 
reads  
\begin{eqnarray} 
  \label{eq:selfsbottomQCDgluino} 
  \left.\Sigma^{ab}(k^2)\right|_{(c)}&=&-i\,16\,\pi\,\alpha_s\,C_F\left[ 
    \delta_{ab} (\Bzt+k^2\,\Bo)\right.\nonumber\\ 
    &&\left.\phantom{\Bzt} 
      -\mg\,\mb\left(R_{1b}^{(b)}\,R_{2a}^{(b)}+R_{2b}^{(b)}\,R_{1a}^{(b)}\right) \Bz 
    \right](k,\mg,\mb)\,\,, 
\end{eqnarray} 
which contributes to the wave function renormalization constants as 
\begin{eqnarray} 
  \label{eq:wavesbottomQCDgluino} 
  \left.\delta Z^{a}\right|_{(c)}&=&-i\,16\,\pi\,\alpha_s\,C_F\left[ 
    \Bo+\msbas\,\Bop+ 
    (\mgs-2\,\mg\,\mb\,R_{1a}^{(b)}\,R_{2a}^{(b)})\Bzp\right](\msba,\mg,\mb) 
  \nonumber\\ 
  \left.\delta Z^{ab}\right|_{(c)}&=&i\,16\,\pi\,\alpha_s\,C_F 
   \,\frac{\mg\,\mb}{\msbbs-\msbas}\left(R_{1b}^{(b)}\,R_{2a}^{(b)}+R_{2b}^{(b)}\,R_{1a}^{(b)}\right) 
    \Bz(\msbb,\mg,\mb)\,\,.  
  \end{eqnarray} 
Finally the squark loop contribution from Fig.\,\ref{diag:autoenersbqcd} (d) is 
\begin{eqnarray} 
  \label{eq:selfsbottomQCDsquark} 
  \left.\Sigma^{ab}(k^2)\right|_{(d)}&=&- i\,4\,\pi\alpha_s\,C_F\, 
  \left(R_{1b}^{(b)}R_{1a}^{(b)}-R_{2b}^{(b)}R_{2a}^{(b)}\right)\left\{ 
    \left[(R_{11}^{(b)})^2-(R_{21}^{(b)})^2\right]\,\Az(\msbo) \right.\nonumber\\ 
&+&\left. \left[(R_{12}^{(b)})^2-(R_{22}^{(b)})^2\right]\,\Az(\msbt) 
    \right\}\,\,, 
\end{eqnarray} 
which only contributes to the mass counterterm and the mixed 
self-energy~(\ref{eq:defsbottomcounter}). From the 
Lagrangian~(\ref{eq:scalarQCDLagrangian}) it could 
seem that other squarks could give rise to similar contributions, however these 
are proportional to traces of Gell-Mann matrices, and as a consequence they are 
identically zero.



%% file: tbh.tex

\chapter{Quantum effects on  
$t\rightarrow H^{+}\,b$ in the \MSSM}
\label{cap:tbh}

\section{Introduction}
\label{sec:tbhintro}

In this chapter we analyze the one-loop quantum effects (both \QCD\  and \EW)
to the unconventional top quark decay mode  $t \rightarrow \hplus\,b$ in the \MSSM, and their
effect on the Tevatron Collider physics\footnote{The study in the case of
  generic \thdm\  have been presented in\,\cite{Coarasa:1998xy}, see also
  Ref.\,\cite{TesiToni} for a comprehensive study.}. The analysis of this  process at the
quantum level is useful to unravel the potential supersymmetric nature of the
charged Higgs emerging from that decay. This decay has been subject of interest
since very early in the
literature\cite{Bigi:1986jk,Barger:1990rh,Bawa:1990pc,Drees:1991nf,Bullock:1991fd,Bernreuther:1991hy,Kunszt:1992qe,Roy:1992sp},
we wish to emphasize that this decay is not excluded by present data from the Tevatron
(see sections~\ref{sec:tbhtree} and~\ref{sec:limitstanb}). Therefore, we will analyze
both the \QCD\  and the \EW\  corrections to that decay. The conventional gluon-mediated strong (\QCD)
corrections have been computed in~\cite{Czarnecki:1993ig,Czarnecki:1993zm}, the
\SUSY-\QCD\  corrections mediated by gluinos, stops and sbottoms have been
computed in~\cite{Guasch:1995rn}, and a very detailed discussion can
be found 
in\cite{TesiRicci}. Here we will concentrate in the remaining part, to wit: the
electroweak one-loop quantum corrections mediated by squarks, sleptons,
charginos, neutralinos and supersymmetric Higgs bosons. These corrections were
first computed in~\cite{Coarasa:1996qa}, and we will combine all of these to obtain
the full \MSSM\  quantum corrections.

In this study we will concentrate on those regions of the parameter space in
which the partial decay width $t\rightarrow \hplus\,b$ is competitive with the
\sm\  decay width $t\rightarrow W^+\,b$, and where the one-loop \EW\  quantum
corrections are important. We will take the convention that the decay
$t \rightarrow \hplus\,b$ will be interesting whenever its branching ratio is
$BR(t\rightarrow \hplus\,b)> 10\%$. Theoretically this condition is taken for
granted when \tb\  is large enough ($\gsim 30$). Under these conditions the
\SUSY-\QCD\  corrections can be around $50\%$, and the purely \SUSY-\EW\  ones,
induced by the Yukawa couplings $\lambda_t$ and $\lambda_b$ can reach $20\%$, so
both effects could be measured at the Tevatron and/or the LHC.

In section~\ref{sec:tbhtree} we present the tree-level relations, and the
status of the charged Higgs decay of the top quark in view of the Tevatron
data. In section~\ref{sec:tbhoneloop} we describe the process-dependent
renormalization procedure and we write down all
the analytical formulae for the three point irreducible vertices. In
section~\ref{sec:tbhnumeric} we make an exhaustive numerical analysis of the
\MSSM\  (\QCD+\SUSY-\QCD+\SUSY-\EW) corrections. In section~\ref{sec:limitstanb} we make an study of the implications of
these corrections for the Tevatron data. Finally in
section~\ref{sec:tbhconclusions} we present our conclusions.

\section[Tree-level relations and experimental determination of 
$BR (t\rightarrow \hplus\,b)$]{Tree-level relations and experimental determination of \\
$BR (t\rightarrow \hplus\,b)$} 
\label{sec:tbhtree}

We recall here the tree-level interaction Lagrangian~(\ref{LqqH}) in terms of
the mass-eigenstates 
\begin{equation}
{\cal L}_{Hbt}={g\,V_{tb}\over\sqrt{2}M_W}\,\hmin\,\bar{b}\,
[m_t\cot\beta\,\pr + m_b\tb\,\pl]\,t+{\rm h.c.}\,,
\label{eq:LtbH}
\end{equation}
where $V_{tb}$ is the corresponding Cabibbo-Kobayashi-Maskawa matrix element. On
the phenomenological side, one should not dismiss the possibility that  
the bottom-quark Yukawa coupling could play a
momentous role in the physics of the top quark, to the extend of
drastically changing standard expectations on top-quark observables,
particularly on the top-quark width. Of course, this is possible because of the
potential \tb-enhancement of that Yukawa coupling.

From the Lagrangian~(\ref{eq:LtbH}), the tree-level width of the unconventional
top quark decay into a charged Higgs boson reads:
\begin{eqnarray}
\Gamma^{(0)}(t\rightarrow \hplus\,b)
&=&\left({G_F\over 8\pi\sqrt{2}}\right){|V_{tb}|^2\over m_t}
\ \lambda^{1/2}
(1, {m_b^2\over m_t^2},{\mHps\over m_t^2})\nonumber\\
& & \times[(m_t^2+m_b^2-\mHps) (m_t^2\cot^2\beta+m_b^2\tan^2\beta)+4m_t^2m_b^2]\,,
\label{eq:treeH}
\end{eqnarray}    
where
\begin{equation}
\lambda^{1/2} (1, x^2, y^2)\equiv\sqrt{[1-(x+y)^2][1-(x-y)^2]}\,.
\end{equation}
It is useful to compare eq.(\ref{eq:treeH}) with the tree-level 
width of the canonical top quark decay in the SM:
\begin{eqnarray}
\Gamma^{(0)}(t\rightarrow W^+\,b) & = &
\left({G_F\over 8\pi\sqrt{2}}\right){|V_{tb}|^2\over m_t} 
\ \lambda^{1/2} (1, {m_b^2\over m_t^2}, {M_W^2\over m_t^2})
\nonumber\\
& & \times [M_W^2(m_t^2+m_b^2)+ (m_t^2-m_b^2)^2-2M_W^4]\,.
\label{eq:treeW}
\end{eqnarray}
The ratio between the two partial widths becomes more transparent upon
neglecting the kinematical bottom mass
contributions, while retaining all the Yukawa coupling effects: 
\begin{equation}
{\Gamma^{(0)}(t\rightarrow H^+\,b)\over \Gamma^{(0)}(t\rightarrow W^+\,b)}=
{\left(1-{M_{H^{+}}^2\over m_t^2}\right)^2\,
\left[{m_b^2\over m_t^2}\,\tan^2\beta+\cot^2\beta\right]
\over
\left(1-{M_W^2\over m_t^2}\right)^2\,\left(1+2{M_W^2\over m_t^2}\right)}\,.
\label{eq:ratioHW} 
\end{equation}     
We see from it that if \mHp\  is not much heavier than \mw, then
there are two regimes, namely a low and a high \tb\  regime,
where the decay rate of the
unconventional top quark decay becomes sizeable 
as compared to the conventional decay. They can be 
defined approximately as follows:
i) Low \tb\  regime: $\tb<2$, and 
ii) High \tb\  regime: $\tb\geq m_t/m_b\simeq 35$.
The critical regime of the decay
$t\rightarrow \hplus\,b$ occurs at the intermediate value 
$ \tb= \sqrt{m_t/m_b}\sim 6$, where the partial width has a pronounced
dip. Around this value, the canonical decay $t\rightarrow W^+\,b$ is dominant
over the charged Higgs decay; more specifically, for
$3\lsim\tb\lsim 15$
the decay rate of the mode $t\rightarrow \hplus\,b$ is basically irrelevant as
compared to the standard mode: $BR(t\rightarrow \hplus\,b)<10\%$.
Therefore, a detailed study of the 
quantum effects within that interval is of no practical interest.

Even though the approximate perturbative regime
for \tb\  extends over the wide range
\begin{equation}
0.5\lsim\tb\lsim 70\,,
\label{eq:tanbeta}
\end{equation}
we shall emphasize the results obtained in
the phenomenologically interesting high \tb\
region (typically $\tb\gsim 30$). As for the low \tb\  range,
while $BR (t\rightarrow \hplus\,b)$ can also be sizeable it turns out that
the corresponding quantum
effects are generally much smaller than in the high \tb\  case
(Cf. Section~\ref{sec:tbhnumeric}). Still, we find that in the very low
\tb\  segment $0.5\lsim\tb\lsim 1$ these effects
can be of some phenomenological interest and we shall also report on them.  

As a matter of fact, and despite naive expectations, the non-\sm\  branching
ratio 
$BR (t\rightarrow \hplus\,b)$ is not as severely constrained
as apparently dictated by the existing measurements of the \SM\  
branching ratio
at the Tevatron, namely,
$BR (t\rightarrow W^+\,b)\gsim 70\%$\,\cite{Moriond96}.
To assess this fact, notice that the former result
strictly applies only 
under the assumption that the sole source of top quarks in $p\bar{p}$
collisions is the  standard Drell-Yan pair
production mechanism
 $q\,\bar{q}\rightarrow t\,\bar{t}$\,\cite{Yuan:1995ez}. 
Now, as noted in chapter~\ref{cap:Intro} the observed
cross-section is equal to the Drell-Yan 
production cross-section
convoluted over the parton distributions times the
squared branching ratio~(\ref{eqn:productionSM}), whereas in the \MSSM\  one
expects a generalization of the production mechanism through the production and
subsequent decay of $R$-odd particles~(\ref{eq:productionMSSM}).

It should be clear that the observed cross-section on eq.(\ref{eq:productionMSSM})
refers not only to the standard
$bW\,bW$ events, but to all kind of final states that can simulate them. 
Thus, effectively, we should substitute $BR (t\rightarrow W^+\,b)$ in that
formula by $BR (t\rightarrow X\,b)$, and then sum the cross-section
over $X$, where $X$ is any state that leads to an observed 
pattern of leptons and jets similar to those resulting from $W$-decay.
In particular, 
$X=H^{\pm}$ would contribute (see below) to the $\tau$-lepton signature,
if \tb\  is large enough. Similarly, there can be
direct top quark decays into SUSY
particles that could mimic the SM decay of the top quark\,\cite{Guasch:1997dk}.
Notwithstanding, even in the absence of 
the $X$ contributions, eq.(\ref{eq:productionMSSM}) shows
that if there are alternative (non-SM) sources of top quarks subsequently decaying
into the SM final state, $W^+\,b$, one cannot rigorously
place any stringent upper bound on
$BR (t\rightarrow W^+\,b)$ from the present data.
The only restriction being an approximate lower
bound $BR (t\rightarrow W^+\,b)\gsim 40-50\%$ in order to guarantee the
purported standard top quark events at the
Tevatron\,\cite{Abe:1995hr,Abachi:1995iq}. 
Thus, from these considerations it is not excluded
that the non-SM branching ratio of the top quark, 
$BR (t\rightarrow $``new''$)$, could be
as big as the SM one, i.e.\  $\sim 50\%$.

Notice that at present one cannot exclude eq.(\ref{eq:productionMSSM})
since the observed form of the conventional $t\rightarrow W^+\,b$ final state
involves missing energy, as it is also the case for the decays
comprising supersymmetric particles.
A first step to improve
this situation would be to compute some of the 
additional top quark production cross-sections in the \MSSM\  under given hypotheses
on the \SUSY\  spectrum. For instance, the inclusion of the 
$q\,\bar{q}\rightarrow \tilde{g}\,\bar{\tilde{g}}$ mechanism followed by
the $\tilde{g}\rightarrow t\,\bar{\tilde{t}}_1$ decay has been considered
in Ref.\,\cite{Kane:1996ny,Kon:1994uc}, 
where it was claimed that
$BR (t\rightarrow \tilde{t}_1\,\chi_1^0)\simeq 50\%$.
 By the same token, one cannot place any compelling
restriction on $BR (t\rightarrow \hplus\,b)$ from the present FNAL data.
In particular, if \tb\  is large and there exists a relatively light
chargino with a non-negligible higgsino component, the third mechanism suggested on
eq.(\ref{eq:productionMSSM}), namely 
$q\,\bar{q}\rightarrow \tilde{b}_a\,\bar{\tilde{b}}_a$ followed by
$\tilde{b}_a\rightarrow t\,\chi^-_1$, could also be a rather efficient
non-SM source of top quarks. Moreover,
if $100\GeV\lsim M_{H^{\pm}}\lsim 150\GeV$, 
then a sizeable portion of the top quarks will decay into a charged Higgs.
Thus, if either
$m_t+m_{\tilde{t}_1}\lsim m_{\tilde{g}}\lsim 300\GeV$
and/or $m_t+m_{\chi_i^-}\lsim m_{\tilde{b}_a}\lsim 300\GeV$, so that at
least one of the alternative \SUSY\  sources of top quark final states 
contributing to eq.(\ref{eq:productionMSSM}) is available
(and $m_{\tilde{g}}, m_{\tilde{b}_a}$ are not too heavy so that the production
cross-section is not too phase-space suppressed), then 
one may equally argue that
a large branching ratio $BR (t\rightarrow \hplus\,b)\simeq 50\%$ is not incompatible
with the present measurement of the top quark
cross-sections\,\cite{Abe:1995hr,Abachi:1995iq}. 
This could be most likely the case if the frequently 
advocated \SUSY\  decay 
$t\rightarrow \chi_1^0\,\tilde{t}_1$ is kinematically forbidden. Nonetheless,
even if it is allowed, it is non-enhanced 
in our preferential large \tb\  region, in contrast
to $t\rightarrow \hplus\,b$.

Furthermore, it is worth mentioning that the decay mode $t\rightarrow \hplus\,b$
has a distinctive signature 
which could greatly help in its detection, viz.\  the fact that at large
\tb\  the emergent charged Higgs 
would seldom decay into a pair of quark jets, but rather into
a $\tau$-lepton and associated neutrino. This follows from inspecting the
ratio
\begin{eqnarray}  
{\Gamma(H^{+}\rightarrow\tau^{+}\nu_{\tau})\over
\Gamma(H^{+}\rightarrow c\bar{s})}&=&\frac{1}{3}
\left(\frac{m_{\tau}}{m_c}\right)^2
{\tan^2\beta\over
(m^2_s/ m^2_c)\tan^2\beta+\cot^2\beta}\nonumber\\
&\rightarrow & \frac{1}{3}\left(\frac{m_{\tau}}{m_s}\right)^2
>10\ \ \ \ \ \ ({\rm for}\ \tb>\sqrt{m_c/m_s}\gsim 2)\,,
\label{ratiotaucs}
\end{eqnarray}
where we see that the identification of the charged Higgs decay of the top quark
could be a matter of measuring a departure from the universality prediction
for all lepton channels. In practice, $\tau$-identification is possible at the
Tevatron; and the feasibility of tagging the excess of events
with one isolated $\tau$-lepton as compared to events with an additional lepton
has also been substantiated by studies of the LHC collaborations \,\cite{Atlas}.
The experimental signature for $t\bar{t}\rightarrow \hplus\,H^-\,b\,\bar{b}$
would differ from $t\bar{t}\rightarrow W^+\,W^-\,b\,\bar{b}$ by
an excess of final states with two $\tau$-leptons and two b-quarks and large missing
transverse energy.

A study in this direction by the CDF 
collaboration at the Tevatron\,\cite{Conway}
has been able to exclude a large portion of the
$(\tb, M_{H^{\pm}})$-plane characterized by $\tb\gsim 60$
and $M_{H^{\pm}}$ below a given value which varies with \tb.
For extremely high 
$\tb\gsim {\cal O}(100)$, the uppermost excluded mass region is
$M_{H^{\pm}}\lsim 140\GeV$.  
However, within the interval $\tb=60-80$, the allowed upper
limit on $M_{H^{\pm}}$ varies
very fast with \tb.
In particular, the \MSSM\  permissible values $M_{H^{\pm}}\gsim 110\GeV$ 
(compatible with $M_{A^0}\gsim 75\GeV$) are not manifestly excluded for
\tb\  equal or below the perturbative bound $\tb=70$,
eq.(\ref{eq:tanbeta}). On the other hand radiative corrections alter this bounds
in a significant way (see sec.~\ref{sec:limitstanb}).
We shall nevertheless err on the conservative side and assume 
that $\tb\leq 60$ throughout our analysis. Thus,
as far as the high \tb\  regime is concerned, we will for definiteness 
optimize our results in the safe, and phenomenologically interesting, 
high \tb\  segment
\begin{equation}
30\leq\tb\leq 60\,.
\label{eq:tanbeta2}
\label{eq:htanr}
\end{equation}
To round off the $\tau$-lepton business,
it has been shown that it should be fairly easy to  
discriminate between the $W$-daughter $\tau$'s and the 
$H^{\pm}$-daughter $\tau$'s by just taking advantage 
of the opposite states of $\tau$ polarization resulting from the $W^{\pm}$
and $H^{\pm}$ decays; the two polarization states can be 
distinguished by
measuring the charged and neutral contributions to the $1$-prong $\tau$-jet
energy (even without identifying the individual meson states)
\,\cite{Raychaudhuri:1995kv,Raychaudhuri:1996cc}. 

In short, there are good prospects for
detecting the decay $t\rightarrow \hplus\,b$, if it is kinematically accessible.
Unfortunately, on the sole basis of 
computing tree-level effects we cannot find out whether the charged Higgs
emerging from that decay is supersymmetric or not. Quantum effects, however, can.

\section{One-loop corrected $\Gamma(t\rightarrow \hplus\,b)$ in the \MSSM}
\label{sec:tbhoneloop}

First of all let us sketch the renormalization procedure for the
$t\,b\,H^{+}$-vertex. 
The full $t\,b\,H^{+}$ bare Lagrangian
is found by substituting the $t$, $b$ and $H^{+}$ fields by bare fields, and $g$
and the various masses by bare parameters in~(\ref{eq:LtbH}), as defined in
chapter~\ref{cap:Renorm}. There is still another piece to be add, namely the
$W^\pm-H^\pm$ mixing~(\ref{eq:defdzhw}), which must substitute the $W^\pm$ field
in the gauge interaction Lagrangian~(\ref{eq:Lqqw}). 
Proceeding in this way  we find
\begin{equation}
{\cal L}_{Hbt}^0={\cal L}_{Hbt}+{g\over\sqrt{2}\,M_W}\,H^-\,\bar{b}\left[
\delta C_R\ m_t\,\cot\beta\,\,\pr+
\delta C_L\ m_b\,\tb\,\pl\right]\,t
+{\rm h.c.}\,,
\label{eq:LtbH2}
\end{equation}
with
\begin{eqnarray}
\delta C_R &=& {\delta m_t\over m_t}-{\delta v\over v}
+\frac{1}{2}\,\delta Z_{\hplus}+\frac{1}{2}\,\delta Z_L^b+\frac{1}{2}
\,\delta Z_R^t
-{\delta\tb\over\tb}+\delta Z_{HW}\,\tb\,,\nonumber\\
\delta C_L &=& {\delta m_b\over m_b}-{\delta v\over v}
+\frac{1}{2}\,\delta Z_{\hplus}+\frac{1}{2}\,\delta Z_L^t+\frac{1}{2}
\,\delta Z_R^b
+{\delta\tb\over\tb}-\delta Z_{HW}\,\cot\beta\,,
\end{eqnarray}
and where we have set $V_{tb}=1$  ($V_{tb}=0.999$ within
$\pm 0.1\%$, from unitarity of the CKM-matrix under the assumption of
three generations).

As stated in Section~\ref{sec:tbhtree}, the study of the decay $t\rightarrow \hplus\,b$
is worthwhile in the small ($\tb<2$), and most conspicuously in
the high ($\tb\geq 30$) \tb\  region, where the branching 
ratio can be comparable to the one of the standard decay $t\rightarrow W^+\,b$.
These are, therefore, the regions on which we will focus our search for
potentially significant (strong and electroweak like) \SUSY\  quantum
corrections to $t\rightarrow \hplus\,b$. 
As for the strong effects, they can be rather large and
have been evaluated in Refs.\,\cite{TesiRicci,Guasch:1995rn};
here we shall not dwell any longer
on their detailed structure apart from including them in our numerical analysis
and recalling some interesting remarks in section~\ref{sec:tbhnumeric}.

On the electroweak side, one may also expect
sizeable quantum corrections from  enhanced Yukawa couplings
of the type (\ref{eq:Yukawas}).
In the relevant \tb\  regions mentioned above, 
the latters yield the leading electroweak contributions  and
in these conditions we will neglect the
pure gauge corrections from transversal gauge bosons in the Feynman gauge. 
Moreover, as already stressed in section~\ref{sec:tbhtree}, the branching ratio of the 
charged Higgs mode in the intermediate \tb\  region is too small to speak of, 
so that the detailed structure of the radiative corrections in this range
is irrelevant.

\diagtbhtree

In the following we will describe
the relevant electroweak one-loop supersymmetric diagrams entering
the amplitude of  
$t\rightarrow \hplus\,b$ in the \MSSM\@.  At the tree-level, the
only  Feynman diagram is the one in Fig.\,\ref{diag:tbhtree}. 
At the one-loop level, we have the vertex diagrams exhibited in
Figs.\,\ref{diag:tbhSEW}-\ref{diag:tbhHIGGS} and the fermion and Higgs
self-energies of chapter~\ref{cap:Renorm}. 
The computation of the one-loop diagrams requires to use the full structure of
the \MSSM\  Lagrangian. 

Specifically, Fig.\,\ref{diag:tbhSEW} shows the electroweak
\SUSY\  vertices involving squarks, charginos and neutralinos. 
In all these diagrams a sum over all indices is taken for
granted.
The supersymmetric Higgs particles of the \MSSM\  and
Goldstone bosons (in the Feynman gauge) contribute a host of one-loop vertices
as well (see Fig.\,\ref{diag:tbhHIGGS}). 
As for the various self-energies, they will be treated as counterterms
to the vertices. Their structure is dictated by the
Lagrangian (\ref{eq:LtbH2}).

\diagtbhSEW

Although we have displayed only the process dependent diagrams,
the full analysis should also include the \SUSY\  and Higgs/Goldstone
boson contributions to the various
universal vacuum  polarization effects comprised in our counterterms.
However, the
calculation of all these pieces has already been discussed in detail 
long ago in the literature\,\cite{TesiJefe,Grifols:1985xs,Grifols:1984gu,Sola:1991ju,Bertolini:1986ia} 
and thus the
lengthy formulae accounting for these results
will not be explicitly quoted here. Their contribution
is not \tb-enhanced, but since we wish to
compute the full supersymmetric contribution in the relevant regions of
the \MSSM\  parameter space, those 
effects will be included in our numerical code.
Finally,
the smaller --though numerically overwhelming -- subset of strong supersymmetric 
one-loop graphs are displayed in Fig.\,\ref{diag:tbhSEW} of
Ref.\cite{Guasch:1995rn}. 
We will use
the formulae from the latter reference in the present analysis
to produce the total (electroweak+strong) \SUSY\  correction
to our process. 

Next let us report on the contributions from
the various vertex diagrams and counterterms
in the on-shell renormalization scheme.
The generic structure of any
renormalized vertex function, $\Lambda$, in Figs.\,\ref{diag:tbhSEW}-\ref{diag:tbhHIGGS} is composed of two form 
factors $F_L$, $F_R$ plus the counterterms.
Therefore, on making use of~(\ref{eq:LtbH2}) and the formulae of chapter~\ref{cap:Renorm}, one immediately finds:
\begin{equation}
\Lambda = {i\,g\over\sqrt{2}\,M_W}
\,\left[m_t\,\cot\beta\,(1+\Lambda_R)\,\pr
 + m_b\,\tb\,(1+\Lambda_L)\,\pl\right]\,,
\label{eq:AtbH}
\end{equation}
where
\begin{eqnarray}
\Lambda_R & = & F_R+{\delta m_t\over m_t}
+\frac{1}{2}\,\delta Z_L^b+\frac{1}{2}\,\delta Z_R^t-\Delta_{\tau}\nonumber\\
& - & {\delta v^2\over v^2}+\delta Z_{\hplus}+(\tb-\cot\beta)\,\delta Z_{HW}
 \,,\nonumber\\
\Lambda_L &=& F_L+{\delta m_b\over m_b}
+\frac{1}{2}\,\delta Z_L^t+\frac{1}{2}\,\delta Z_R^b
+\Delta_{\tau}\,.
\label{eq:lambdaLR}
\end{eqnarray}
In the following the analytical contributions to the vertex 
form factors and counterterms will be specified diagram by
diagram.

\subsection{\SUSY\  vertex diagrams}


In this section we will make intensive use of the definitions and
formulae of chapter~\ref{cap:MSSM}. We refer the reader there for questions
about notation and conventions. 
Following the labeling of Feynman graphs in Fig.\,\ref{diag:tbhSEW} we write
down the terms coming from virtual \SUSY\  particles.
\begin{itemize}
\item {Diagram (a):}
Making use of the coupling matrices of eqs.~(\ref{V1Apm}) and (\ref{eq:QLQR}) we 
introduce the shorthands\footnote{Lower indices are summed over, whereas
upper indices (some of them within parenthesis) are just for notational 
convenience.}
\begin{equation}
  \label{V1Apmdef}
  \Apm\equiv\Apmit\  \ \mbox{and}\ \ \Apmz\equiv\Apmat\,,
\end{equation}
and define the combinations (omitting indices also for $\QaiL,\QaiR$)
\begin{eqnarray}
  \label{V1matrices}
    \RRLo =\cbt\Apc\HRc\Amz\,,&\ \ \ 
   &\RRRo =\cbt\Amc\HRc\Amz\,,\nonumber\\
    \LRLo =\cbt\Apc\HRc\Apz\,,&\ \ \ 
   &\LRRo =\cbt\Amc\HRc\Apz\,,\nonumber\\
    \RLLo =\sbt\Apc\HLc\Amz\,,&\ \ \ 
   &\RLRo =\sbt\Amc\HLc\Amz\,,\nonumber\\
    \LLLo =\sbt\Apc\HLc\Apz\,,&\ \ \ 
   &\LLRo =\sbt\Amc\HLc\Apz\,.           
\end{eqnarray}

The contribution from diagram (a) to the form factors $F_L$ and $F_R$
is then
\begin{eqnarray}
  \label{V1FF}
  F_L&=&M_L\left[\LLRo\Czt+\right.\nonumber\\
  &+&\mb\,\left(\mt\RRLo+\ma\LRLo+\mb\LLRo+\mi\LLLo\right)\Cot\nonumber\\
  &+&\mt\,\left(\mt\LLRo+\ma\RLRo+\mb\RRLo+\mi\RRRo\right)
     \left(\Coo-\Cot\right)\nonumber\\
  &+&\left.\left(\mt\mb\RRLo+\mt\mi\RRRo+\ma\mb\LRLo+\mi\ma\LRRo\right)
     \Cz\right]\,,\nonumber\\
  F_R&=&M_R\left[\RRLo\Czt+\right.\nonumber\\
  &+&\mb\,\left(\mt\LLRo+\ma\RLRo+\mb\RRLo+\mi\RRRo\right)\Cot\nonumber\\
  &+&\mt\,\left(\mt\RRLo+\ma\LRLo+\mb\LLRo+\mi\LLLo\right)
     \left(\Coo-\Cot\right)\nonumber\\
  &+&\left.\left(\mt\mb\LLRo+\mt\mi\LLLo+\ma\mb\RLRo+\mi\ma\RLLo\right)
     \Cz\right]\,,
\end{eqnarray}
where the overall coefficients $M_L$ and $M_R$ are the following:
\begin{equation}
  \label{MLMR}
  M_L=-\frac{ig^2\mw}{\mb\tb}\ \ \ \ M_R=-\frac{ig^2\mw}{\mt\ctb}\,.
\end{equation}
The notation for the various $3$-point functions is summarized in 
Appendix~\ref{ap:pointfun}. On eq.~(\ref{V1FF}) they must be evaluated with arguments:
\begin{equation}
  \label{V1Cs}
  C_{*}=C_{*}\left(p,p',\msta,\ma,\mi\right)\,.
\end{equation}

\item {Diagram (b):}
For this diagram --which in contrast to the
others is finite-- we also use the matrices on
eqs.~(\ref{V1Apm}) and (\ref{gdef}), and introduce the shorthands
\begin{equation}
  \label{V2Apmdef}
   \Apmb\equiv A_{\pm b\alpha}^{(b)}\ \ \mbox{and}\ \ \Apmt\equiv\Apmat\,,
\end{equation}
to define the products of coupling matrices
\begin{eqnarray}
  \label{V2matrices}
  \RLt =\Gba\Apbc\Amt\,,&\ \ \ &\RRt =\Gba\Ambc\Amt\,,\nonumber\\
  \LLt =\Gba\Apbc\Apt\,,&\ \ \ &\LRt =\Gba\Ambc\Apt\,.           
\end{eqnarray}
The contribution to the form factors $F_L$ and $F_R$ from this diagram is
\begin{eqnarray}
  \label{V2FF}
  F_L&=&\frac{M_L}{2\mw}\left[
  \mb\LLt\Cot+\mt\RRt\left(\Coo-\Cot\right)-\ma\LRt\Cz\right]\,,\nonumber\\
  F_R&=&\frac{M_R}{2\mw}\left[
  \mb\RRt\Cot+\mt\LLt\left(\Coo-\Cot\right)-\ma\RLt\Cz\right]\,,
\end{eqnarray}
the coefficients $M_L$, $M_R$ being those of eq.~(\ref{MLMR}) and the 
scalar $3$-point functions now evaluated with arguments
\begin{equation}
  \label{V2Cs}
  C_{*}=C_{*}\left(p,p',\ma,\msta,\msbb\right)\,.
\end{equation}

\item {Diagram (c):}
For this diagram  we will need
\begin{equation}
  \label{V3Apmdef}
  \Apm\equiv\Apmib\ \ \mbox{and}\ \ \Apmz\equiv\Apmab\,,
  \end{equation}
and again omitting indices we shall use
\begin{eqnarray}
  \label{V3matrices}
    \RRLth =\cbt\Apzc\HRc\Am\,,&\ \ \ 
   &\RRRth =\cbt\Amzc\HRc\Am\,,\nonumber\\
    \LRLth =\cbt\Apzc\HRc\Ap\,,&\ \ \ 
   &\LRRth =\cbt\Amzc\HRc\Ap\,,\nonumber\\
    \RLLth =\sbt\Apzc\HLc\Am\,,&\ \ \ 
   &\RLRth =\sbt\Amzc\HLc\Am\,,\nonumber\\
    \LLLth =\sbt\Apzc\HLc\Ap\,,&\ \ \ 
   &\LLRth =\sbt\Amzc\HLc\Ap\,.           
\end{eqnarray}
{}From these definitions the contribution of diagram (c) to the form
factors can be obtained by performing the following changes in that of
diagram (a), eq.~(\ref{V1FF}): 
\begin{itemize}
\item Everywhere on eqs.~(\ref{V1FF}) and (\ref{V1Cs}) replace 
$\mi\leftrightarrow\ma$ and $\msta\leftrightarrow\msba$.
\item Replace on eq.~(\ref{V1FF}) couplings from (\ref{V1matrices}) 
with those of (\ref{V3matrices}).
\item Include a global minus sign.
\end{itemize}

\end{itemize}

\subsection{Higgs vertex diagrams}

\diagtbhHIGGS

Now we consider contributions arising from the exchange of virtual
Higgs particles
and Goldstone bosons in the Feynman gauge,
as shown in Fig.\,\ref{diag:tbhHIGGS}. We follow the vertex
formula for the form factors by the value of the
overall coefficient $N$ 
and by the arguments of the corresponding $3$-point functions.
\begin{itemize}
\item {Diagram (a):}
\begin{eqnarray*}
  F_L&=&N\,[\mbs (\Cot-\Cz)+\mts \ctbs (\Coo-\Cot) ]\,,\\
  F_R&=&N\mbs [\Cot-\Cz +\tbs (\Coo-\Cot) ] \,,\\
  N&=&\mp\frac{ig^2}{2}\left(1-\frac{\{\mHs,\mhs\}}{2\mws}\right)
        \frac{\{\ca , \sa \}}{\cbt}\{\cbma , \sbma\}\,, \\
  C_{*}&=&C_{*}\left(p,p',\mb,\mHp,\{\mH,\mh\}\right)\,.
\end{eqnarray*}

\item {Diagram (b):}
\begin{eqnarray*}
  F_L&=&N\ctb[\mts (\Coo -\Cot )+\mbs(\Cz -\Cot )]\,,\\
  F_R&=&N\mbs\tb (2\Cot -\Coo -\Cz ) \,,\\
  N&=&\frac{ig^2}{4} \frac{\{\ca , \sa \}}{\cbt}\{\sbma  , \cbma \}
        \left(\frac{\mHps}{\mws}-\frac{\{\mHs  , \mhs \}}
{\mws}\right)\,, \\
  C_{*}&=&C_{*}\left(p,p',\mb,\mw,\{\mH,\mh\}\right)\,.
\end{eqnarray*}

\item {Diagram (c):}
\begin{eqnarray*}
  F_L&=&N\mts [\ctbs\Cot +\Coo -\Cot -\Cz ] \,,\\
  F_R&=&N\,[\mbs\tbs\Cot  +\mts (\Coo -\Cot-\Cz )]\,, \\
  N&=&-\frac{ig^2}{2}\frac{\{\sa , \ca \}}{\sbt}\{\cbma  , \sbma \}
        \left(1-\frac{\{\mHs  ,\mhs \}}{2\mws} \right) \,,\\
  C_{*}&=&C_{*}\left(p,p',\mt,\{\mH,\mh\},\mHp\right)\,.
\end{eqnarray*}

\item {Diagram (d):}
\begin{eqnarray*}
  F_L&=&N\mts (2\Cot-\Coo+\Cz)\ctb \,,\\
  F_R&=&N\,[-\mbs\Cot +\mts (\Coo-\Cot-\Cz)]\tb\,, \\
  N&=&\mp\frac{ig^2}{4}\frac{\{\sa , \ca \}}{\sbt}\{\sbma  , \cbma \}
        \left(\frac{\mHps}{\mws}-\frac{\{\mHs  ,\mhs \}}
{\mws}\right) \,,\\
  C_{*}&=&C_{*}\left(p,p',\mt,\{\mH,\mh\},\mw\right)\,.
\end{eqnarray*}

\item {Diagram (e):}
\begin{eqnarray*}
  F_L&=&N\,[\mbs(\Cot+\Cz)+\mts(\Coo-\Cot)]\,, \\
  F_R&=&N \mbs \tbs (\Coo+\Cz)\,, \\
  N&=&-\frac{ig^2}{4}\left(\frac{\mHps}{\mws}-\frac{\mAs}
{\mws}\right)\,, \\
  C_{*}&=&C_{*}\left(p,p',\mb,\mw,\mA\right)\,.
\end{eqnarray*}

\item {Diagram (f):}
\begin{eqnarray*}
  F_L&=& N\mts\ctbs (\Coo+\Cz)\,, \\
  F_R&=& N\,[\mbs\Cot +\mts (\Coo-\Cot+\Cz)]\,, \\
  N&=&-\frac{ig^2}{4}\left(\frac{\mHps}{\mws}-\frac{\mAs}
{\mws}\right) \,,\\
  C_{*}&=&C_{*}\left(p,p',\mt,\mA,\mw\right)\,.
\end{eqnarray*}

\item {Diagram (g):}
\begin{eqnarray*}
  F_L&=&N\,[(2\mbs\Coo+\Czt+2(\mts-\mbs)(\Coo-\Cot ))\ctbs
        +2\mbs(\Coo+2\Cz)]\mts\,, \\
  F_R&=&N\,[(2\mbs\Coo+\Czt+2(\mts-\mbs)(\Coo-\Cot ))\tbs
        +2\mts(\Coo+2\Cz)]\mbs\,, \\
  N&=&\pm\frac{ig^2}{4\mws}\frac{\sa \ca}{\sbt \cbt}\,, \\
  C_{*}&=&C_{*}\left(p,p',\{\mH,\mh\},\mt,\mb\right)\,.
\end{eqnarray*}

\item {Diagram (h):}
\begin{eqnarray*}
  F_L&=& N\mts\ctbs\,\Czt\,,  \\
  F_R&=& N\mbs\tbs\,\Czt\,,   \\
  N&=&\mp \frac{ig^2}{4\mws}\,, \\
  C_{*}&=&C_{*}\left(p,p',\{\mA,\mz\},\mt,\mb\right)\,.
\end{eqnarray*}

\end{itemize}

In the equations above, 
it is understood that the CP-even mixing angle, $\alpha$, is renormalized 
into $\alpha_{\rm eff}$ by the one-loop Higgs mass
relations\,\cite{Ellis:1991zd,Brignole:1991pq,Haber:1991aw,Haber:1993an,Haber93}.

The evaluation of  $\Delta_{\tau}$, the process dependent term of the \tb\
renormalization,  on 
eq.(\ref{eq:deltatau}), 
yields similar bulky analytical formulae, which follow
after computing diagrams akin to those in
Figs.\,\ref{diag:tbhSEW}-\ref{diag:tbhHIGGS} and the corresponding counterterms
to the $\tau$, $\nu_\tau$ and $H^+$ external legs from chapter~\ref{cap:Renorm}
for the \MSSM\  corrections to $\hplus\rightarrow \tau^+\,{\nu}_{\tau}$.
We refrain from quoting them explicitly. The numerical
effect, though, will be explicitly used, and isolated (Fig.\,\ref{fig:tbhdeltadeltataumult0}), in our computation. 

We are now ready to furnish the corrected width of $t\rightarrow \hplus\,b$
in the \MSSM\@. It just follows after computing the interference between
the tree-level amplitude and the one-loop amplitude. It is convenient to
express the result  as a relative correction with respect to
the tree-level width  both
in the $\alpha$-scheme and in the $G_F$-scheme.  In the former we obtain
the relative \MSSM\  correction
\begin{eqnarray}
\delta_{\alpha}^{MSSM} & = & {\Gamma -\Gamma^{(0)}_{\alpha}
\over \Gamma^{(0)}_{\alpha}}\nonumber\\ 
&=&\frac{N_L}{D}\,[2\,Re(\Lambda_L)]+
\frac{N_R}{D}\,[2\,Re(\Lambda_R)]+\frac{N_{LR}}{D}\,
[2\,Re(\Lambda_L+\Lambda_R)]\,,
\label{eq:deltaalpha}
\end{eqnarray}
where the corresponding lowest-order width is
\begin{equation}
\Gamma^{(0)}_{\alpha}=\left({\alpha\over s^2_W}\right)
\,{D\over 16\, M_W^2\,m_t}\,
\lambda^{1/2} (1, {m_b^2\over m_t^2},{\mHps\over m_t^2})\,,
\label{treeHalpha}
\end{equation}
with
\begin{eqnarray}
D &=& (m_t^2+m_b^2-\mHps)\,(m_t^2\cot^2\beta+m_b^2\tan^2\beta)
+4m_t^2m_b^2\,,\nonumber\\
N_L & = & (m_t^2+m_b^2-\mHps)\,m_b^2\tan^2\beta\,,\nonumber\\
N_R & = & (m_t^2+m_b^2-\mHps)\,m_t^2\cot^2\beta\,,\nonumber\\
N_{LR} & = & 2m_t^2m_b^2\,.
\end{eqnarray} 
From these equations it is obvious that at low \tb\  the relevant 
quantum effects basically come from the contributions to the form factor $\Lambda_R$
whereas at high \tb\  they come from $\Lambda_L$.   

Using  eq.(\ref{eq:DeltaMW}) we find that the relative \MSSM\  
correction in the $G_F$-parametrization reads
\begin{equation}
\delta_{G_F}^{MSSM}= {\Gamma-\Gamma^{(0)}_{G_F}
\over \Gamma^{(0)}_{G_F}}=\delta_{\alpha}^{MSSM}-\Delta r^{MSSM}\,,
\label{eq:alphagf1}
\end{equation}
where the tree-level width in the $G_F$-scheme,
 $\Gamma^{(0)}_{G_F}$, is given by eq.(\ref{eq:treeH}) and is related to
eq.(\ref{treeHalpha}) through
\begin{equation}
\Gamma^{(0)}_{\alpha}=\Gamma^{(0)}_{G_F}\,(1-\Delta r^{MSSM})\,.
\label{eq:alphagf}
\end{equation} 

\section{Numerical analysis and discussion }
\label{sec:tbhnumeric}

Quantum effects should be able to
discriminate whether the charged
Higgs emerging from the decay $t\rightarrow \hplus\,b$ is supersymmetric or not,
for the \MSSM\  provides a well defined prediction of the size of
these effects for given values of the sparticle masses. 
Some work on radiative corrections to the decay width of $t\rightarrow \hplus\,b$ has
already appeared in the literature. In particular, the conventional 
\QCD\  corrections have been evaluated\,\cite{Czarnecki:1993ig,Czarnecki:1993zm}  
and found to significantly
reduce the partial width. The \SUSY-\QCD\  corrections are also substantial and have
been analyzed, only in part in Refs.\cite{Konig:1994xf,Li:1992ga},  
and in more detail
in Refs.\cite{TesiRicci,Guasch:1995rn}. 
The electroweak corrections produced by the roster of genuine ($R$-odd)
sparticles was first computed in Ref.\,\cite{Coarasa:1996qa}.
As for the virtual effects mediated by the Higgs bosons,
a first treatment is given in Refs.\cite{Mendez:1991gx} 
and \cite{Li:1993gd}.
However, these references disagree in several parts of the calculation,
and moreover they are both incomplete
calculations on their own, for they fully ignore the Higgs effects  
associated to the bottom quark
Yukawa coupling, which could in principle be significant in 
the large \tb\  region. On the other hand, even though the latter kind of 
Higgs effects have been discussed
in the literature in other renormalization schemes based on alternative definitions
of \tb\,\cite{Dabelstein:1995hb,Dabelstein:1995js,Chankowski:1994er,Yamada:1994kj,Haber:1993an,Yamada:1991ih,Brignole:1992uf,Diaz:1992ki,Diaz:1993kn,Yamada:1996hn}, 
a detailed analysis including the genuine \SUSY\  effects themselves has never
been attempted.  
Thus, if only for completeness, we are providing here
not only a dedicated treatment of the $R$-odd 
contributions mediated by the sparticles of the \MSSM, but also
the fully-fledged pay-off of the supersymmetric Higgs effects.

Before presenting the results of the complete numerical
analysis, it should be clear that the bulk of the high \tb\  corrections
to the decay rate of $t\rightarrow \hplus\,b$ in the \MSSM\  is expected
to come from \SUSY-\QCD\@. This could already be foreseen from what is known in
\SUSY\  GUT models\,\cite{Carena:1994bv,Hall:1994gn,Rattazzi:1996gk}; 
in fact, in this context 
a non-vanishing sbottom mixing  (which we also assume
in our analysis) may lead to important \SUSY-\QCD\  quantum effects 
on the bottom mass, $m_b= m_b^{GUT}+\Delta m_b$, where $\Delta m_b$ is 
proportional to $M_{LR}^b\rightarrow -\mu\tb $ at sufficiently high
\tb.
These are finite threshold effects that one has to include
when matching the SM and \MSSM\  renormalization group equations (RGE) at the
effective supersymmetric threshold
scale, $T_{SUSY}$, above which the RGE evolve according to the \MSSM\  
$\beta$-functions in the $\overline{MS}$ scheme\,\cite{Carena:1993ag}. 
In our case, since the bottom mass is an input parameter for the on-shell
scheme, these effects are just fed into
the mass counterterm $\delta m_b/m_b$
on eq.(\ref{eq:lambdaLR}) and contribute to it with opposite sign
($\delta m_b/m_b=-\Delta m_b/m_b +\ldots$).

\diagdeltambEW

Explicitly, when viewed in terms of diagrams of the 
electroweak-eigenstate basis, the relevant finite corrections from
the bottom mass counterterm 
are generated by mixed LR-sbottoms and gluino loops (Cf.\
Fig.\,\ref{diag:deltambEW}(a))\,\cite{TesiRicci}:
\begin{eqnarray}
\left({\delta m_b\over m_b}\right)_{\rm SUSY-QCD} &=&
{2\alpha_s(m_t)\over 3\pi}\,m_{\tilde{g}}\,M_{LR}^b\,
I(m_{\tilde{b}_1},m_{\tilde{b}_2},m_{\tilde{g}})\nonumber\\
&\rightarrow & -{2\alpha_s(m_t)\over 3\pi}\,m_{\tilde{g}}\,\mu\tb\,
I(m_{\tilde{b}_1},m_{\tilde{b}_2},m_{\tilde{g}}) \,,
\label{eq:dmbQCD}
\end{eqnarray}
where the last result holds for sufficiently large \tb\  and for
$\mu$ not too small as compared to $A_b$. We have introduced 
the positive-definite function (Cf.\  Appendix~\ref{ap:pointfun})
\begin{equation}
I(m_1,m_2,m_3)\equiv 16\,\pi^2 i\,C_0(0,0,m_1,m_2,m_3)=
{m_1^2\,m_2^2\ln{m_1^2\over m_2^2}
+m_2^2\,m_3^2\ln{m_2^2\over m_3^2}+m_1^2\,m_3^2\ln{m_3^2\over m_1^2}\over
 (m_1^2-m_2^2)\,(m_2^2-m_3^2)\,(m_1^2-m_3^2)}\,.
\label{eq:I123}
\end{equation}
In addition, we could also foresee potentially large (finite)
\SUSY\  electroweak effects
from $\delta m_b/m_b$. They are induced by
\tb-enhanced Yukawa couplings of the type (\ref{eq:Yukawas}). 
Of course, these effects
have already been fully included in the calculation presented in
Section~\ref{sec:tbhoneloop} that we have performed in the mass-eigenstate basis, but it is illustrative
of the origin of the leading contributions to pick them 
up again directly from the diagrams in the
electroweak-eigenstate basis. In this case, from loops involving mixed LR-stops 
and mixed charged higgsinos (Cf. Fig.\,\ref{diag:deltambEW}(b)), one finds:
\begin{eqnarray}
\left({\delta m_b\over m_b}\right)_{\rm SUSY-Yukawa} &=&
-{h_t\,h_b\over 16\pi^2}\,\, {\mu\over m_b}\,m_t\,M_{LR}^t
I(m_{\tilde{t}_1},m_{\tilde{t}_2},\mu)\nonumber\\
&\rightarrow &
-{h_t^2\over 16\pi^2}\,\mu\tb\,A_t\,
I(m_{\tilde{t}_1},m_{\tilde{t}_2},\mu)\,,
\label{eq:dmbEW}
\end{eqnarray}
where again the last expression holds for large enough \tb.

Notice that, at variance with eq.(\ref{eq:dmbQCD}), the Yukawa coupling
correction (\ref{eq:dmbEW}) dies away with increasing $\mu$. 
Setting $h_t\simeq 1$ at high \tb, and assuming that
there is no large hierarchy between the sparticle masses, the ratio
between (\ref{eq:dmbQCD}) and (\ref{eq:dmbEW}) is given, in good approximation,
by $4\,{m_{\tilde{g}}/ A_t}$
times a slowly varying function of the masses of
order $1$, where the (approximate) proportionality
to the gluino mass reflects the very slow decoupling rate of
the latter\,\cite{TesiRicci,Guasch:1995rn}. 

In view of the present bounds on the gluino mass\,\cite{Moriond97}, and
since $A_t$ (as well as $A_b$) cannot increase
arbitrarily, we expect that the \SUSY-\QCD\  effects
can be dominant and even overwhelming for sufficiently heavy gluinos.
Unfortunately, in contradistinction to the \SUSY-\QCD\  case, there are also plenty of 
additional vertex contributions  both from the Higgs sector 
and from the stop-sbottom/gaugino-higgsino sector
where those Yukawa couplings enter once again the game.
So if one wishes to trace the origin of the leading
contributions in the electroweak-eigenstate
basis, a similar though somewhat more involved 
exercise has to be carried out also for vertex functions. Of course,
all of these effects are automatically included in our
calculation of section~\ref{sec:tbhoneloop} within the framework of
the mass-eigenstate basis\footnote{The mass-eigenstate basis
is extremely convenient to carry out the numerical analysis, but it does not
immediately provide a ``physical interpretation'' of the results.
The electroweak-eigenstate basis, in contrast, is a better bookkeeping device
to trace the origin of the most relevant effects, but as a drawback the
intricacies of the full analytical calculation
can be (in general) abhorrent.}. 
It is worth noticing that these effects do {\bf not}
decouple when we send the scale of the sparticle masses to infinity.
Indeed, if we scale all \SUSY\  parameters in eqs.\,(\ref{eq:dmbQCD}) and~(\ref{eq:dmbEW}) by the
same factor, this factor drops off from these formulas;  and of course
this factor could be sent to infinity! Therefore, we have here a
dramatic  example of non-decoupling phenomenon which is intrinsically
associated to the breaking of \SUSY\@. Indeed, in a truly supersymmetric
theory these threshold corrections would (obviously) be zero! 
Another matter would be to judge what sort of weird
fine-tunings among the parameters would entail letting the scale factor to
infinity, but what cannot be denied is the bare fact that it really drops
out automatically!
One could work in an alternative framework (as in Ref.\cite{Guasch:1995rn})
assuming no mixing in the sbottom and/or stop mass matrix. 
Nonetheless, the typical size of the radiative corrections does not change
as compared to the present approach (in which we do assume a non-diagonal
sbottom and stop matrix) the
reason being that in the absence of mixing, i.e. $M_{LR}^{\{b,t\}}=0$,
the contribution $\delta m_b/m_b\propto -\mu\tan\beta$ at large $\tan\beta$
is no longer possible but, in contrast, the vertex correction
(Cf. diagram~\ref{diag:tbhSEW}(b) and its \SUSY-\QCD\  analog)
does precisely inherits this dependence and
compensates for it (see eq.\,(\ref{eq:MLRdefinition})).
The drawback of an scenario based on
$M_{LR}^b=0$, however, is that when it is combined with a large value
of $\tan\beta$ it may lead to a value of
$A_b$ which overshoots the natural range expected for this parameter --eq.(\ref{eq:necessary}).

We may now pass on to the numerical analysis of the over-all quantum effects.
After explicit computation of the various loop diagrams, the results are 
conveniently cast in terms of the relative
correction with respect to the tree-level width:
\begin{equation}
\delta={\Gamma_H-\Gamma^{(0)}_H\over \Gamma^{(0)}_H}
\equiv{\Gamma (t\rightarrow \hplus\,b)-\Gamma^{(0)}(t\rightarrow \hplus\,b)
\over \Gamma^{(0)}(t\rightarrow \hplus\,b)}\,.
\label{eq:pito}
\end{equation}
In what follows we understand that $\delta$ defined by eq.(\ref{eq:pito})
is  $\delta_{\alpha}$ -- Cf.\  eq.(\ref{eq:deltaalpha}) -- i.e.\
we shall always give our corrections with respect to the tree-level width
$\Gamma_{\alpha}^0$ in the $\alpha$-scheme. The corresponding correction with
respect to the tree-level width in the $G_F$-scheme is simply given by 
eq.(\ref{eq:alphagf1}), where $\Delta r^{MSSM}$ was object of a particular
study \,\cite{TesiJefe,Garcia:1994sb,Chankowski:1994eu} 
and therefore it can be easily incorporated, if necessary.
Notice, however, that $\Delta r^{MSSM}$ is already tightly
bound by the experimental data on $M_Z$ 
at LEP 
and the ratio
$M_{W}/M_Z$ in
$p\bar{p}$, which lead
 to $M_W$~(\ref{eq:EWprecisiondata}). 
Therefore, even without doing the exact theoretical calculation
of $\Delta r$ within the \MSSM, we
already know from
\begin{equation}
\Delta r=1-{\pi\alpha\over \sqrt{2}\,G_F}\,{1\over M_W^2\,(1-M_W^2/M_Z^2)}\,,
\label{eq:deltar}
\end{equation}
that $\Delta r^{MSSM}$ must lie in the experimental interval 
$\Delta r^{\rm exp}\simeq 0.037\pm 0.014$. 

Now, since the corrections computed
in Section~\ref{sec:tbhoneloop} can typically be about one order of magnitude larger than 
$\Delta r^{\rm MSSM}$, the bulk of the quantum effects 
on $t\rightarrow \hplus\,b$ is already comprised in the relative
correction (\ref{eq:pito}) in the $\alpha$-scheme\footnote{For the standard decay $t\rightarrow W^+\,b$, the
situation is quite different since the SM electroweak
corrections\,\cite{Jezabek:1989iv,Jezabek:1989ja,Li:1991qf,Eilam:1991iz,Denner:1990tx,Denner:1991ns,Irwin:1991ka,Yuan:1991av} 
and the maximal \SUSY\  electroweak 
corrections\,\cite{Garcia:1994rq} 
in the $\alpha$-scheme are much smaller than for the
decay $t\rightarrow \hplus\,b$ , namely they are of the order
of $\Delta r$. Therefore, for the standard decay
$t\rightarrow W^+\,b$ there is a significant cancellation between the corrections 
in the $\alpha$-scheme and $\Delta r$ in most of the \tb\  range 
resulting in a substantially diminished
correction in the $G_F$-scheme.}.
Furthermore, in the conditions under study, 
only a small fraction of $\Delta r^{MSSM}$ is
supersymmetric\,\cite{TesiJefe,Garcia:1994sb,Chankowski:1994eu}, 
and
we should not be dependent on isolating this universal, relatively small,
part of the total \SUSY\  correction to
$\delta$. To put in a nutshell:
if there is to be any hope to measure supersymmetric quantum effects on the
charged Higgs decay of the top quark, they should better come from the potentially
large, non-oblique, corrections computed in Section~\ref{sec:tbhoneloop}. The \SUSY\  effects contained
in $\Delta r^{MSSM}$\,\cite{TesiJefe,Garcia:1994sb,Chankowski:1994eu}, 
instead,  will be measured in a much more efficient 
way from a high precision  ($\delta M_W^{\rm exp}=\pm 40\,MeV$) determination of 
$M_W$ at LEPII.   
 
Another useful quantity is the branching ratio
\begin{equation}
B_H\equiv BR (t\rightarrow \hplus\,b) ={\Gamma_H\over
\Gamma_W +\Gamma_H+\Gamma_{SUSY}}\,,
\label{eq:BR}
\end{equation}
where $\Gamma_W\equiv\Gamma (t\rightarrow W^+\,b)$ and 
$\Gamma_{SUSY}$ stands for decays of the top quark into \SUSY\  particles.
In particular, the potentially important \SUSY-\QCD\  mode
$t\rightarrow \tilde{t}_1\,\tilde{g}$
is kinematically forbidden in most part of our analysis where 
we usually assume $m_{\tilde{g}}={\cal O} (300)\GeV$.
There may also be the competing electroweak \SUSY\  decays
$t\rightarrow \tilde{t}_1\,\chi^0_{\alpha}$ and
$t\rightarrow \tilde{b}_1\,\chi^+_i$ for some $\alpha=1,\ldots,4$ and
some $i=1,2$.  The latter, however, 
is also phase space obstructed in most of
our explored parameter space, since we
typically assume $m_{\tilde{b}_1}=150\GeV$.
The decay $t\rightarrow \tilde{t}_1\,\chi^0_{\alpha}$,
instead, is almost always open, but it is not \tb-enhanced in our
favourite segment (\ref{eq:tanbeta2}).  
However, when studying the branching ratio (\ref{eq:BR})
as a function of the squark and gluino masses, we do include the effects from all
these supersymmetric channels whenever they are kinematically open. Thus in
general $\Gamma_{SUSY}$ on eq.(\ref{eq:BR}) is given by
\begin{equation}
\Gamma_{SUSY}=\Gamma (t\rightarrow \tilde{t}_1\,\tilde{g})+
\sum_{\alpha}\Gamma (t\rightarrow \tilde{t}_1\,\chi^0_{\alpha}) +
\sum_i\Gamma (t\rightarrow \tilde{b}_1\,\chi^+_i)\,.
\label{eq:gsusy}
\end{equation}  
The various terms contributing to this equation are computed at the
tree-level. Recently, the
\SUSY-\QCD\  corrections to some of these supersymmetric modes have been
evaluated and in some cases may be important\,\cite{Djouadi:1996pi,Li:1996fc}. 
Similarly, we treat the computation of the partial width of the standard mode 
$t\rightarrow W^+\,b$ at the tree-level. 
This is justified since, as shown in 
Refs.\cite{Garcia:1994rq,Dabelstein:1995jt,TesiRicci,Grzadkowski:1992nj,Denner:1993vz,Yang:1994ra}, 
this decay 
cannot in general 
develop large supersymmetric radiative corrections, or at least as large as to be 
comparable to those affecting the charged Higgs mode (for the same value of the
input parameters). 
The reason for it stems from the very different
structure of the counterterms for both decays; in particular, the
standard decay mode of the top quark does not involve the mass renormalization
counterterms for the external fermion lines, and as a consequence the
aforementioned large quantum effects associated
to the bottom quark self-energy at high \tb\  are not possible.

Figures \ref{fig:tbhgammatbmult0}-\ref{fig:tbhdeltadeltataumult0} and \ref{fig:tbhdeltamsbmugt0}-\ref{fig:tbhgammatbmugt0} display a clear-cut r{\'e}sum{\'e} of our numerical results. 
We wish to point out that 
they have been thoroughly checked. Scale independence of
$\delta$, eq.(\ref{eq:pito}),
and cancellation of UV-divergences have been explicitly verified, both
analytically and numerically. 
In all our numerical evaluations we have imposed the various restrictions of sec.~\ref{sec:limits}.

To start with, we concentrate on the case $\mu<0$, which we study in
Figs.\,\ref{fig:tbhgammatbmult0}-\ref{fig:tbhdeltadeltataumult0}. (The case $\mu>0$ is studied apart in Figs.\,\ref{fig:tbhdeltamsbmugt0}-\ref{fig:tbhgammatbmugt0} and will be
commented later on.)    
We observe that, for negative $\mu$, the leading \SUSY-\QCD\  effects on
$\delta$ are positive. This means that
in these circumstances the potentially large strong supersymmetric effects
are in frank competition with the conventional 
\QCD\  corrections, which are also very large and stay always negative
as will be discussed later on.

\figtbhgammatbmultz

Needless to say, a crucial parameter to be investigated is \tb. 
In Fig.\,\ref{fig:tbhgammatbmult0} we plot the tree-level width, $\Gamma_{0} (t\rightarrow \hplus\,b)$, and 
the total partial width, $\Gamma_{MSSM} (t\rightarrow \hplus\,b)$,
comprising all the \MSSM\  effects, as a function of \tb.
A typical set of parameters is chosen
well within canonical expectations (see below); the individual influence of each one
of them is tested in Figs.\,\ref{fig:tbhdeltamgmult0} to \ref{fig:tbhdeltadeltataumult0}.
Also shown in Fig.\,\ref{fig:tbhgammatbmult0} is the (tree-level) partial width
of the standard 
top quark decay $t\rightarrow W^+\,b$, which is (as noted above) 
far less sensitive to quantum corrections.
For convenience, we have included in Fig.\,\ref{fig:tbhgammatbmult0} a plot of
$\Gamma_{\rm QCD} (t\rightarrow \hplus\,b)$, i.e.\  the partial width that would be
obtained in the presence of only the standard \QCD\  corrections.
Recently it has been shown that \EW\  corrections in the absence of \SUSY\
particles, i.e.\  2HDM effects, can also be large, though it happens for a Higgs
spectrum very different of the \MSSM\  one\,\cite{TesiToni,Coarasa:1998xy}. On the other
hand by simple inspection of Fig.\,\ref{fig:tbhdeltatbmult0} we see that the Higgs sector
corrections are indeed small, so the \QCD\  corrected curve is in fact the
partial width that would be obtained in a non-\SUSY\  2HDM under the conditions
of the present study.

\figtbhdeltatbmultz

From eq.(\ref{eq:deltaalpha}) it is clear that,
for large (resp.\  small) \tb, the renormalized
form factor yielding the bulk of the \SUSY\  contribution is 
$\Lambda_L$ (resp. $\Lambda_R$).  
To appraise the relative importance of the 
various types of \MSSM\  effects on $\Gamma (t\rightarrow \hplus\,b)$, in Figs.\,\ref{fig:tbhdeltatbmult0}(a)-\ref{fig:tbhdeltatbmult0}(b) we
provide plots for the correction 
to the partial width, eq.(\ref{eq:pito}), and to
the branching ratio, eq.(\ref{eq:BR}), as a function of \tb, reflecting the
various individual contributions.
Specifically, we show in Fig.\,\ref{fig:tbhdeltatbmult0}(a):
\begin{itemize}
\item{(i)} The supersymmetric electroweak
contribution from genuine ($R$-odd) sparticles
 (denoted $\delta_{\rm SUSY-EW}$), i.e.\  
from sfermions (squarks and sleptons), charginos and neutralinos;
\item{(ii)} The electroweak contribution from non-supersymmetric ($R$-even) 
particles ($\delta_{EW}$). It is composed of two distinct types
of effects, namely, those from Higgs 
and Goldstone bosons (collectively called
``Higgs'' contribution, and denoted $\delta_{\rm Higgs}$) plus
the leading  SM effects\,\cite{Bohm:1986rj,Hollik:1990ii,Hollik1995Llibre} 
from conventional fermions ($\delta_{\rm SM}$):
\begin{equation}
\delta_{EW}=\delta_{\rm Higgs}+\delta_{\rm SM}\,;
\end{equation}
The remaining non-supersymmetric electroweak effects
are subleading and are neglected.
\item{(iii)} The strong supersymmetric contribution
(denoted by $\delta_{\rm SUSY-QCD}$) from squarks and gluinos; 
\item{(iv)}
The strong contribution from conventional quarks
and gluons (labeled $\delta_{\rm QCD}$);
 and
\item{(v)}
The total \MSSM\  contribution, $\delta_{\rm MSSM}$,
namely, the net sum of all the previous contributions:
\begin{equation}
\delta_{\rm MSSM}=\delta_{\rm SUSY-EW}+\delta_{EW}+
\delta_{\rm SUSY-QCD}+\delta_{\rm QCD}.
\label{eq:individ}
\end{equation}
\end{itemize}
In Fig.\,\ref{fig:tbhdeltatbmult0}(b) we reflect the impact of the \MSSM\  on the branching ratio,  
as a function of \tb; also shown are the tree-level value of
the branching ratio and the latter quantity after including the
(non-supersymmetric) QCD corrections. 
A typical common set of inputs has been chosen
in Figs.\,\ref{fig:tbhdeltatbmult0}(a)-\ref{fig:tbhdeltatbmult0}(b) such that the supersymmetric electroweak
corrections reinforce the
strong supersymmetric effects (\SUSY-\QCD). For this set of inputs, the total
\MSSM\  
correction to the partial width of $t\rightarrow \hplus\,b$ 
is positive for $\tb> 20$ (approx.).
Remarkably enough, this is so in spite of
the huge negative effects induced by QCD\@. In fact, we see that 
the gluon effects are overridden by the gluino effects provided \tb\  
is sufficiently large,
to be concrete for $\tb\geq 30$. Beyond this value, the strength of the
supersymmetric loops
becomes rapidly overwhelming; e.g.\  at the representative value $\tb=m_t/m_b=35$ 
we find $\delta_{\rm MSSM}\simeq +27\%$; 
and at $\tb\simeq 50$, which
is the preferred value claimed by $SO(10)$ Yukawa coupling unification
models\,\cite{Carena:1994bv,Hall:1994gn,Rattazzi:1996gk}, 
the correction is already $\delta_{\rm MSSM}\simeq +55\%$.
Quite in contrast, at that \tb\  one would expect, in the
absence of \SUSY\  effects, a (\QCD) correction
of about $-57\%$, i.e.\  virtually of the same size but opposite in sign! 

Coming back to Fig.\,\ref{fig:tbhgammatbmult0}, we see that,
after including the \SUSY\  effects, the partial width of $t\rightarrow \hplus\,b$
equals the partial width of the standard decay $t\rightarrow W^+\,b$
near the ``$SO(10)$'' point
$\tb=50$. (The meeting point is actually a bit earlier in \tb, 
after taking into account the known\,\cite{Garcia:1994rq,Dabelstein:1995jt}, 
negative, \SUSY\  corrections to
$t\rightarrow W^+\,b$, but this effect is
not shown in the figure since it is relatively small.)
Now, for the typical set of parameter values introduced
in Fig.\,\ref{fig:tbhgammatbmult0}, the top quark decay width into \SUSY\  particles,
eq.(\ref{eq:gsusy}), is rather tiny. Thus it is not surprising
that in these conditions the branching ratio
of the charged Higgs mode can be remarkably high:
 $BR (t\rightarrow \hplus\,b)\simeq 50\%$,
i.e.\  basically $50\%-50\%$ versus the standard decay mode.
In contrast, the branching ratio without \SUSY\  effects (i.e.\  essentially the
QCD-corrected branching ratio) is much
smaller: at the characteristic $SO(10)$
value, $\tb=50$, it barely reaches $20\%$.  
Clearly, if the \SUSY\  quantum effects are there,
they could hardly be missed!
 
As noted before, even though the dominant \MSSM\  effects are, by far, 
the \QCD\  and \SUSY-\QCD\  ones, they have opposite signs.
Therefore, there is a crossover point of the
two strongly interacting dynamics, where the conventional \QCD\  
loops are fully counterbalanced by the \SUSY-\QCD\  loops.
This leads to a funny situation, namely,
that at the vicinity of that point the total \MSSM\  correction is given by just the 
subleading, albeit non-negligible, electroweak supersymmetric 
contribution: $\delta_{\rm MSSM}\simeq \delta_{\rm SUSY-EW}$.
The crossover point occurs at $\tb\gsim 32\simeq m_t/m_b$,
where $\delta_{\rm SUSY-EW}\gsim 20$.
For larger and larger \tb\  beyond $m_t/m_b$,
the total (and positive) \MSSM\  correction grows very fast,
as we have said, since the
\SUSY-\QCD\  loops largely over-compensate the standard \QCD\  corrections.
As a result, the net effect on the partial width
appears to be opposite in sign to what might
naively be ``expected'' (i.e.\  the \QCD\  sign).   
Of course, this is not a general
result since it depends on the actual values of the \MSSM\  parameters. In the
following we wish to explore the various parameter dependences and
in particular we want to assess whether a favourable
situation as the one just described is likely to happen in an ample portion of
the \MSSM\  parameter space. 
In particular, the value $\tb=m_t/m_b=35$ will be chosen
in all our plots where
that parameter must be fixed. We consider it as representative of
the low end of the  high \tb\  segment, eq.(\ref{eq:htanr}).
Thus $\tb=m_t/m_b=35$ behaves as
a sort of threshold point beyond which
the \MSSM\  quantum effects on $t\rightarrow \hplus\,b$ take off so fast that
they should have indelible experimental consequences on top quark physics.

As regards to the non-supersymmetric electroweak
corrections, $\delta_{\rm EW}$, it is apparent from Fig.\,\ref{fig:tbhdeltatbmult0}(a)  that 
they are very small, especially 
in the high \tb\  segment. Also in the very low \tb\  
segment, $0.5\lsim\tb\lsim 1$,
$\delta_{EW}$ is relatively small; and this is so not only because both
$\delta_{\rm Higgs}$ and $\delta_{\rm SM}$ become never too large 
in absolute value, but also because in that region
there is a cancellation between $\delta_{\rm Higgs}<0$  and
$\delta_{\rm SM}>0$. 
As it happens, we end up with the fact
that the complicated Higgs effects result in a 
very tiny contribution, except in the 
very low \tb\  end, where e.g.\  they
can reach $-15\%$ at $\tb\simeq 0.5$. In this corner
of the parameter space, $\delta_{\rm Higgs}$
becomes the dominant part of $\delta_{\rm MSSM}$,
being even larger than the \QCD\  effects, which stay at the level of $-8\%$, and
also larger than the \SUSY-\QCD\  and \SUSY-\EW\  corrections, which remain below $+4\%$ 
and $-1\%$, respectively.

We have treated in detail the very low \tb\  segment by
including the one-loop
renormalization of the Higgs
masses\,\cite{Ellis:1991zd,Brignole:1991pq,Haber:1991aw,Haber:1993an,Haber93}. 
This is necessary in order to avoid that
the lightest CP-even Higgs mass either vanishes at $\tb=1$ or becomes lighter
than the phenomenological bounds near that value. In passing, we have checked 
that the one-loop shift of the masses, as well as of the CP-even
mixing angle, $\alpha$,
has little impact on the partial width of $t\rightarrow \hplus\,b$ in the
entire range of \tb, eq.(\ref{eq:tanbeta}).
They entail 
at most an additional $5\%$ negative shift of $\delta_{\rm MSSM}$ in the very 
low \tb\  region\footnote{To perform that check, we have
included both the stop and sbottom contributions to the one-loop Higgs mass
relations. A set of $7$ independent parameters has been used
to fully characterize these effects, viz.
$(M_{H^\pm}, \mu, \tb, m_{\tilde{b}_1}, m_{\tilde{t}_1}, A_b, A_t)$.
We refrain from writing out the cumbersome
formulae\,\cite{Ellis:1991zd,Brignole:1991pq,Haber:1991aw,Haber:1993an,Haber93}.}.   
It is precisely in this region where the Higgs effects 
could be expected of some relevance, and thus
where the renormalization of the  CP-even mixing angle
could have introduced some noticeable change in the neutral Higgs couplings.
Quite on the contrary, at high \tb\  the corresponding effect
is found to be of order one per mil and is thus negligibly small. 
On the other hand, a simple inspection of Figs.\,\ref{fig:tbhgammatbmult0} and \ref{fig:tbhdeltatbmult0}(b) shows that
even in the very low \tb\  ballpark, where there may be some ten
percent effect from
the Higgs sector, the rising of the tree-level width is so fast that
it becomes very hard to isolate these corrections.
We conclude that, despite the rather large number of
diagrams involved, the over-all yield 
from the Higgs sector of the \MSSM\  on $t\rightarrow \hplus\,b$
is rather meagre in the whole \tb\  range (\ref{eq:tanbeta}). 
This fact is somewhat surprising and was not obvious
a priori, due to the presence of enhanced Yukawa couplings
(\ref{eq:Yukawas}) in the whole plethora of Higgs diagrams.
The cancellations involved are reminiscent of the scanty \SUSY\  Higgs effects
obtained for the standard top quark decay $t\rightarrow
W^+\,b$\,\cite{Grzadkowski:1992nj,Denner:1993vz}. 

We come now to briefly discuss the standard \QCD\  effects up to ${\cal O}(\alpha_s)$,
which involve one-loop gluon corrections and gluon
bremsstrahlung\,\cite{Czarnecki:1993ig,Czarnecki:1993zm}. 
As it is plain from Fig.\,\ref{fig:tbhdeltatbmult0}(a), $\delta_{\rm QCD}$
is negative-definite and very important in the high \tb\
segment. It quickly saturates for $\tb\gsim 10$ at a large value 
of order $-60\%$.
Therefore, the \QCD\  effects need to be considered in order to isolate the
virtual \SUSY\  signature\,\cite{Czarnecki:1993zm,Czarnecki:1993ig}. 
The leading behaviour of the standard \QCD\  component in
the relative correction (\ref{eq:pito})  
can be easily assessed by considering the
following asymptotic formula
\begin{equation}
\delta_{QCD}=-\frac{2\, \alpha_s}{3\pi}\;\frac{\frac{8\pi^2-15}{12}
             (m^2_b\tan^2\beta+m^2_t\cot^2\beta)+
             3(4+\tan^2\beta-2\frac{M^2_{\hplus}}{m^2_t}\cot^2\beta)
             m^2_b\ln\left(\frac{m^2_t}{m^2_b}\right)}
             {m^2_b\tan^2\beta+m^2_t\cot^2\beta}\,,
\label{eq:QCD1}
\end{equation}  
which is 
obtained by expanding the exact one-loop formula up to
${\cal O}(m_b^2/m_t^2, M_{\hplus}^2/m_t^2)$.
Here $\alpha_s\equiv \alpha_s(m_t^2)$, normalized as $\alpha_s(M_Z^2)\simeq 0.12$.
The big log factor 
$\ln(m^2_t/m^2_b)$ originates from the running b-quark mass evaluated 
at the top quark scale. The correction is seen to be always negative. 
We point out that while we have used the exact 
${\cal O}(\alpha_s)$ formula for the
numerical evaluation, the approximate expression given above is
sufficiently accurate to convey the general features
to be expected
both at low and at high \tb.
In particular, for $m_b\neq 0$ and 
\tb\  in the relevant high segment (\ref{eq:tanbeta2}), 
the \QCD\  correction becomes very large and saturates at the value
\begin{equation}
\delta_{QCD}=-\frac{2\, \alpha_s}{\pi}\;\left(\frac{8\pi^2-15}{36}
            + \ln\frac{m^2_t}{m^2_b}\right)\simeq -62\%\,\
 \ \ (\tb>>\sqrt{m_t/m_b}\simeq 6)\,.
\label{eq:QCD2}
\end{equation} 
(The exact ${\cal O}(\alpha_s)$ formula gives slightly below $-60\%$.)
At low values of \tb, the corrections are much smaller, as it follows from
the approximate expression 
$\delta_{\rm QCD}\simeq (-\alpha_s/\pi)(8\pi^2-15)/18\simeq -12\%$. 
We remark that for $m_b=0$ the dependence on \tb\
totally disappears from eq.(\ref{eq:QCD1}), so that 
one would never be able to suspect the large contribution (\ref{eq:QCD2}) in
the high \tb\  regime. The limit $m_b=0$, nevertheless, has been
considered for the standard \QCD\  corrections
in some places of the literature but,
as we have seen, it is untenable unless one concentrates on values 
of \tb\  of order $1$, in which case the 
relevance of our decay for \SUSY\  is doomed to oblivion.
This situation is similar to the one
mentioned above concerning the \SUSY-\QCD\  corrections in the
limit $m_b=0$, 
which leads to an scenario totally blind to
the outstanding supersymmetric quantum effects obtained for 
$m_b\neq 0$ at high \tb\,\cite{TesiRicci,Guasch:1995rn}. 
We stress that in spite of the respectable size of the standard \QCD\  effects,
they become fast stuck at the saturation value (\ref{eq:QCD2}),
which is independent of \tb. On the contrary, the
\SUSY-\QCD\  effects grow endlessly  with \tb\  and thus rapidly overtake 
the standard \QCD\  prediction.

\figtbhdeltamgmultz

Worth noticing is the evolution of the quantities (\ref{eq:pito}) and (\ref{eq:BR})
as a function of the gluino mass (Cf. Figs.\,\ref{fig:tbhdeltamgmult0}(a)-\ref{fig:tbhdeltamgmult0}(b)). Of course, only
the \SUSY-\QCD\  component is sensitive to $m_{\tilde{g}}$.
Although the \SUSY-\QCD\  effects have been object of
a particular study in Ref.\cite{TesiRicci,Guasch:1995rn}, 
we find it convenient, to ease comparison, to display the corresponding 
results in the very same
conditions in which the electroweak supersymmetric corrections are presented.
The steep falls in Fig.\,\ref{fig:tbhdeltamgmult0}(a) are associated to the presence of
threshold effects occurring at points satisfying
$m_{\tilde{g}}+m_{\tilde{t}_1}\simeq m_t$. 
An analogous situation was observed in Ref.\cite{Garcia:1994rq,Dabelstein:1995jt,TesiRicci} 
for the SUSY
corrections to
the standard top-quark decay. Away from the threshold points, the behaviour of
$\delta_{SUSY-QCD}$ is smooth and perfectly consistent with perturbation theory.
In Fig.\,\ref{fig:tbhdeltamgmult0}(b), where the branching ratio (\ref{eq:BR}) is plotted, the steep falls
at the threshold points are no longer present since they are compensated for by the
simultaneous opening of the two-body supersymmetric mode
 $t\rightarrow \tilde{t}_1\,\tilde{g}$, for $m_{\tilde{g}}<m_t-m_{\tilde{t}_1}$.

We emphasize that the relevant gluino mass region for the decay $t\rightarrow \hplus\,b$ is 
not the light gluino region, but the heavy one, the reason being that the
important self-energy correction mentioned above, eq.(\ref{eq:dmbQCD}), 
involves a  gluino mass insertion. 
As a consequence, virtually for any set of \MSSM\  parameters, there is a well sustained
\SUSY\  correction for any gluino mass above a certain value, in our case
$m_{\tilde{g}}\gsim 250-300\GeV$.  
The correction raises with the gluino mass up to a long flat maximum  before
bending --very gently -- into the decoupling regime (far beyond $1\mbox{ TeV}$)\,\cite{TesiRicci,Guasch:1995rn}. 
The fact that the decoupling rate of the gluinos appears to be 
so slow has an obvious phenomenological interest.

Next we consider in detail the sensitivity of our decay
on the higgsino-gaugino parameters $(\mu,M)$ characterizing 
the chargino-neutralino
mass matrices (Cf. section~\ref{sec:MSSMlagrangian}). 
We start with the supersymmetric Higgs mixing 
mass, $\mu$. As already stated above, we will largely concentrate on
the $\mu<0$ case. 
Together with $A_t>0$ this yields $A_t\,\mu<0$,
which is a sufficient
condition\,\cite{Rattazzi:1996gk,Barbieri:1993av,Garisto:1993jc,Diaz:1994fc,Borzumati:1994zg,Bertolini:1995cv,Carena:1995ax,Bertolini:1991if}
(see also section \ref{sec:limitstanb}) 
for the \MSSM\  prediction on
$BR(b\rightarrow s\,\gamma)$ to be compatible with
experiment in the presence of a relatively light
charged Higgs boson  (as the one participating in the
top decay under study). In fact, it is known that charged Higgs
bosons of ${\cal O}(100)\GeV$ interfere constructively with the
SM amplitude and would render
a final value of $BR(b\rightarrow s\,\gamma)$ exceedingly high. 
Fortunately, this situation
can be remedied in the \MSSM\  since the alternative contribution from charginos
and stops tends to cancel the Higgs contribution provided that
$A_t\,\mu<0$. Furthermore, one must also
require relatively light values for the masses of the
lightest representatives of these sparticles, as well
as high values of
\tb\,\cite{Rattazzi:1996gk,Barbieri:1993av,Garisto:1993jc,Diaz:1994fc,Borzumati:1994zg,Bertolini:1995cv,Carena:1995ax,Bertolini:1991if};  
hence one is led to
a set of conditions 
which fit in with nicely to build up a favourable scenario
for the decay $t\rightarrow \hplus\,b$. 

\figtbhdeltamumultz

The evolution of the individual
contributions (\ref{eq:individ}), together with the total \MSSM\  yield,
as a function of $\mu<0$, is shown
in Figs.\,\ref{fig:tbhdeltamumult0}(a)-\ref{fig:tbhdeltamumult0}(b) for given values of the other parameters.
We immediately gather from these figures
that the \SUSY-\QCD\  correction is extremely
sensitive to $\mu$. In fact, $\delta_{\rm SUSY-QCD}$ grows rather
fast with $|\mu|$.
This is already patent at the level of the leading $\delta m_b/m_b$ 
effect given by eq.(\ref{eq:dmbQCD}).
In all figures where a definite $\mu<0$ is to be
chosen, we have taken the moderate value $\mu=-150\GeV$.

Concerning the electroweak contribution,
we noted above that the component $\delta m_b/m_b$, eq.(\ref{eq:dmbEW}),
actually decreases with $\mu$. However,
the $\mu$ dependences in the full $\delta_{\rm SUSY-EW}$ are more complicated than
in $\delta_{\rm SUSY-QCD}$ and cannot be read off
 eq.(\ref{eq:dmbEW}).
This is evident from  Fig.\,\ref{fig:tbhdeltamumult0}(a) where the total  $\delta_{\rm SUSY-EW}$
is fairly insensitive to $\mu$; $\delta_{\rm MSSM}$, therefore, inherits
its marked $\mu$-dependence basically from the \SUSY-\QCD\  component.
As for the sensitivity of the corrections on the $SU(2)_L$-gaugino soft
SUSY-breaking parameter, $M$, Fig.\,\ref{fig:tbhdeltaMmult0} conveys
immediately that it is virtually non-existent.  
   
\figtbhdeltaMmultz

\figtbhdeltaAmultz

There is some slight evolution of the corrections with $A_b$ (Fig.\,\ref{fig:tbhdeltaAmult0}(a)),
mainly on the \SUSY-\QCD\  side.
We realize that $\delta_{\rm SUSY-QCD}$ is not perfectly
symmetric with respect to the sign of $A_b$. 
Once the sign $\mu<0$ is chosen, the correction is 
larger for negative 
values of $A_b$ than for positive values. We have erred on the 
conservative side by choosing $A_b=+300\GeV$ wherever this parameter
is fixed.
As far as $A_t$ is concerned, $\delta_{\rm SUSY-QCD}$ can only evolve as
a function of that parameter through vertex corrections,
which are proportional to $A_t\,\cot\beta$ (Cf. section~\ref{sec:sfmas}); however, at large
\tb\  these are very depressed. The electroweak correction
$\delta_{\rm SUSY-EW}$, instead, is very much dependent on $A_t$; indeed,
a typical component exhibiting this behaviour 
is given by eq.(\ref{eq:dmbEW}),
which is linear in $A_t$. The full dependence, however, is not linear and is
recorded in Fig.\,\ref{fig:tbhdeltaAmult0}(b). We realize that $\delta_{\rm SUSY-EW}$ and $\delta_{\rm MSSM}$
change sign with $A_t$. The shaded vertical band in Fig.\,\ref{fig:tbhdeltaAmult0}(b) is excluded by our
choice of parameters in Fig.\,\ref{fig:tbhgammatbmult0}.

\figtbhdeltamsbmultz

Another very crucial parameter to be investigated is the value of
$m_{\tilde{b}_1}$. 
This is because the \SUSY-\QCD\  correction hinges a great deal on the value of the
sbottom masses, as it is plain from eq.(\ref{eq:dmbQCD}).
As  a matter of fact, a too large a value of $m_{\tilde{b}_1}$  may upside down
the leadership of the \SUSY-\QCD\  effects.
As a typical mass value for all squarks other than the stop we use
$m_{\tilde{q}}\geq 150-200\GeV (\tilde{q}\neq \tilde{t})$.
From Fig.\,\ref{fig:tbhdeltamsbmult0}(a) we see that provided
$m_{\tilde{b}_1}\lsim 300\GeV$ the \SUSY-\QCD\  effects remain dominant, but
they steadily go down the larger is $m_{\tilde{b}_1}$. The
electroweak correction $\delta_{SUSY-EW}$, on the other hand, is quite sustained 
with increasing $m_{\tilde{b}_1}$ and there are parameter configurations where for
sufficiently heavy sbottoms the supersymmetric electroweak effects are larger
than the \SUSY-\QCD\  effects. However, this is not the most likely situation.
The behaviour of the branching ratio is plotted in Fig.\,\ref{fig:tbhdeltamsbmult0}(b).

\figtbhdeltamstmultz

Obviously, the evolution of the \SUSY-\QCD\
corrections with the stop masses is basically flat (Fig.\,\ref{fig:tbhdeltamstmult0}) since the leading
contribution is independent of $m_{\tilde{t}_1}$. Therefore, for definiteness
we fix $m_{\tilde{t}_1}\simeq 100\GeV$.
Nonetheless, if we wish to keep $\delta_{\rm MSSM}>0$,
we cannot go too far with  $m_{\tilde{t}_1}$,
for the electroweak correction is also seen to 
decrease with $m_{\tilde{t}_1}$.
Indeed, whereas for $m_{\tilde{t}_1}=65-100\GeV$ one has
$\delta_{\rm SUSY-EW}\gsim 20\%$, for
 $m_{\tilde{t}_1}\gsim 250\GeV$ one finds 
$\delta_{\rm SUSY-EW}\lsim 10\%$. For heavier stop masses, $\delta_{\rm MSSM}$
becomes zero or slightly negative. In this situation, the imprint of SUSY
lies in the fact that the total quantum effect is not as negative as predicted
by standard QCD, eq.(\ref{eq:QCD2}).

\figtbhdeltamsupmultz

The influence from the sleptons and the other squarks is practically irrelevant
as it is borne out by Figs.\,\ref{fig:tbhdeltamsupmult0}(a)-\ref{fig:tbhdeltamsupmult0}(b). 
They enter the correction through oblique (universal) quantum
corrections. The only exception are the $\tau$-sleptons $\tilde{\tau}_a$ 
(``staus''), since they are involved in the process-dependent (non-oblique) 
contribution eq.(\ref{eq:deltatau}), where the $\tau$-lepton Yukawa coupling becomes
enhanced at large \tb.
For this reason, $\delta_{\rm SUSY-EW}$ in Fig.\,\ref{fig:tbhdeltamsupmult0}(b) is
somewhat larger the smaller is the
$\tau$-sneutrino mass (assumed to be degenerate with the other sneutrinos).
In all our calculation we have fixed the common
sneutrino mass at $200\GeV$.

\figtbhdeltamtmultz

\figtbhdeltambmultz

We have also tested the variation of our results as a function of the
external particle masses, namely the masses of the
top quark, bottom quark and charged Higgs. 
As for the external fermion masses,
the corrections themselves are not very sensitive (see Figs.\,\ref{fig:tbhdeltamtmult0}(a) and \ref{fig:tbhdeltambmult0}(a)). 
Among the \SUSY\  corrections, the most sensitive
one on $m_t$ (respectively on $m_b$) is $\delta_{\rm SUSY-EW}$
(resp. $\delta_{\rm SUSY-QCD}$).
The branching ratios also show 
some dependence (Figs.\,\ref{fig:tbhdeltamtmult0}(b) and \ref{fig:tbhdeltambmult0}(b)), especially on $m_b$.
This effect is mainly due to the variation of 
the tree-level partial widths
as a function of $m_t$ and $m_b$.
As for the charged Higgs mass, $M_{\hplus}$, up to now it has been
fixed at $M_{\hplus}=120\GeV$, which is the preferred value for this mass at
large \tb\,\cite{HollikQEMSSM}.
We confirm from  Fig.\,\ref{fig:tbhdeltamhcmult0}(a) that
there is nothing special in the chosen value for that parameter since the
sensitivity of the correction is generally low, except near the
uninteresting boundary of the phase space where the 
branching ratio (Fig.\,\ref{fig:tbhdeltamhcmult0}(b)) boils down to zero. 

\figtbhdeltamhcmultz

\figtbhdeltadeltataumultz

We close our study of the corrections in the $\mu<0$ case
by plotting $\delta_{\tau}$ as a function of \tb\
(see Fig.\,\ref{fig:tbhdeltadeltataumult0}). By definition,  $\delta_{\tau}$ is that part of
$\delta_{\rm MSSM}$ originating from
the full process-dependent term $\Delta_{\tau}$, eq.(\ref{eq:deltatau}),
which stems from our  definition of \tb\  on eq.(\ref{eq:tbetainput}).
This piece of information is relevant enough.
In fact, it should be recalled that the quantum corrections described in the previous
figures are scheme dependent. In particular, they rely on our definition of
\tb\  given on eq.(\ref{eq:tbetainput}). What is {\it not} scheme
dependent, of course, is the predicted
value of the width and branching ratio (Figs.\,\ref{fig:tbhgammatbmult0} and \ref{fig:tbhdeltatbmult0}(b))
after including all the radiative corrections.
Now, from Fig.\,\ref{fig:tbhdeltadeltataumult0} it is clear that 
the  $\Delta_{\tau}$-term is not negligible, 
and so there is a process-dependence in our
definition of \tb, as it was announced in chapter~\ref{cap:Renorm}. 
At first sight, the $\delta_{\tau}$-effects are not
dramatic since they are small as compared to
$\delta_{\rm SUSY-QCD}$, but since the latter is canceled out by
standard \QCD\  we end up with 
$\delta_{\tau}$ being of the order (roughly half the size) of the 
electroweak correction $\delta_{\rm SUSY-EW}$.


The main source of process-dependent $\delta_{\tau}$-effects 
lies in the corrections generated by the $\tau$-mass counterterm,
$\delta m_{\tau}/ m_{\tau}$, and can be easily picked out in the 
electroweak-eigenstate basis (see Fig.\,\ref{diag:deltamtauEW}) much in the same way as we did for the
$b$-mass counterterm. There are, however, some differences, as can be
appraised by comparing the diagrams in Figs.\,\ref{diag:deltambEW} and \ref{diag:deltamtauEW}, where we see that
in the latter case the effect derives from diagrams
involving $\tau$-sleptons with gauginos or mixed gaugino-higgsinos.  
An explicit computation
of the diagrams (a) $+$ (b) in Fig.\,\ref{diag:deltamtauEW} yields
\begin{eqnarray}
{\delta m_{\tau}\over m_{\tau}} &=&
{g'^2\over 16\pi^2}\,\mu\,M'\,\tb\,
I(m_{\tilde{\tau}_1},m_{\tilde{\tau}_2},M')\nonumber\\
& + & {g^2\over 16\pi^2}\,\mu\,M\,\tb\,
I(\mu,m_{\tilde{\nu}_{\tau}},M)\,,
\label{eq:dmtau12}
\end{eqnarray}
where $g'=g\,s_W/c_W$ and $M', M$ (Cf. section~\ref{sec:cnmas}) are the Soft-\SUSY-Breaking
Majorana masses
associated to the bino $\tilde B$ and winos $\tilde{W}^{\pm}$, respectively,
and the function $I(m_1,m_2,m_3)$ is again given by eq.(\ref{eq:I123}).
In the formula above
we have projected, from the bino diagram in Fig.\,\ref{diag:deltamtauEW}(a), only the leading piece which is
proportional to \tb.
Even so, the contribution from the wino-higgsino diagram in Fig.\,\ref{diag:deltamtauEW}(b) is much larger. 
Numerical evaluation of the sum of the two contributions on
eq.(\ref{eq:dmtau12}) indeed shows that it reproduces to within
few percent the full numerical result (Cf. Fig.\,\ref{fig:tbhdeltadeltataumult0}) previously obtained in
the mass-eigenstate basis, thus 
confirming that eq.(\ref{eq:dmtau12}) gives the leading contribution. 
In practice, for a typical choice of parameters as in Fig.\,\ref{fig:tbhgammatbmult0},
this contribution is approximately canceled out by part of the electroweak
supersymmetric corrections associated to the original process $t\rightarrow \hplus\,b$,
and one is effectively left with eq.(\ref{eq:dmbEW}) as being the main source   
of electroweak supersymmetric quantum effects at high \tb. 

\diagdeltamtauEW

\figtbhdeltamsbmugtz

Finally, the corrections corresponding to the case  $\mu>0$ are 
studied in Figs.\,\ref{fig:tbhdeltamsbmugt0}(a) and \ref{fig:tbhdeltamsbmugt0}(b).  
The problem with $\mu>0$ is that, then, the large \SUSY-\QCD\  corrections
have the same (negative) sign as the conventional \QCD\  effects, and as
a consequence the total \MSSM\  correction can easily blow up above $100\%$,
the branching ratio becoming negative! 
To avoid this disaster (from the point of view of perturbation
theory), we enforce the \SUSY-\QCD\  correction to be smaller
than in the $\mu<0$ case
by assuming an ``obese \SUSY\  scenario''
characterized by  very large values for the
sbottom mass ($m_{\tilde{b}_1}=600\GeV$) and the
gluino mass ($m_{\tilde{g}}=1000\GeV$).
We also choose $A_t<0$ so that the electroweak \SUSY\  
correction becomes opposite
in sign to the \SUSY-\QCD\  correction (a feature that also applies in the
$\mu<0$ case, see Fig.\,\ref{fig:tbhdeltaAmult0}(b)) and in this way the total \SUSY\  correction is
further lessened in absolute value. Incidentally, we
remark that the simultaneous sign change of both
$\mu$ and $A_t$ is also necessary in order to
keep  $A_t\,\mu<0$; as noted above, this is required in order
that the \MSSM\  can be
compatible with $BR(b\rightarrow s\,\gamma)$ in the presence of
a relatively light charged Higgs.
In Fig.\,\ref{fig:tbhgammatbmugt0} we bring forward the
effect of the new situation on the total partial width. 
In the present instance, the physical signature would
be to measure a  partial width significantly
smaller than the one predicted by QCD\@.
Clearly, the $\mu>0$ ($A_t\,\mu<0$) scenario is not
as appealing as the $\mu<0$ ($A_t\,\mu<0$) one.

\figtbhgammatbmugtz

\input{tanbeta.tex}

\section{Conclusions}
\label{sec:tbhconclusions}
From the explicit numerical analysis (section~\ref{sec:tbhnumeric}), we have
confirmed 
our expectations that the \SUSY-\QCD\  contribution
to $\Gamma (t\rightarrow \hplus\,b)$ is generally dominant.
This conclusion would not hold only in some (unlikely) cases, e.g.\
if the gluino is very light
and/or the lightest bottom squark is ``obese'' as compared to
lightest top squark, i.e.\  if the former is
unusually much heavier than expected. 
Furthermore, by restricting ourselves to the case $\mu<0$ ($A_t\,\mu<0$)
we confirm that at large \tb\  and for typical values of the parameters
the total (standard plus supersymmetric) \QCD\  correction largely 
cancels out, leaving a remainder
on the \SUSY-\QCD\  side (Figs.\,\ref{fig:tbhgammatbmult0}-\ref{fig:tbhdeltatbmult0}).
In all circumstances the virtual Higgs effects remain comparatively very small.   
Around $\tb=m_t/m_b\simeq 35$,
one is left with basically
the electroweak supersymmetric correction, $\delta_{SUSY-EW}$,
which can be sizeable enough to be pinned down by experiment.
However, as stated above, there is in general
a strong remainder, $\delta_{SUSY-QCD}+\delta_{QCD}>0$,
which grows very fast with \tb\  and it has the same sign as 
$\delta_{SUSY-EW}$. 
In this favourable scenario, 
the virtual \SUSY\  effects could be spectacular.
This is true not only because in the
relevant window of parameter space the 
\SUSY\  quantum corrections are by themselves rather
large, but also because they push into the opposite direction than the 
``expected'' standard \QCD\  corrections. 
As a result, the relative deviation
between the \MSSM\  prediction and the \QCD\  prediction effectively
``doubles'' the size of the observable effect, a fact which is
definitely welcome from the experimental point of view.

From all the previous discussion there is one fact
standing out which can be 
hardly overemphasized: If the charged Higgs decay mode of the top quark,
$t\rightarrow \hplus\,b$, does show up  
with a branching ratio of order $10\%$ or above (perhaps even as
big as $50\%$), a fairly rich event 
statistics will be collected at the Tevatron
and especially at the LHC e.g.\  by making use
of the identification methods described in Section~\ref{sec:tbhtree}.  
If, in addition, it comes out that the dynamics underlying that decay is truly
supersymmetric, then the valuable quantum signatures that our calculation has
unveiled over an ample portion of the \MSSM\  parameter
space should eventually become manifest and, for sure, we could not miss them. 

\diagtbhfusion

At present all the collected event statistics basically relies on
our experimental ability to recognize the top quark decays originating from
standard patterns (angular distribution, energy spectrum, 
jet topology etc.) associated to
the usual Drell-Yan production mechanism. 
Notwithstanding, we wish to point out that it should in principle be possible to   
clutch at the supersymmetric virtual corrections associated to the
vertex $t\,b\,H^{\pm}$ also
through an accurate measurement of the various 
inclusive top quark and Higgs boson production cross
sections in hadron colliders. As an example, in Fig.\,\ref{diag:tbhfusion} we sketch a few
alternative mechanisms which
would generate top quark production patterns  
heavily hinging
on the properties of the interaction
$t\,b\,H^{\pm}$-vertex\,\cite{Jimenez:1996wf,Coarasa:1996yg}. 
Thus, while this vertex could be 
responsible in part for the decay
of the top quark once it is produced,   
it might as well be at the root of the production process itself
at LHC energies, where it could take over from 
Drell-Yan production\,\cite{Yuan:1995ez}.

We observe that
in some of these mechanisms a Higgs boson is produced in
association, but in some others (fusion processes)
the Higgs boson enters as a virtual particle. 
Now, however different these production processes might be,
all of them are sensitive to the effective structure
of the $t\,b\,H^{\pm}$-vertex.
The top quark-charged Higgs associated production can be used to search for a
charged Higgs boson at 
hadron colliders, using the single top quark production as a signature, and
searching for an excess of $\tau\nu_\tau$ lepton pairs in this processes, in a
kind of analysis similar to that of section~\ref{sec:limitstanb}. Work
is currently in progress to compute these effects, and the discovery limit of a
charged Higgs particle at the Tevatron Run II and at the LHC.
 Similar mechanisms can of course be depicted
involving the neutral Higgs bosons of the \MSSM\  interacting with $t\,\bar{t}$ and
$b\,\bar{b}$  via enhanced Yukawa
couplings\,\cite{Jimenez:1996wf,Coarasa:1996yg}. 
While it goes beyond the scope of this Thesis
to compute the \SUSY\  corrections to the production processes
themselves, we have at least faced the detailed analysis of a partial
decay width which involves one of the
relevant production vertices. 
In this way, a definite prediction is made on the
properties of a physical observable and, moreover,
this should suffice both
to exhibit the relevance of the \SUSY\  quantum effects
and to  demonstrate the necessity to incorporate these corrections 
in a future, truly 
comprehensive, analysis of the cross-sections, namely, an analysis where
one would include the quantum effects on all the relevant
production mechanisms within the framework 
of the \MSSM\@.  
For this reason we think that   
in the future a precise measurement of the various (single and double) top
quark production cross-sections\,\cite{Willenbrock:1995rm,Quigg:1995bv} 
should be able to detect or to exclude
the $t\,b\,H^{\pm}$-vertex as well as the vertices $q\,\bar{q}\, A^0 (h^0,H^0)$
involving the neutral Higgs particles
of the \MSSM\  and the third generation quarks $q=t,b$. Notice the fact that
finding a light $A^0$ either at LEP or at the Tevatron and not finding a $H^+$
below $\mt$ through $t\rightarrow H^+\,b$ would not exclude the \MSSM\  at all,
provided
$(A_t<0,\mu>0)$. In this latter scenario the $b\,\bar{b}\,h$ and
$b\,h$ channels production cross-sections will suffer from the same large 
negative corrections than the $t\,b\,H^+$ vertex itself (see below). Then the most
plausible production process 
of $A^0$, $H^0$ and $h^0$ should be the associated production with a gauge boson
(e.g. $W^+$)\,\cite{Carena:1998gk}. In fact, the    
observation of this latter channel and non-observation of the
bottom-Higgs coupling would point toward the \MSSM\  nature of this Higgs    
boson, with a definite prediction on the sign of the $\mu$ parameter:
$\mu>0$.

The finite threshold effects to the bottom quark mass
--eqs.\,(\ref{eq:dmbQCD}) and (\ref{eq:dmbEW})-- will be
on top of any observable proportional to the bottom quark Yukawa coupling. At
one-loop, and maintaining only the leading finite contributions, this Yukawa
coupling reads --see eq.\,(\ref{eq:Yukawasgeneric})--
\begin{equation}
  \label{eq:Yukawabottom1loop}
  h_b=\frac{g\,\mb}{\sqrt{2}\,\mw\,\cos\beta}\left(1+\frac{\delta\mb}{\mb}\right)\,\,,
\end{equation}
where the factor $\delta\mb/\mb$ is the net sum of expr.\,\,(\ref{eq:dmbQCD})
and (\ref{eq:dmbEW}). Thus we could define an ``effective'' bottom Yukawa
coupling~(\ref{eq:Yukawabottom1loop}) which would give us a first estimation of
the supersymmetric radiative corrections to any process in the bottom-Higgs
supersector, such as $A^0\rightarrow
b\bar{b}$\,\cite{Dabelstein:1995js,Coarasa:1996yg} 
or the production processes
quoted above. However a full computation will always be needed to be sure this
estimation really gives the bulk of the correction (see
chapter~\ref{cap:sbdecay}).

Whereas, on the one hand, one expects that some top quark partial widths
will be determined with an accuracy
of $10\%$ at the upgraded Tevatron
and perhaps better than $10\%$ 
at LHC\,\cite{Moriond97}, on the other hand we believe that
from the point of view of an {\sl inclusive} model-independent
measurement of  
the {\sl total} top-quark width, $\Gamma_t$, the future $e^+\,e^-$ supercollider
should be a better suited machine\,\cite{Fujii:1994mk,Frey:1995ai}. 
For, in an inclusive measurement,
all possible non-SM effects will appear on top
of the corresponding SM effects already computed in the
literature\,\cite{Jezabek:1989iv,Jezabek:1989ja,Li:1991qf,Eilam:1991iz,Denner:1990tx,Denner:1991ns,Irwin:1991ka,Yuan:1991av}.  
Moreover, 
as shown in Ref.\cite{Fujii:1994mk}, 
one hopes to be able to measure the total top-quark
width in $e^+\,e^-$ supercolliders at an unmatched precision
of $\sim 4\%$ on the basis of a
detailed analysis of the threshold effects in the cross-section, in particular
of the top momentum distribution and the
resonance contributions to the forward-backward
asymmetry in the $t\bar{t}$ threshold region. Under the assumption that
$\Gamma_H\simeq \Gamma_W$,  and that the
\SUSY\  effects on $\Gamma_t$ are purely virtual effects, 
it follows that a large \SUSY\  correction of, say $50\%$, to $t\rightarrow \hplus\,b$ 
translates into a $20\%$ correction to $\Gamma_t$. 
This effect could not escape detection.
Thus, the combined information
from a future $e^+\,e^-$ supercollider
and from present and medium term hadron machines
can be extremely useful to pin down the nature of the observed effects.

From the study of the quantum effects on the top quark
decay channel into charged Higgs particles  we arrive
at the conclusion that the
recently presented $\tau$-lepton analyses by the CDF Collaboration 
at Fermilab are in general model-dependent and could be significantly
altered by potentially underlying new physics.
In particular, since  in the absence of new interactions the results
from radiative $B$-meson decays generally preclude
the existence of charged Higgs bosons below the top quark mass, it is 
reasonable to link the existence of the decay $t\rightarrow H^+\,b$ to the 
viability of the leading candidate for physics beyond the \SM, viz.\  the \MSSM\@.
In this framework we find that, depending on the sign of the higgsino
mixing parameter, $\mu$, the recent $\tau$-lepton exclusion plots in 
($\tan\beta, \mHp)$-space presented by CDF could
either be further strengthened or on the contrary be greatly weakened.
This dual situation could
only be decided from additional experimental information unambiguously 
favouring a given sign of $\mu$ in other physical processes. 

In mSUGRA models in the literature one usually claims $A_t<0$. This would imply
a positive value for the $\mu$ parameter in order to comply with the
$b\rightarrow s\,\gamma$ constraint. Thus it seems that the prediction 
from these models would be that there is no actual limit to the charged Higgs boson
mass, and that there is no hope of seeing the $t\rightarrow H^+\,b$ at collider
experiments even if \tb\  were large. However the explicit prediction of mSUGRA models
is\,\cite{Carena:1997km,Carena:1997mb}
$$
A_t = (1-r) A_0-2\,m_{1/2}\,\,,
$$
$r$ being the ratio of the Yukawa coupling squared with respect its value at the
RGE fixed point. Its value at large \tb\  is $r\simeq3/4$. We can see that for 
$m_{1/2}\simeq 100-200 \GeV$ and $A_0\gsim 1 \mbox{ TeV}$ these models already
predict $A_t>0$. Together with $\mu<0$ we already see that a large branching
ratio for $t\rightarrow H^+\,b$ is not excluded at all.

We remark that although for brevity sake we have presented our numerical
analysis for a given choice of the \MSSM\   parameters, we have checked that
our conclusions hold basically unaltered in ample regions of parameter space
involving typical sparticle masses of a few hundred
GeV. 
While the details of the exclusion plot in $(\tan\beta,\mHp)$-space may
depend on the particular channel used to tag a potential
excess of $\tau$-leptons, all of these plots (and of course also 
the one from the inclusive measurement) should undergo significant changes.
Finally, it is clear that similar considerations 
apply to experiments of the same nature being planned for
the future at the LHC and at the LC\@. Thereby a general conclusion 
seems to consolidate\,\cite{Sola:1997ds}: 
In contrast to gauge boson observables, the \MSSM\  
quantum effects on Higgs boson physics can be rather large and
should not be ignored in future searches at the Tevatron,
at the LHC and at the LC.


%% file: tanbeta.tex
\section{Implications for the Tevatron data}
\label{sec:limitstanb}
The results presented in previous sections indicate that the \SUSY\  corrections
to the decay under study could have a great impact on the search for charged Higgs
bosons at the Tevatron. 
The CDF Collaboration at the Tevatron
has carried out direct searches for charged Higgs production 
in $p\,\bar{p}$ collisions at
$\sqrt{s}=1.8 \mbox{ TeV}$\,\cite{Conway,Abe:1997rk,Abe:1996pj}. 
In these studies one is concerned with the final configurations
$t\bar{t}\rightarrow H^+\,H^-\,b\,\bar{b}, 
W^+\,H^-\,b\,\bar{b},H^+\,W^-\,b\,\bar{b}$.
The latter would differ from that of the standard model,
$t\bar{t}\rightarrow W^+\,W^-\,b\,\bar{b}$, by
an excess of states with one (or two) $\tau$-lepton ``jets'' 
(i.e.\  usually tagged in the hadronic decay mode) and two b-quarks 
and large missing transverse energy associated to
the decays $H^+\rightarrow \tau^+\,\nu_{\tau}$ and/or 
$H^-\rightarrow \tau^-\,\bar\nu_{\tau}$.

As stated in section~\ref{sec:tbhtree}, 
if $\mHp$ is not too close to
the phase space limit, then 
there are two regimes, namely a low and a high $\tan\beta$ regime,
where the partial width of the
unconventional top quark decay becomes sizeable 
as compared to the standard decay $t\rightarrow W^+\,b$.
Nevertheless we shall focus only on the high $\tan\beta$
regime as it is this case that is correlated with the Higgs maximum rate into
the $\tau$-mode versus the hadronic mode (Cf.\  eq.(\ref{ratiotaucs})).
Clearly, the identification of the charged Higgs decay of the top quark
could be a matter of observing a departure from the universality prediction
for all the lepton channels through the measurement of an excess 
of inclusive (hadronic) $\tau$-events. 
However, from the non-observation of any $\tau$-lepton surplus,
one may determine
an exclusion region in the $(\tan\beta, M_{H})$-plane
\,\cite{Conway,Abe:1997rk,Abe:1996pj,Guchait:1997fh} 
for any
(Type II) $2HDM$\,\cite{Hunter}.
The region highlighted in this plane
consists of a sharply edged area forbidding too high values
of $\tan\beta$ in correlation with $\mHp$.
In the relevant \SUSY\  region $\mHp>100\GeV$ (see below) 
the most recent analysis would imply that values in the range
$\tan\beta\gsim 40$ would be excluded\,\cite{Abe:1997rk,Abe:1996pj}.

In spite of its foreseeable importance, the impact of the \SUSY\  
quantum corrections on the dynamics of $t\rightarrow H^+\,b$ 
was not included in any of the aforementioned
analysis\,\cite{Conway,Abe:1997rk,Abe:1996pj,Guchait:1997fh}. 
And this is especially significant
in a decay like $t\rightarrow H^+\,b$ whose sole existence could, in a sense, 
already be an indirect sign of \SUSY\@. For, as is well-known, the CLEO
data\,\cite{Alam:1995aw} 
on the radiative decays $\bar{B}^0\rightarrow X_s\,\gamma$ 
(viz. $b\rightarrow s\,\gamma$) set a
rather stringent lower bound on the mass of any
charged Higgs scalar belonging to a generic $2HDM$, to wit: $\mHp>240\GeV$. 
Therefore, with only the $W^\pm$ and $H^\pm$ electroweak corrections, 
the charged Higgs mass is forced to lie in a range
where the decay $t\rightarrow H^+\,b$ becomes kinematically blocked up. 
Of course, this is so because the virtual Higgs effects go in the same
direction  
as the \SM\  contribution. Fortunately, this situation can be remedied in the
\MSSM\  
where the complete formula for the $b\rightarrow s\,\gamma$
branching ratio reads (see the extensive
literature\,\cite{Rattazzi:1996gk,Barbieri:1993av,Garisto:1993jc,Diaz:1994fc,Borzumati:1994zg,Bertolini:1995cv,Carena:1995ax,Bertolini:1991if}
for details):
\begin{equation}
BR (b\rightarrow s\,\gamma)\simeq BR (b\rightarrow c\,e\,\bar{\nu})\,
{(6\,\alpha_{\rm em}/\pi)\,\left(\eta^{16/23}\,A_{\gamma}+C\right)^2\over
I(m_c/m_b)\,\left[1-\frac{2}{3\pi}\,\alpha_s(m_b)f_{\rm QCD}(m_c/m_b)\right]}
\label{eq:bsg}\,,
\end{equation}
with
\begin{equation}
A_{\gamma}=A_{\rm SM}+A_{H^-}+A_{\chi^-\tilde{q}}
\label{eq:AS}
\end{equation} 
being the sum of the
SM, charged Higgs and chargino-squark
amplitudes, respectively. (The contributions from the neutralino
and gluino amplitudes are in this case generally smaller as they 
enter through \FCNC.) Now,
the important feature here is that the unwanted charged Higgs effects
could to a large extent be compensated for by the chargino-stop contributions.
And in this case a relatively light charged Higgs particle would perfectly be 
allowed in the \MSSM\  for the decay $t\rightarrow H^+\,b$ to occur.

In our renormalization framework, we use $H^{+}\rightarrow\tau^{+}\nu_{\tau}$ 
to define the parameter $\tan\beta$ through eqs.\,(\ref{eq:tbetainput})
and~(\ref{eq:deltabeta}), 
which allows to renormalize the $t\,b\,H^+$-vertex in perhaps the most
convenient way to deal with our physical process
$t\rightarrow H^+\,b\rightarrow \tau^+\,\nu_{\tau}\,b$. 
Apart from the full set of electroweak and strong corrections
from the roster of \SUSY\  particles (squarks, 
gluinos, chargino-neutralinos and higgses), we of course include the
standard \QCD\  correction\,\cite{Czarnecki:1993zm}.

The results are presented in Figs.\,\ref{fig:tanbtop1}-\ref{fig:tanbtop4}.
We point out that in the present work we have locked together
the \MSSM\  parameter space regions for
the two decays $b\rightarrow s\,\gamma$ and $t\rightarrow H^+\,b$
in order to find compatible solutions. In doing so
we have used the full structure involved
in eqs.\,(\ref{eq:bsg}), (\ref{eq:AS}). 
Notice that recently the NLO \QCD\  effects in the \SM\  amplitude
have been computed and the total error has diminished
from roughly $30\%$ to $15\%$ (including the error in $m_b/m_c$)
\,\cite{Chetyrkin:1997vx}. 
Also NLO order 
corrections to $b \to s \gamma$ in extensions of the
\SM\  have become available~\cite{Paolobsg,Ciuchinibsg,Borzumatibsg}, but, on
the other hand,  new data from the CLEO
collaboration makes the upper limit on this decay grow up~\cite{NEWCLEO}. The
inclusion of these new information would be complicated and would not
change our results. 

\figtanbtopu

In Fig.\,\ref{fig:tanbtop1} we determine the permitted region in the
$(\tan\beta,A_t)$-plane in accordance with
the CLEO data on radiative $\bar{B}^0$ decays at $2\,\sigma$. For fixed
$\mu<0$, we find that $A_t\,\mu<0$ in the allowed region. This piece of
information could be important since, as it is patent in that figure, the
trilinear coupling $A_t$ -- entering the
\SUSY\  electroweak corrections -- becomes strongly correlated with $\tan\beta$. 
This correlation depends slightly upon the value of the charged Higgs
mass, $\mHp$, and it is built-in
for the rest of the plots (Figs.\,\ref{fig:tanbtop2}-\ref{fig:tanbtop4}).
From the \SM\  result mentioned above, we have made allowance for
an uncertainty of order $30\%$ stemming from the
non-included NLO corrections within the \MSSM\@.

For definiteness, and to ease comparison with the non-supersymmetric
results, we will normalize our analysis with respect
to Ref.\cite{Guchait:1997fh}. 
Here the $(l,\tau)$-channel, with $l$ a light
lepton, is used to search for
an excess of $\tau$-events. This should suffice to illustrate the
potential impact of the \MSSM\  effects on this kind of physics.
To be precise, we are interested in the
$t\,\bar{t}$ cross-section for the $(l,\tau)$-channel, $\sigma_{l\tau}$,
i.e.\  for the final states caused by the decay sequences  
$t\,\bar{t}\rightarrow H^+\,b,W^-\,\bar{b}$ and 
$H^+\rightarrow \tau^+\,\nu_{\tau}$, $W^-\rightarrow l\,\bar{\nu}_l$ 
(and vice versa). The relevant quantity can be
easily derived from the measured value
of the canonical cross-section $\sigma_{t\bar{t}}$ for the standard
channel $t\rightarrow b\,l\,\nu_l$, $\bar{t}\rightarrow b\,q\,q'$, after
inserting appropriate branching fractions, namely 
\begin{equation}
\sigma_{l\tau}=\left[\frac4{81}\,\epsilon_1+\frac49\,
{\Gamma (t\rightarrow H\,b)\over
\Gamma (t\rightarrow W\,b)}\,\epsilon_2\right]\,\sigma_{t\bar{t}}\,,
\label{eq:bfrac}
\end{equation} 
where the first term in the bracket comes from the \SM\  decay, and 
for the second term we assume (at high $\tan\beta$) $100\%$ branching 
fraction of $H^+$ into $\tau$-lepton, as explained before. Finally,
$\epsilon_i$ are detector efficiency factors\,\cite{Guchait:1997fh}, 
which we quote in table~\ref{tab:eficiencies}. 
Notice that the
use of the measured value of $\sigma_{t\bar{t}}$\,\cite{Tipton}, instead 
of the predicted value within the standard NLO \QCD\  
approach\,\cite{Berger:1995xz,Berger:1996ad}, 
allows a model-independent treatment
of the result. In this respect, we note that there could be \MSSM\  effects
on the standard mechanisms for $t\,\bar{t}$ production\,\cite{Hollik:1997hm} 
(viz. Drell-Yan $q\,\bar{q}$ annihilation and gluon-gluon fusion) as well
as corrections in the subsequent top quark
decays\,\cite{Garcia:1994rq,Dabelstein:1995jt}. 
To be concrete, we use the following value for the top pair production
cross-section\,\cite{Tipton}:
$$
\sigma_{t\bar{t}}=7.5 \pm 1.5\ {\rm pb}\,\,.
$$
The number of events found in the $(l,\tau)$-channel up to an integrated luminosity
of $110\,{\rm pb}^{-1}$ is $4$~\cite{PDB,Abe:1997uk}, with an expected background of $\sim
2$ events and $\sim 1$ event expected in the \SM. This implies an
upper limit of $7.7$ events at $95\%$ C.L., that is
$$
\sigma_{l\tau} < 70\ {\rm fb}\ \  (95\%\ {\rm C.L.})\,\,\,.
$$

\tableficiencies 

Therefore, proceeding in this way the bulk 
of the \MSSM\  pay-off stems from the $t\rightarrow H^+\,b$ contribution 
in eq.(\ref{eq:bfrac}). Specifically,
in Fig.\,\ref{fig:tanbtop2} we determine, as a function of $\tan\beta$
and for a given Higgs mass and fixed set of \SUSY\  parameters,
the cross-section for the $(l,\tau)$ final state.
There we show the tree-level ($\sigma_0$), \QCD-corrected ($\sigma_{QCD}$)
and fully \MSSM-corrected ($\sigma_{MSSM}$)
results. Of course, $\sigma_{MSSM}$ includes both the \SUSY-\QCD\  and standard
\QCD\  effects, plus the \MSSM\  electroweak corrections.
Note that the \QCD\  curve is similar to the one
in Ref.\cite{Guchait:1997fh}
\footnote{There
is, however, a small difference due to the fact that we use the top quark
scale, instead of the Higgs mass scale, to compute the \QCD\  corrections.}, but
as it is also patent the full \MSSM\  curve is quite different from the QCD
one: in fact, the two
curves lie mostly on opposite sides with respect to the tree-level curve!. This
is the same effect we have seen in Fig.\,\ref{fig:tbhgammatbmult0} translated
to the cross-section determination.

\figtanbtopdos

The horizontal line in Fig.\,\ref{fig:tanbtop2} gives the cross-section
for the number of events expected in the $(l,\tau)$-configuration
at the $95\%$ C.L. after correcting for the
detector efficiencies.
Hence the crossover points of the three curves with this
line determine (at $95\%$ C.L.) the maximum allowed value of
 $\tan\beta$ for the given set
of parameters. It is plain that the \MSSM\  curve crosses that line
much earlier than the \QCD\  curve, so that the
$\tan\beta$ bound is significantly tighter than in the
non-supersymmetric case. Notice that for this particular set of 
parameters the \MSSM\  and tree-level curves turn out to meet the horizontal line
at about the same point, which means that the \SUSY\  effects fully 
counterbalance the standard \QCD\  correction. We remark that this
feature may occur for negative values of the higgsino mixing
parameter (in Fig.\,\ref{fig:tanbtop2}, $\mu=-90\GeV$). 
The situation with $\mu>0$, with its very different corrections,  is different
and it will be discussed below.  

\figtanbtoptres

In Fig.\,\ref{fig:tanbtop3} we present our results in the $(\tan\beta, \mHp)$-plane, by
iterating the procedure followed in Fig.\,\ref{fig:tanbtop2} for $\mu<0$ and for charged Higgs
masses comprised in the relevant kinematical range $100\GeV<\mHp<m_t$.
The lower bound from the LEP constraint
$M_{A^0}\gsim 75\GeV$
and the \SUSY\  Higgs mass relations implies $\mHp\gsim 110 \GeV$.
We also show the three exclusion
curves for the tree-level, \QCD\  and \MSSM\  corrected 
cross-sections. The excluded region in each case is the one below the curves.
By simple inspection of Fig.\,\ref{fig:tanbtop3},
it can hardly be overemphasized that the \MSSM\  quantum effects
can be dramatic. Thus e.g.\  while for $\mHp=110\GeV$ the maximum allowed
value of $\tan\beta$ is about $50$  according to the \QCD\  contour,
it is only about $35$ according to the \MSSM\  contour. 
We have also checked that, after all, the modulation of the latter by the 
$b\rightarrow s\,\gamma$ constraint is not too significant even
when including the $30\%$ uncertainty mentioned above.
For, it turns out that although the branching ratio
for the $b\rightarrow s\,\gamma$ decay severely limits the set of possible 
values of $A_t$ for each $\tan\beta$ (Cf. Fig.\,\ref{fig:tanbtop1}), the corresponding impact
on $t\rightarrow H^+\,b$ is really minor. This is due in part to
the fact that the supersymmetric electroweak corrections are {\it not} the
dominant component 
in $t\rightarrow H^+\,b$, and also
in part to the observed 
stabilization of its contribution within the region of parameter
space allowed by $b\rightarrow s\,\gamma$. 

\figtanbtopquatre

The above picture may undergo a significant qualitative change when we
move to the $\mu>0$ scenario, as can be easily guessed from the very different
corrections obtained in it.
This can be appraised in Fig.\,\ref{fig:tanbtop4},
where we plot the excluded region in the $(\tan\beta, \mHp)$-plane again
for the same cases as before. Although not shown, 
we have also determined the
portion of the $(\tan\beta,A_t)$-plane permitted by $b\rightarrow s\,\gamma$
for $\mu>0$ (implying that $A_t<0$), and checked that also in this
case the influence on our 
top quark analysis is not dramatic.   
The point with the $\mu>0$ scenario is that the \MSSM\  curve is,
in contradistinction to the $\mu<0$ case, the less restrictive one. 
As a matter of fact it is even less restrictive than the original CDF 
curve for the inclusive $\tau$ channel! (Cf. Ref.\cite{Conway}). 
The reason being that for
$\mu>0$, the \SUSY\  corrections have the same (negative) sign as the standard
\QCD\  corrections (Fig.\,\ref{fig:tbhdeltamsbmugt0}) and, therefore, the
cross-section for the $\tau$-lepton 
signal becomes extremely depleted. In Fig.\,\ref{fig:tanbtop4} we have chosen a heavier
\SUSY\  spectrum than in the previous figures in order to keep the total
correction within the limits of perturbation theory. We see that
for squark masses of several hundred GeV and a gluino mass of 
$1\mbox{ TeV}$ the excluded area can be enforced to withdraw to a corner of 
parameter space. However, in this corner one cannot be precise any
more since further reduction would make also the Higgs sector
non-perturbative (see below).
Hence the conclusion emerging for the case $\mu>0$ is quite remarkable, to
wit: 
relatively light ($\gsim 110-120\GeV$) charged Higgs masses
within the kinematical range of $t\rightarrow H^+\,b$ could be allowed
for essentially any admissible value of $\tan\beta$ within 
perturbation theory (i.e. $\tan\beta<60-70$).
In other words, within this scenario one could not disprove the existence
of relatively light supersymmetric charged higgses by the current methods
of $\tau$-lepton analysis at the Tevatron.

It is also interesting to compare our results with the bounds obtained 
from semileptonic and semitauonic $B$-meson decays. 
In Ref.\cite{Grossman:1995yp,Grossman:1994ax} 
the excluded
region in the $(\tan\beta,\mHp)$-plane is computed for a general $2HDM$
whereas in Ref.\cite{Coarasa:1997uw} 
the corresponding \MSSM\  analysis is performed
and it is also confronted with the (uncorrected) top quark decay exclusion
region.  
However, in the presence of the corrected results, we may compare Fig.\,\ref{fig:tanbtop2} of
the present work with Fig.\,3 of Ref.\cite{Coarasa:1997uw} 
(both for $\mu<0$).
We realize that the supersymmetric results on the top quark decay greatly
improve the bound from semitauonic $B$ decays across 
the crucial region defined by $30\lsim\tan\beta\lsim 65$ and
Higgs masses ranging between $100-150\GeV$. Even though for
$\tan\beta>65$ the semitauonic $B$-meson decays are more restrictive,
it should be pointed out that this range is already ruled out on sound
theoretical grounds, namely by the breakdown of perturbation theory; for 
instance, the top quark Yukawa coupling with the CP-odd Higgs boson $A^0$
would become $g\,m_b\,\tan\beta/2\,M_W>1$. On the other hand, the 
$\mu>0$ region is not so favoured by $B$-meson decays, but it is still
compatible with experimental data at the $1\sigma$ level for
$\tan\beta\lsim 40$\,\cite{Coarasa:1997uw}. 


%% file: fcnctop.tex

\chapter{FCNC top decays into Higgs bosons in the \MSSM}
\label{cha:fcnc}

\section{Introduction}

In this chapter we perform the computation of the Flavour Changing Neutral Current
(\FCNC) decay of the top quark into a charm quark and a neutral Higgs particle in
the framework of the \MSSM, $\tch$ where $h$ is any
of the neutral Higgs particles of the \MSSM. We
compute the contributions from the \SUSY\
electroweak, Higgs, and \SUSY-\QCD\  sectors, in a sparticle mass model motivated by model
building and Renormalization Group Equations (RGE). However, we neither restrict
ourselves
to a spectrum of any \SUSY-GUT model (such as SUGRA) --which would constrain
the masses in a narrow range--, nor to a generic, phenomenological motivated,
spectrum --which would have too many parameters to play with.

There exist some computations of \FCNC\  top quark decays, both in the \SM\  and in
the \MSSM\cite{Eilam:1991zc,Hou:1991un,Agashe:1996qm,Hosch:1997gz,Li:1994mg,Couture:1995rr,Lopez:1997xv,Couture:1997ep,deDivitiis:1997sh,Yang:1994rb}.
The Standard Model branching ratio $BR(t\rightarrow c\,H)$ is 
$\sim 10^{-13}$ for Higgs boson mass around $80\GeV$, and it decreases with the
Higgs mass\cite{Mele:1998ag}.
There has
been some work concerning the decay channel into gauge bosons ($t\rightarrow
c\,V$, $V\equiv \gamma,Z,g$), see for example
Refs.\cite{Li:1994mg,Couture:1995rr,Lopez:1997xv,Couture:1997ep,deDivitiis:1997sh}
for some works on
the subject. The conclusion of these works is that the branching ratio of this
decay is at most $10^{-5}$, maybe a bit larger in the gluon channel. However,
to our knowledge, there are not so many works on the \FCNC\  decay of the top quark
into Higgs in the \MSSM\cite{Yang:1994rb}, 
and they are not so complete as in the case
of the gauge bosons. For example in~\cite{Yang:1994rb} 
it is concluded that the
branching ratio for 
the decay channel
$\tch$ in the \MSSM\  is at most of $10^{-9}$, for the \SUSY\  electroweak
contributions, and 
$10^{-5}$ for the \SUSY-\QCD\  contributions. However we think that the work
of~\cite{Yang:1994rb} 
is not complete. They do not include effects of the Higgs particles
in the loops, and they do not take into account the $\tilde q_L\,\tilde q_R\,h$
vertices, so they miss the potentially large contributions coming from the
trilinear soft-SUSY-breaking terms $A_{t,b}$~(\ref{eq:MLRdefinition}), and from the higgsino mass
parameter $\mu$. We find that a full treatment
of the \SUSY-\QCD\  contributions may greatly enhance the \FCNC\   width by some orders
of magnitude. Therefore, a more general and  rigorous computation of the
decay $\tch$ is mandatory. 

In section~\ref{sec:gen} we make a summary of the technics of the
computation. In sections~\ref{sec:ew} and~\ref{sec:qcd} we present our results
for the \SUSY\  electroweak and the \SUSY-\QCD\  contributions to the decay width $\tch$
respectively. Finally we present the conclusions.

\section{One-loop \FCNC\  decays}
\label{sec:gen}

The computation of \FCNC\  processes at one-loop, unlike the other calculations
presented in this Thesis,
does not involve renormalization of parameters or wave functions, so one is
left only with the computation of the different diagrams that contribute to the
process. The generic type one-loop Feynman diagrams contributing to the decay
under study are in Fig.~\ref{diag:fcncgenerics}. The vertex diagram $V$ follows
after a straightforward calculation. As for the diagrams $S_t$ and $S_c$ we
define a mixed self-energy, 
\begin{equation}
  \label{eq:automixta}
   \Sigma_{tc}(k) \equiv
  \slas{k}\,\Sigma_L(k^2)\,\pl+\slas{k}\,\Sigma_R(k^2)\,\pr+ 
  m_t\,\left(\Sigma_{Ls}(k^2)\,\pl+\Sigma_{Rs}(k^2)\,\pr\right)
\end{equation}
--where the $m_t$ factor multiplying the scalar part is arbitrary, 
put there only to maintain the same units between the different $\Sigma_i$.

To present the expressions of this computation we shall introduce a notation
that allows to treat the three 
possible decays in an unified way. To this end we introduce  a  vector of neutral Higgs
fields 
\begin{equation}
  \label{eq:defIndexH0}
\Phi^0=(H^0,h^0,A^0)\,\,,
\end{equation}
as a function of which  interaction Lagrangian with up-type quarks reads
\begin{equation}
  \label{eq:LneutralHiggsGenericup}
  {\cal L}_{\Phi u} =  -\frac{g\,\mup}{2\mw\sbt} \sum_{r=1,3} \Phi^0_r\,
  \bar{u}\,(K^{0u}_{r}\pl+(K^{0u}_{r})^*\pr)\,u\,\,,
\end{equation}
 the $K$ matrix being
\begin{equation}
  \label{eq:Kdefinitionup}
  K^{0u}_{r}=\pmatrix{\sa \cr \ca \cr i\cbt}\,\,.
\end{equation}

Now we are ready to give a general expression of the effects of $\Sigma_i$ to
the amplitude $t\rightarrow c\,\Phi^0_r$:
\begin{eqnarray}
  -i\, T_{S_c}^r&=& \frac{-i\,g\,\mt}{2\mw\sbt}\frac{1}{\mcs-\mts} \bar{u}_c(p) \left\{\phantom{ \pr\,(K^{0u}_{r})^*\,\left[ L\leftrightarrow R \right] }\right.\nonumber\\
&&   \pl\,K^{0t}_{r}\,\left[\mcs \Sigma_R(\mcs) +
    \mc\,\mt\,\left(\Sigma_{Rs}(\mcs)+\Sigma_L(\mcs)\right) +
    \mts\,\Sigma_{Ls}(\mcs)\right]\nonumber\\
&&+ \left. \pr\,(K^{0t}_{r})^*\,\left[ L\leftrightarrow R \right]  \right\}
u_t(k)\nonumber\\
  -i\, T_{S_t}^r&=& \frac{-i\,g\,\mc}{2\mw\sbt}\frac{\mt}{\mts-\mcs} \bar{u}_c(p)
  \left\{ \pl\,K^{0c}_{r}\,\left[\mt \left(\Sigma_L(\mts) +
        \Sigma_{Rs}(\mts)\right)\right.\right.\nonumber\\
&&  \left.\left.+
      \mc\,\left(\Sigma_R(\mts) + \Sigma_{Ls}(\mts)\right) \right]
+\pr\,(K^{0c}_{r})^*\,\left[  L\leftrightarrow R \right] \right\}u_t(k)
  \label{eq:fcncmixamp}
\end{eqnarray}

\fcncgenerics

After adding up the vertex contributions from diagram $V$
(Fig.\,\ref{diag:fcncgenerics}) to the 
expressions~(\ref{eq:fcncmixamp}) we can define an ``effective'' vertex
\begin{equation}
\label{eq:effvertex}
 -i\,T\equiv -i\,g\, \bar{u}_c(p)\, \left( F_L\,P_L+ F_R\,P_R\right)\,
  u_t(k)\,\,\,.
\end{equation}

We have taken into account all three generations of quarks and squarks, and have
performed the usual checks of
the computation, in particular that the form factors $F_L$ and $F_R$  are free
of divergences before adding up 
the three quark generations, both analytically and numerically in the
implementation of the code.

After squaring the matrix element~(\ref{eq:effvertex}), and multiplying by the
phase space 
factor, we can compute the decay width, 
\begin{eqnarray}
  \label{eq:fcncdecaywidth}
  \Gamma(\tch)&=&\frac{g^2}{32\,\pi\,\mt^3} \lambda^{1/2}(\mts,m_h^2,\mcs)  \nonumber\\
&\times&  \left[
  (-m_h^2+\mts+\mcs) (|F_L|^2+|F_R|^2)
    +2\,\mt\,\mc\,(F_L\,F_R^*+F_L^*\,F_R)
\right]
\end{eqnarray}
and define the ratio
\begin{equation}
\label{eq:defbr}
 B(\tch)\equiv\frac{\Gamma
          (t\rightarrow c\,h)}{\Gamma(t \rightarrow b\,W^+)}
\end{equation}
which will be the main object under study. This ratio is not the total branching
fraction of this decay mode, as there are many other channels that should be
added up to the denominator of (\ref{eq:defbr}) in the MSSM, such as the two and
three body 
decays of the top quark into \SUSY\  particles, and also the decay channel $t\rightarrow
H^+\,b$\,\nocite{IWQEMSSM}\cite{Coarasa:1996qa,Guasch:1997dk,Sola:1998kh}. For the mass spectrum used in the numerical
analysis (see sections~\ref{sec:ew} and~\ref{sec:qcd}) the former are phase
space closed, whereas the latter could have a large branching ratio.

\section{\SUSY-\EW\  contributions}
\label{sec:ew}

\fcncEWvertex

\fcncEWmix

For the electroweak contributions to the decay channel $\tch$ we work in the so
called Super-\ckm\  basis, that is, we take the
simplification that the squark mass matrix diagonalizes as the quark mass
matrix, so that \FCNC\  processes appear at one-loop through the charged sector (charged
Higgs and charginos) with the same mixing matrix elements as in the Standard
Model (the CKM matrix). 

We have taken into account the contributions from charginos ($\cplus_i$) and
down type squarks ($\tilde d_\alpha$, $\alpha=1,2,\ldots,6\equiv
\tilde d_1,\tilde d_2,\ldots,\tilde b_2$, the mass eigenstates down squarks),
and from charged Higgs and 
Goldstone bosons ($H^+,G^+$) and down type quarks ($d,s,b$). The various
diagrams contributing to this decay can be seen in
Figs.\,\ref{diag:fcncEWvertex} and~\ref{diag:fcncEWmix}. We have not
included the diagrams with gauge bosons ($W^+$) as the largest contributions
will come from the Yukawa couplings of the top and (at large $\tb$) bottom
quarks. However, the leading terms from longitudinal $W^+$ are included through
the inclusion of Goldstone bosons. 

\subsection{Vertex and self-energy functions}

To write down the concrete form of the various contributions to the form
factors~(\ref{eq:effvertex}) we generalize the compact notation introduced in
section~\ref{sec:gen}, thus we define a vector of charged Higgs and Goldstone particles
\begin{equation}
  \label{eq:defIndexHplus}
  \Phi^+=(H^+,G^+)\,\,,
\end{equation}
and write down the interaction Lagrangian of down-type squarks with Higgs
particles analogously to~(\ref{eq:LneutralHiggsGenericup})
\begin{eqnarray}
  \label{eq:LneutralHiggsGenericdown}
  {\cal L}_{\Phi d}&=& -\frac{g\,\md}{2\mw\cbt} \sum_{r=1,3} \Phi^0_r\,
  \bar{d}\,(K^{0d}_{r}\pl+(K^{0d}_{r})^*\pr)\,d  \nonumber\\
  &+& \frac{g}{\sqrt{2}\mw} V_{ud} \sum_{r=1,2} \left( \Phi^-_r \bar{d}(K^{+ud}_{rL}\pl+K^{+ud}_{rR}\pr)
    u+ \mbox{ h.c.}\right)\,\,,
\end{eqnarray}
where the $K$ matrices are 
\begin{equation}
  \label{eq:Kdefinitiondown}
  K^{0d}_{r}=\pmatrix{\ca \cr -\sa \cr i\sbt}\,\,,
  \,\, K^{+ud}_{rL}=\md \pmatrix{\tb \cr -1} \,\,,\,\, K^{+ud}_{rR}=\mup
  \pmatrix{\ctb \cr 1}\,\,.
\end{equation}
The interaction Lagrangian between up-type quarks,
down-type squarks and charginos can be read directly
from the expressions~(\ref{V1Apm}), multiplying
by the appropriate element of the CKM matrix. In a compact notation we write it as
$$
{\cal  L}_{u\,\sd\,\cplus}= -g\,V_{ud}\,\sd_a^*\,
\bar{\psi}^+_i\left(\Apaidu\,\pl+\Amaidu\,\pr\right)\,u +{\mbox h.c.}
$$
with $u$ (\sd) up-type quarks (down-type squarks) of any generation, the
coupling matrices being
$$
\Apaidu=R_{1 a}^{(d)*} V_{i1}^*-\lambda_d\,R_{2a}^{(d)*}V_{i2}^*
\,\,,\,\,\Amaidu=- R_{1a}^{(d)*} \lambda_u U_{i2}\,\,.
$$
We similarly define a 3-dimensional matrix containing the triple Higgs
interactions as\footnote{Note that elements $B_{ij3}$ are complex and $B_{ii3}=0$.}
\begin{equation}
  \label{eq:tripleHiggsGeneric}
  {\cal L}_{\Phi\Phi\Phi}= -g \sum_{r,s,t}B_{rst}\,\Phi^+_r\,\Phi^-_s\,\Phi^0_t\,\,,
\end{equation}
and the chargino couplings to neutral Higgs
\begin{equation}
  \label{eq:charcharHiggsgeneric}
 { \cal L}_{\cplus \cplus \Phi}=-g \sum_{r,i,j} \Phi^0_r \bar{\chi}_i^+
 (W_{ijL}^{r}\,\pl+W_{ijR}^{r}\,\pr) \cplus_j\,\,,
\end{equation}
These $B_{rst}$ and $W_{ij}^{r}$ matrices are the corresponding Feynman rules
(divided my $-ig$) of
the respective processes and can be found in~\cite{Hunter} --they can also be
found in ~\cite{Tesina} where there is also a detailed explanation of how to
obtain them\footnote{We have generated all the Feynman rules derived the scalar
  potential (Higgs particles self-couplings and squark-Higgs couplings) by
  means of a {\em Mathematica}\cite{Mathematica} code.}.

The contributions from the various diagrams in
Fig.\,\ref{diag:fcncEWvertex} can be written generically as,
\begin{eqnarray}
  \label{eq:FCNCV2Ssgeneric}
  F_L&=& N_A \left[ (\Cot-\Coo) \mt\, A^{(1)}_R \,A^{(2)}_R -
    \Cot\,\mc\, A^{(1)}_L\,A^{(2)}_L +\Cz\, m_A\, 
     \,A^{(1)}_R \,A^{(2)}_L\right] \nonumber\\
  F_R &=&F_L\,(A_L^{(*)} \leftrightarrow A_R^{(*)})
\end{eqnarray}
for diagrams\,\ref{diag:fcncEWvertex}(a) and
(d). Diagrams\,\ref{diag:fcncEWvertex}(b) and (c)
contributions can be written as 
\begin{eqnarray}
  \label{eq:FCNCV1Ssgeneric}
  F_L&=&N_D\,\left[
  C_0\,(D^{(1)}_L\,D^{(2)}_L\,D^{(3)}_R\,\mc\,\mt+D^{(1)}_L\,D^{(2)}_L\,D^{(3)}_L\,\mc\,m_{D1} \right.\nonumber\\
&&\ \ \ \ +D^{(1)}_R\,D^{(2)}_L\,D^{(3)}_R\,\mt\,m_{D2}+D^{(1)}_R\,D^{(2)}_L\,D^{(3)}_L\,m_{D1}\,m_{D2})\nonumber\\
  &&+
  C_{12}\,\mc\,(D^{(1)}_R\,D^{(2)}_R\,D^{(3)}_L\,\mc+D^{(1)}_L\,D^{(2)}_L\,D^{(3)}_R\,\mt\nonumber\\
&&\ \ \ +D^{(1)}_L\,D^{(2)}_L\,D^{(3)}_L\,m_{D1}+D^{(1)}_L\,D^{(2)}_R\,D^{(3)}_L\,m_{D2})\nonumber\\
  &&+
  (C_{11}-C_{12})\,\mt\,(D^{(1)}_L\,D^{(2)}_L\,D^{(3)}_R\,\mc+D^{(1)}_R\,D^{(2)}_R\,D^{(3)}_L\,\mt\nonumber\\
&&\left.\ \ \
  +D^{(1)}_R\,D^{(2)}_R\,D^{(3)}_R\,m_{D1}+D^{(1)}_R\,D^{(2)}_L\,D^{(3)}_R\,m_{D2})+
  \Czt\,D^{(1)}_R\,D^{(2)}_R\,D^{(3)}_L \right] \nonumber\\ 
  F_R&=& F_L(D^{(*)}_L\leftrightarrow D^{(*)}_R)\,\,.
\end{eqnarray}
Now we are ready to write down the different contributions, by giving the values
of the different matrices and masses appearing in the expressions above, thus to
obtain the vertex functions of the decay $t\rightarrow c\,\Phi^0_r$
we must apply the following rules
\begin{itemize}
  \item diagram\,\ref{diag:fcncEWvertex}(a): take~(\ref{eq:FCNCV2Ssgeneric}) and
    assign
    \begin{eqnarray*}
      &&   A^{(1)}_L=\Apbidc\,\,,\,\,A^{(1)}_R=\Ambidc\,\,,\,\,A^{(2)}_L=\Apaidt\,\,,\,\,A^{(2)}_R=\Amaidt\\
      && m_A=\mi\,\, ,\,\,N_{A}=i\,g^{2}V_{td}V_{cd}R_{ea}^{(d)}\,(R_{fb}^{(d)})^{*}\,G_{fe}^{r}\,\,, \\
      && C_{*}=C_{*}(k,-p^\prime,\mi,\msdal,\msdbe)\,\,,
    \end{eqnarray*}
    where  $G_{fe}^{r}$ 
    is the Feynman rule for $\Phi^0_r \rightarrow
    \sd_f^{\prime}\,\sd_e^{\prime *}$ divided by $-ig$, $\sd^{\prime}_{1,2}$ are
    the electroweak eigenstates 
    of down 
    type squarks\,\cite{Hunter,Tesina}\footnote{We recall that our convention for the
  $\mu$ parameter is opposite in sign to that of\,\cite{Hunter}.}.
  \item diagram\,\ref{diag:fcncEWvertex}(b): substitute
    in~(\ref{eq:FCNCV1Ssgeneric})
    \begin{eqnarray*}
      &&   D^{(1)}_L=\Apajdc\,\,,\,\,D^{(1)}_R=\Amajdc\,\,,\,\,
      D^{(2)}_L=W_{ijL}^r\,\,,\,\,D^{(2)}_R=W_{ijR}^r\\
      &&   D^{(3)}_L=\Apaidt\,\,,\,\,D^{(3)}_R=\Amaidt\\
      &&m_{D1}=\mi\,\,,\,\,m_{D2}=\mj\,\,,\,\, N_D=i\,g^2\, V_{td}\,V_{cd} \\
      &&  C_{*}=C_{*}(k,-p^\prime,\msdal,\mi,\mj)\,,
    \end{eqnarray*}
  \item diagram\,\ref{diag:fcncEWvertex}(c): the following assignations must be
    performed to~(\ref{eq:FCNCV1Ssgeneric})
    \begin{eqnarray*}
      &&   D^{(1)}_L=\kpcl\,\,,\,\,D^{(1)}_R=\kpcr\,\,,\,\,
      D^{(2)}_L=\kz\,\,,\,\,D^{(2)}_R=(\kz)^{*}\\
      &&   D^{(3)}_L=\kptr\,\,,\,\,D^{(3)}_R=\kptr\\
      && m_{D1}=m_{D2}=\mb\,\,,\,\,N_D=i\frac{g^2\,\md}{4\mw^3\,\cbt} V_{td} V_{cd}\\
      &&  C_{*}=C_{*}(k,-p',m_{\Phi^+_i},\md,\md)\,,
    \end{eqnarray*}
  \item  diagram\,\ref{diag:fcncEWvertex}(d): make the following substitutions
    to~(\ref{eq:FCNCV2Ssgeneric})
    \begin{eqnarray*}
      &&   A^{(1)}_L=\kpclj\,\,,\,\,A^{(1)}_R=\kpcrj\,\,,\,\,A^{(2)}_L=\kptl\,\,,\,\,A^{(2)}_R=\kptr\\
      &&m_A=\md\,\,,\,\, N_A=i\frac{g^2}{2\,\mws}\,B_{ijk}\,V_{td}\,V_{cd} \\
      &&C_{*}(k,-p',\md,m_{\Phi^+_i},m_{\Phi^+_j})\,\,.
    \end{eqnarray*}
\end{itemize}
As can be  noted from above expressions the form factors induced by Higgs
mediated diagrams -Fig.\,\ref{diag:fcncEWvertex}(c)
and~(d)- have the property $F_L=F_R$ for $H^0$ and $h^0$, and $F_L=-F_R$ for $A^0$. This only
form factor for each one of the different Higgs mediated diagrams can be written
in a more convenient form, but then we should write down 16 different expressions!

The one-loop mixing Feynman diagrams between the two mass-eigenstates quarks $t$
and $c$ can be seen in Fig.\,\ref{diag:fcncEWmix}, their contribution to the
mixing self-energies~(\ref{eq:automixta}) can be written as follows:
\begin{eqnarray}
\left.\Sigma_{R}(k^2)\right|_{(a)}&=& i\,g^2\,V_{td}\,V_{cd}
         \,\Amaidc\,\Amaidt\,B_1(k,\mi,\msdal)\nonumber\\
\left.\Sigma_{L}(k^2)\right|_{(a)}&=&  i\,g^2\,V_{td}\,V_{cd}
         \,\Apaidc\,\Apaidt\,B_1(k,\mi,\msdal)\nonumber\\
\mt\,\left.\Sigma_{Rs}(k^2)\right|_{(a)}&=& i\,g^2\,\mi\,V_{td}
         \,V_{cd}\,\Apaidc\,\Amaidt\,B_0(k,\mi,\msdal)\nonumber\\
\mt\,\left.\Sigma_{Ls}(k^2)\right|_{(a)}&=& i\,g^2\,\mi\,V_{td}
         \,V_{cd}\,\Amaidc\,\Apaidt\,B_0(k,\mi,\msdal)\nonumber\\
\left.\Sigma_{R}(k^2)\right|_{(b)}&=&\frac{i\,g^2\,\mc\,\mt}{2\,\mws}
         V_{td}\,V_{cd}
         \left[
           \ctbs \,\left(B_0+B_1\right)(k,\mHp,\mda)\right.\nonumber\\
   &&\left.\phantom{\ctbs}+\left(B_0+B_1\right)(k,\mw,\mda)\right] 
         \nonumber\\
\left.\Sigma_{L}(k^2)\right|_{(b)}&=&\frac{i\,g^2\,\mdas}{2\,\mws}
         V_{td}\,V_{cd}
         \left[\tbs\,\left(B_0+B_1\right)(k,\mHp,\mda)\right.\nonumber\\
   &&\left.\phantom{\tbs}+\left(B_0+B_1\right)(k,\mw,\mda)\right]\nonumber\\
\mt\,\left.\Sigma_{Rs}(k^2)\right|_{(b)}&=&\frac{i\,g^2\,\mt\,\mdas}{2\,\mws}
         V_{td}\,V_{cd}
         \left[ B_0(k,\mHp,\mda)-B_0(k,\mw,\mda)\right]\nonumber\\
\mt\,\left.\Sigma_{Ls}(k^2)\right|_{(b)}&=&\frac{i\,g^2\,\mc\,\mdas}{2\,\mws}
         V_{td}\,V_{cd}
         \left[ B_0(k,\mHp,\mda)-B_0(k,\mw,\mda)\right]\,\,.
\end{eqnarray}
The {\it compact} form of these self-energies allows to avoid the use of the cumbersome
notation we used for the vertex factors.

\subsection{Numerical analysis}
With all these expressions we are now ready to look at the numerical results. We
plug in all these contributions in~(\ref{eq:effvertex}) and~(\ref{eq:fcncmixamp})
and evaluate numerically the expression~(\ref{eq:defbr}).
The input parameters chosen to illustrate the results in 
Figs.\,\ref{fig:formew}-\ref{fig:resultsew} are:
\begin{eqnarray}
  \label{eq:inputew}
  &&\tb=35 \,\,,\,\,
  \mu=-200 \GeV  \,\,,\,\,
  M=150 \GeV \,\,,\,\,
  M_{A^0}=80 \GeV \,\,,\nonumber\\
  &&m_{\tilde t_1}=150 \GeV \,\,,\,\,
  m_{\tilde b_1}=m_{\tilde q}=200 \GeV \,\,,\nonumber\\
  &&A_t=A_q=300 \GeV \,\,,\,\,
  A_b=-300 \GeV
\end{eqnarray}
where $m_{\stopp_1},m_{\sbottom_1}$ are the lightest $\stopp$ and $\sbottom$
mass, and all the 
masses are above present experimental
bounds. 
This somewhat light value of the pseudoscalar Higgs mass is not essential in the
results, as can bee seen in Fig.\,\ref{fig:resultsew}~(d). We have chosen a
\SUSY\  
mass spectrum around $200\GeV$, which 
is not too light, so the results will not be artificially optimized. We have
also checked all 
through the numerical analysis that other bounds on experimental parameters
(such as $\delta\rho$) are fulfilled.

\figfcncdos

In Fig.\,\ref{fig:formew} we have plotted the different form factors
of~(\ref{eq:effvertex}) as a function of $\tb$ for the channel with the lightest
scalar Higgs ($h^0$). We can see that the contributions from the Higgs sector
and the contributions from the chargino sector are of the same order. It
turns out that they can be either of the same sign, or of opposite sign. The
chosen negative value for $A_b$ is to make the two contributions of the same
sign. It is also clear that in both cases $F_R \gg F_L$. This can be easily
understood by looking at the interaction Lagrangians involving higgsino-sbottom-charm and
Higgs-bottom-charm:
\begin{eqnarray}
  \label{eq:mainew}
  {\cal L}_{\tilde h\,\tilde b\,c}&=&- g\,V_{cb}\,\bar{c} \left({R_{1a}^{(b)}\,\lambda_c}\, P_L+
    {R_{2a}^{(b)}\,\lambda_b}\, P_R \right) \chi^+ \tilde{b}_a+\mbox{ h.c.} \nonumber\\
  {\cal L}_{H\,b\,c}&=&\frac{g}{\sqrt{2}M_W}
  V_{cb}\,\bar{c}\left({m_c\,\cot\beta}\,P_L+{m_b\,\tan\beta}\,
    P_R\right)\,b\,H^+ + \mbox{ h.c.}\,\,\,,
\end{eqnarray}
we can see that in both of them the contribution to the right-handed form factor
will be enhanced by the Yukawa coupling of the bottom quark, compared with the
charm Yukawa coupling that will contribute to the left-handed form factor. On the other
hand we have checked that the inclusion of the first two generations of quarks
and squarks only has an effect of a few percent on the total result.

\figfcnctres

In Fig.\,\ref{fig:resultsew} we can see the evolution of the
ratio~(\ref{eq:defbr})
with various parameters of the MSSM, by taking into account only the electroweak
contributions. The growing of the width with  $\tb$ 
(Fig.\,\ref{fig:resultsew}~(a)) shows that the bottom Yukawa coupling plays a
central role in these contributions. The evolution with the trilinear coupling
$A_b$ and the higgsino mass parameter $\mu$ --the two parameters that appear in
the trilinear coupling $\sbottom_L\,\sbottom_R\,h$-- displayed in Figs.\,\ref{fig:resultsew}~(b)
and~(c) shows that these parameters can 
enhance the width some orders of magnitude. We have artificially let $A_b$ grow up to
large scales (that are not allowed if one wants that squarks do not develop
vacuum expectation values) in order to emphasize the dependence on $A_b$. The various spikes
in these figures reflect the points where the form factors change sign, whereas
the shaded region in Fig.\,\ref{fig:resultsew}~(c) reflects the exclusion region of
$\mu$ by present LEP bounds on the chargino mass. 

In all these figures the ratio~(\ref{eq:defbr}) is smaller for the heaviest
scalar Higgs ($H^0$) because with the parameters~(\ref{eq:inputew}) the CP-even
Higgs mixing angle $\alpha$ is near $-\pi/2$, so making the couplings of $H^0$ with down
quarks and squarks much weaker, but in fig.~\ref{fig:resultsew}~(d) it can be seen
that when the pseudoscalar Higgs mass grows (and this shifts $\alpha$ far away
from $-\pi/2$) the two scalar Higgs bosons change roles.

We conclude that the typical value of the ratio~(\ref{eq:defbr}), at large
$\tan\beta\lsim 50$ and for a \SUSY\  
spectrum around $200\GeV$, is
\begin{equation}
  \label{eq:conew}
  B^{\rm SUSY-EW}(\tch) \simeq {\cal O}(10^{-8})\,\,\,.
\end{equation}
In favorable regions of the parameter space it can grow up to $10^{-7}$. The
maximum value is found to be in the ballpark of several $10^{-6}$, for sufficiently large $A_b$.
This is an improvement of the previous result\cite{Yang:1994rb}, specially in
the $A^0$ 
channel, by two orders of magnitude. 

\section{SUSY-QCD contributions}
\label{sec:qcd}

The gluino-mediated supersymmetric strong interactions in the \MSSM\   can also
produce \FCNC\  
processes. This occurs when the squark 
mass matrix does not diagonalize with the same matrix as the one for the
quarks. We introduce then intergenerational mass terms for the squarks, but in
order to prevent the number of parameters from being too large, we have allowed (symmetric)
mixing mass terms only for the left-handed squarks. This simplification is often used in the
MSSM, and is justified by RGE analysis\cite{Duncan:1983iq}. 

The mixing terms are introduced through the parameters $\delta_{ij}$ defined as
\begin{equation}
  \label{eq:defdelta}
(M^2_{LL})_{ij}=m_{ij}^2\equiv\delta_{ij}\,m_i\,m_j\,\,,
\end{equation}
where $m_i$ is the mass of the left-handed $i$ squark, and $m^2_{ij}$ is the mixing
mass matrix element between the generations $i$ and $j$. Thus we must
diagonalize two $6\times6$ mass matrices in order to obtain the mass-eigenstates
squark fields. Following a notation similar to the standard one  we introduce the
mixing matrices as follows (Cf. Sec.\,\ref{sec:sfmas})
\begin{eqnarray}
  \label{eq:definicioR6gen}
  \sq^{\prime}_{\alpha}&=&\sum_\beta R^{(q)}_{\alpha\beta}\sq_{\beta}^{} \nonumber\\
  R^{(q)\dagger}{\cal M}^{2}_{\sq} R&=&{\cal M}_{\sq D}^{2}={\rm diag}\{m^2_{\sq_1},\ldots,m^2_{\sq_6}\}\,\,,\,\,q\equiv u,\,d  \,\,,
\end{eqnarray}
where ${\cal M}^{2}_{(\su,\sd)}$ is the $6\times6$ square mass matrix for up-type (or
down-type) squarks
in the \EW\  basis, with indices
$\alpha=1,2,3,\ldots,6\equiv\su_L,\su_R,\sch_L,\ldots,\stopp_R$ for up-type
squarks, and an equivalent choice for down-type squarks. In this study we are
only interested in the up-type quarks-squarks system, so we will drop out the
$(q)$ super-index in the forthcoming expressions. The rotation matrix $R$ introduces gluino mediated tree-level
\FCNC\  between quarks and squarks, the corresponding interaction Lagrangian can
be deduced using the very same formalism of the ``ordinary'' \SUSY-\QCD~(\ref{eq:Lqsqglui})
interactions, but using the more general rotation matrix~(\ref{eq:definicioR6gen}),
\begin{eqnarray}
  \label{eq:SUSYQCDFCNClagrangian}
  {\cal L}_{\rm SUSY-QCD}&=& -\frac{g_s}{\sqrt{2}}\, \bar{\psi}_c^{\sg} \left[
    R_{5\alpha}^*\,\pl-R_{6\alpha}^*\,\pr \right] \sq^*_{\alpha,i}\, \lambda_{ij}^c\, t_j\nonumber\\
  &-& \frac{g_s}{\sqrt{2}}\, \bar{\psi}_c^{\sg} \left[
    R_{3\alpha}^*\,\pl-R_{4\alpha}^*\,\pr \right] \sq^*_{\alpha,i}\, \lambda_{ij}^c\, c_j\nonumber\\
  &-& \frac{g_s}{\sqrt{2}}\, \bar{\psi}_c^{\sg} \left[
    R_{1\alpha}^*\,\pl-R_{2\alpha}^*\,\pr \right] \sq^*_{\alpha,i}\, \lambda_{ij}^c\, u_j\,\,.
\end{eqnarray}

\subsection{Vertex and self-energy functions}
 
Using the Lagrangian~(\ref{eq:SUSYQCDFCNClagrangian}) one can find the
\SUSY-\QCD\   one-loop contributions to 
the process under study, which Feynman diagrams are depicted in
Figs.\,\ref{diag:fcncQCDvertex} and~\ref{diag:fcncQCDmix}. The vertex diagram
contributions to the form factors~(\ref{eq:effvertex}) in Fig.\,\ref{diag:fcncQCDvertex} can be written as
\fcncSQCDvertex
\fcncSQCDmix
\begin{eqnarray}
  \label{eq:fcncQCDvertex}
  F_L&=&N\,\left[\mt\,R_{4\beta}\,R_{6\alpha}^*\,(\Coo-\Cot)\right.\nonumber\\
  &&\left.+\mc\,R_{3\beta}\,R_{5\alpha}^*\,\Cot+\mg\,R_{4\beta}\,R_{5\alpha}^*\Cz\right]\nonumber\\
  F_R&=&F_L\left(3\leftrightarrow4\,\,,\,\,5\leftrightarrow 6 \right)\nonumber\\
  N&=&i\,8\,\pi\,\alpha_s\,C_F\, R_{i\beta}^*\,G_{ij}^{r}\,R_{j\alpha}\nonumber\\
  C_{*}&=&C_{*}(-k,p',\mg,\msual,\msube)\,\,,
\end{eqnarray}
where $G_{ij}^r$ is the Feynman rule of  $\Phi^0_r \rightarrow \su_i^{\prime}
\su_j^{\prime *}$
divided by $-ig$, with $\su^{\prime}_{1,2}$ the electroweak eigenstates of up-type
squarks\cite{Hunter,Tesina}. 

The one-loop mixing self-energy in Fig.\,\ref{diag:fcncQCDmix} takes the
following form
\begin{eqnarray}
  \label{eq:fcncQCDmixing}
  \Sigma_L(k^2)&=&-i\,2\,\pi\,\alpha_s\,C_F\,R_{3\alpha}\,R_{5\alpha}^*\, B_1(-k,\mg,\msual)\nonumber\\
  \Sigma_R(k^2)&=&-i\,2\,\pi\,\alpha_s\,C_F\,R_{4\alpha}\,R_{6\alpha}^*\, B_1(-k,\mg,\msual) \nonumber\\
  \mt\,\Sigma_{Ls}(k^2)&=&-i\,2\,\pi\,\alpha_s\,C_F\,\mg\,R_{4\alpha}\,R_{5\alpha}^*\,B_0(-k,\mg,\msual) \nonumber\\
  \mt\,\Sigma_{Rs}(k^2)&=&-i\,2\,\pi\,\alpha_s\,C_F\,\mg\,R_{3\alpha}\,R_{6\alpha}^*\,B_0(-k,\mg,\msual)
\end{eqnarray}
where $C_F=(N_c^2-1)/2N_c=4/3$ is a well known colour factor. Now we can
introduce these expressions in~(\ref{eq:fcncmixamp}) and~(\ref{eq:effvertex}) to
obtain the relevant branching ratio under study~(\ref{eq:defbr}).

\subsection{Numerical Analysis}

For the numerical analysis we must provide as input parameters, apart from that
of the Higgs sector, the various squark masses and mixings, i.e. the
$\delta$ parameters of~(\ref{eq:defdelta}).
These $\delta$ parameters are
constrained by low energy data on
FCNC\cite{Gabbiani:1996hi,Misiak:1997ei}. 
The bounds have been
computed using some approximations, so they must be taken as order of magnitude
limits. We use the following bounds\cite{Gabbiani:1996hi,Misiak:1997ei}
\begin{eqnarray}
  \label{eq:limdelta}
  |\delta_{12}|&<&.1 \,\sqrt{m_{\tilde u}\,m_{\tilde c}}/{500 \GeV} \nonumber\\
  |\delta_{13}|&<&.098\,\sqrt{m_{\tilde u}\,m_{\tilde t}}/{500 \GeV}\nonumber\\
  |\delta_{23}|&<&8.2 \,m_{\tilde c}\,m_{\tilde t}/{(500 \GeV)^2}\,\,\,.
\end{eqnarray}
For the various parameters that are common to the \EW\   analysis we use
the same input parameters~(\ref{eq:inputew}) 
to which we must add the specific parameters of the \SUSY-\QCD\   sector,
namely
\begin{eqnarray}
  \label{eq:inputqcd}
  m_{\tilde g}&=&150 \GeV\nonumber\\
  \delta&=&\left(\begin{array}{ccc}
      0 &0.03 &0.03\\
      &0&0.6\\
      &&0
    \end{array}\right)\,\,\,.
\end{eqnarray}

A comment is in order for the present set of inputs: we have introduced
in~(\ref{eq:inputew}) the lightest stop mass as an input, and this stop is mostly a
$\stopp_R$. However, in this new parametrization we introduce this mass as the lightest
$\tilde u_\alpha$ mass, which will be mostly a $\stopp_R$. 

Again the largest contribution comes from the right-handed form factor
of~(\ref{eq:effvertex}), but this is only because we have chosen not to
introduce mixing between right-handed squarks. 

\figfcncquatre

We have plotted the evolution of the ratio~(\ref{eq:defbr}) with some parameters
of the MSSM in Fig.\,\ref{fig:resultsqcd}. As can be easily guessed, the most
important parameter for these contributions is the mixing mass parameter between
the 2nd and 3rd generation of left-handed squarks, the
less restricted one of the three (eq.~(\ref{eq:limdelta})). In
Fig.\,\ref{fig:resultsqcd}~(a) it is shown that changing $\delta_{23}$ by 3
orders of magnitude, the ratio~(\ref{eq:defbr}) can increase by 7 orders of
magnitude! We can see in Fig.\,\ref{fig:resultsqcd}~(b) that the $\mu$ parameter
also plays an important role, like in the electroweak contributions
(Fig.\,\ref{fig:resultsew}~(c)), and for the same reasons, bringing the
ratio~(\ref{eq:defbr}) up to values of $10^{-4}$. Notice that the central region
of $|\mu|\lsim 90\GeV$ is excluded by present LEP bounds on the chargino
mass. 

The evolution with the
gluino mass (Fig.\,\ref{fig:resultsqcd}~(c)) is asymptotically quite stable, showing a
slow decoupling. Finally in
Fig.\,\ref{fig:resultsqcd}~(d) we have plotted the evolution with the pseudoscalar Higgs
mass, it is also quite stable, until near the kinematic limit for $A^0$ and
$H^0$.

We conclude that the typical value of the SUSY-QCD contributions
to~(\ref{eq:defbr}), with a SUSY spectrum around $200\GeV$, is
\begin{equation}
  \label{eq:conqcd}
    B^{\rm SUSY-QCD}(\tch) \simeq {\cal O}\,(10^{-5})\,\,\,,
\end{equation}
but in favourable regions of the parameter space (i.e.\  large $\mu$, or
relatively light gluino) it can easily reach values of $10^{-4}$. The upper
bound is at several $10^{-4}$. This is 1-2
orders of magnitude larger than the previous estimate\cite{Yang:1994rb}.

\section{Conclusions}
\label{sec:con}

We have computed the \SUSY-\EW, Higgs, and SUSY-QCD contributions to the
\FCNC\  top quark decay $\tch$ ($h=h^0,H^0,A^0$) in the MSSM, using a mass spectrum
motivated, but not fully restricted, by model building and Renormalization Group
Equations.

We have found that with a SUSY mass spectrum around $200\GeV$, which is well above
present bounds, 
the different contributions to this decay
are typically of the order 
\begin{eqnarray}
  \label{eq:confinal}
  B^{\rm SUSY-EW}(\tch) &\simeq& 10^{-8}\,\,,\nonumber\\
  B^{\rm SUSY-QCD}(\tch)&\simeq& 10^{-5} - 10^{-4}\,\,\,.
\end{eqnarray}

The difference of at least two orders of magnitude between the two contributions
makes unnecessary to compute the 
interference between the two contributions, but if the limits on $\delta_{23}$
(eq.~(\ref{eq:limdelta})) improve, it should be necessary to make the full
computation.

The results~(\ref{eq:confinal}) are an improvement of the previous
estimate\,\cite{Yang:1994rb}, specially in the $A^0$ channel, thanks to the inclusion of
the $\tilde q_L\,\tilde q_R\,h$ vertex. 

It would probably be difficult that this decay can be measured either at the Tevatron,
or at the  NLC, but there exists a possibility for  LHC. As an example
to assess the
discovery reach of these 
accelerators the \FCNC\  top quark decays into a vector boson are\cite{Frey:1997sg}
\begin{eqnarray}
  \label{eq:limtop}
  {\rm \bf LHC:}&B(t\rightarrow c\,V)& > 5\times 10^{-5} \,\,,\nonumber\\
  {\rm \bf NLC:}&B(t\rightarrow c\,V)& > 10^{-3}-10^{-4} \,\,\,,
\end{eqnarray}
where the lack of sensitivity of the NLC is due to the lower
luminosity\footnote{\textbf{Note added}: This estimate is for a $50\, {\rm fb}^{-1}$
  integrated luminosity\,\cite{Frey:1997sg}. Present studies for \textbf{TESLA} future Linear
  Collider expect to reach $500\, {\rm fb}^{-1}$\cite{LC98}, then it would be possible to
  measure this ratio at the LC\cite{tchpaper}\label{pagenoteLC}.}. 
So,
if the discovery reach for \FCNC\  Higgs processes are not very different from that
of the gauge bosons, there is a possibility to measure this decay channel at the
LHC even if \SUSY\  particles are not seen at the LEPII.


%% file: sbdecay.tex
\newcommand{\Apbao}{\ensuremath{A_{+a1}^{(b)}}}
\newcommand{\Ambao}{\ensuremath{A_{-a1}^{(b)}}}
\newcommand{\Apbbo}{\ensuremath{A_{+b1}^{(b)}}}
\newcommand{\Ambbo}{\ensuremath{A_{-b1}^{(b)}}}

\newcommand{\RRR}{\ensuremath{A^{(8)}}}
\newcommand{\RRL}{\ensuremath{A^{(7)}}}
\newcommand{\RLR}{\ensuremath{A^{(6)}}}
\newcommand{\RLL}{\ensuremath{A^{(5)}}}
\newcommand{\LRR}{\ensuremath{A^{(4)}}}
\newcommand{\LRL}{\ensuremath{A^{(3)}}}
\newcommand{\LLR}{\ensuremath{A^{(2)}}}
\newcommand{\LLL}{\ensuremath{A^{(1)}}}

\chapter{One-loop corrections to scalar quark decays}
\label{cap:sbdecay}

\section{Introduction}

Sparticles not much heavier than a few hundred $\GeV$
could be produced in significant numbers already at the Tevatron. For instance, 
selectron production was advocated in Ref.\cite{Ambrosanio:1996zr} 
to explain a
purported non-SM event in the Collider Detector at Fermilab (CDF). 
Subsequently, in Refs.\cite{Kane:1996ny,Djouadi:1996pi} 
it was argued that half of the 
top quarks at the Tevatron might come from gluino decays into
top and stop, $\tilde{g}\rightarrow t\,\bar{\tilde{t}}_1$. Similarly, as discussed in
the introduction, we may
envision the possibility that sbottom squarks are pair produced by the
usual Drell-Yan mechanism and then decay into top quark and charginos:
$\tilde{b}_a\rightarrow t\,\cmin_i$. Indeed, this would be the leading
two-body decay if gluinos are heavy enough that the strong decay mode
$\tilde{b}_a\rightarrow \tilde{g}\,b$ is kinematically
blocked up\footnote{Squark decays have been discussed at the tree-level
in several places of the literature. See
e.g. Refs.\cite{Baer:1990sc,Bartl:1991ie,Bartl:1994bu} 
for some relatively recent references on the subject.}.
The observed
cross-section would then be the one of eq.\,(\ref{eq:productionMSSM}).
We shall assume thorough present study that gluinos are much heavier than squarks,
so that their contribution to this cross-section
through $q\,\bar{q}\rightarrow 
\tilde{g}\,\tilde{g}$ followed by
 $\tilde{g}\rightarrow t\,\bar{\tilde{t}}_1$ is negligible.
We could have non-\SM\  top quark decay modes, such as e.g.
$t\rightarrow \tilde{t}_a\,\neut_{\alpha}$\,\cite{Kane:1996ny,Djouadi:1996pi}
and 
$t\rightarrow H^+\,b$\,\cite{Guasch:1998jc,Coarasa:1996qa}, 
that could
serve, pictorially, as a ``sinkhole''  to compensate (at least in part) for 
the unseen source of extra top quarks produced at the Tevatron from sbottom 
pair production (Cf. eq.(\ref{eq:productionMSSM})).
As stated in section~\ref{sec:tbhtree} one cannot exclude that then non-\SM\
branching ratio
$BR (t\rightarrow \mbox{``new''})$,
could be as big as the SM one, i.e. $\sim 50\%$.

If $\tan\beta$ is large and there exists a relatively light
chargino with a non-negligible higgsino component, the alternative mechanism
suggested in eq.(\ref{eq:productionMSSM}) could be a rather efficient
non-\SM\  source of top quarks that could compensate for the depletion in
the \SM\  branching ratio. 

While the squark production cross-section has already  received some
attention in the literature at the level of NLO radiative corrections
\,\cite{Beenakker:1995fp,Beenakker:1995an,Beenakker:1997ch}, 
an accurate treatment of the
decay mechanisms is also very important to provide a solid basis
for experimental analysis of the top quark production in the \MSSM\@. 
Thus in this chapter we consider the computation of the 
\QCD\  and leading supersymmetric electroweak (\SUSY-\EW)
quantum effects on $\tilde{b}_a\rightarrow t\,\cmin_i$, namely the
ones induced by potentially large Yukawa-couplings from the top and bottom
quarks~(\ref{eq:Yukawasgeneric}).

\section{Vertex renormalization}
\label{sec:sbdecayrenorm}
The one-loop Lagrangian of the $\sbottom\,t\,\cmin$ interaction follows after
substituting the one-loop counterterms, that of \sbottom, $t$, and \cmin\  from
chapter~\ref{cap:Renorm} into the bare Lagrangian~(\ref{eq:Lqsqcn}),
\begin{equation}
  {\cal L}^0_{\cmin\,t\,\sbottom}=-g\, \sbottom^*_a \bar{\chi}^+_i  \left[
    \left(\Apbai+\delta C^{ai}_{+}\right)\pl + \epsilon_i\left(\Ambai
+\delta C^{ai}_{-}\right)\pr\right] t+{\rm h.c.}\,,
\end{equation}
with
\begin{eqnarray}
\delta C^{ai}_{+}&=&\Apbai\left(
\frac{\delta g}{g}+\frac{1}{2}\delta Z^{a} +\frac{1}{2}
      \delta Z_R^i +\frac{1}{2}\delta Z_L^t\right)
+\delta \Apbai+\delta Z^{ba} \Apbai\,,\nonumber\\
\delta C^{ai}_{-}&=&\Ambai\left(
\frac{\delta g}{g}+\frac{1}{2}\delta Z^{a} +\frac{1}{2}
      \delta Z_L^i +\frac{1}{2}\delta Z_R^t\right)
+\delta \Ambai+\delta Z^{ba} \Ambai\,.
\label{eq:sbtchardcounter}
\end{eqnarray}
where we have introduced the shortcuts
\begin{eqnarray}
\label{eq:sbtchardeltaAs}
\delta \Apbai&=&\delta R_{1a}^{(b)*}\, V_{i1}^*-(\delta R_{2a}^{(b)*}\,\lambda_b +R_{2a}^{(b)*}\, \delta \lambda_b)\,V_{i2}^*\nonumber\\
\delta \Ambai&=&-(\delta R_{1a}^{(b)*}\,\lambda_t +R_{1a}^{(b)*}\, \delta \lambda_t)\,U_{i2}\,.
\end{eqnarray}

We have not introduced the mixing self-energies between the two charginos 
($\delta Z^{ij}$) as they do not contribute. In the case of the \QCD\
corrections the chargino gets no correction, whereas the Yukawa coupling
approximation implies no mixing between charginos (see
section~\ref{sec:sbtcharYukawa}). For the very same reasons we do not show the
shift that the chargino mixing matrices $U$ and $V$ would develop at one-loop.

The full structure of the four on-shell renormalized decay amplitudes 
for $\tilde{b}_a\rightarrow t\,\cmin_i\, (a=1,2; i=1,2)$
follows from the previous Lagrangian after including the contributions 
from the (LH and RH)
one-loop vertex form factors $F_{L,R}^{ai}$:
\begin{equation}
i\,T(\sbottom_a\rightarrow t\, \cmin_i )    = i\,g\,\bar
  u_t\,\left[\epsilon_i\left(\Ambai+\Lambda_{L}^{ai}\right)\pl
+\left(\Apbai+\Lambda_R^{ai}\right)\pr\right]
  \,v_i
\end{equation}
where
\begin{equation}
\label{eq:sbdecayLambdes}
\Lambda_{L}^{ai}=\delta C_{-}^{ai}+\epsilon_i\,F_{L}^{ai}\,,\ \ \ \ \
\Lambda_{R}^{ai}=\delta C_{+}^{ai}+F_{R}^{ai}\,.
\end{equation}

\diagsbtchartree

Let now $\Gamma_0^{ai}$ be the
tree-level partial width of the decay $\tilde{b}_a\rightarrow t\,\cmin_i$, the
only Feynman diagram contributing to this process in depicted in
figure~\ref{diag:sbtchartree}, and from the Lagrangian~(\ref{eq:Lqsqcn}) we can
obtain: 
\begin{eqnarray}
\Gamma_0^{ai}&=&\frac{g^2}{16\,\pi\,m_{\tilde{b}_a}^3}\,\lambda^{1/2}(a,i,t)\,
\left\{\left[(\Apbai)^2+(\Ambai)^2\right]\,(m_{\tilde{b}_a}^2-\mis-\mts)\right.
\nonumber\\
&&\left.\phantom{\left[(\Apbai)^2+(\Ambai)^2\right]}-4\,\Apbai\,\Ambai\,\mt\,\epsilon_i\,\mi\right\}\,,
\label{eq:sbtchartree}
\end{eqnarray} 
with $\lambda (a, i, t)\equiv\lambda(m_{\tilde{b}_a}^2, \mis, \mts)$ the usual
K{\"a}llen function for the given arguments.
The quantum correction to $\Gamma_0^{ai}$ can be
described in terms of the quantities
$\delta^{ai}=(\Gamma^{ai} - \Gamma_0^{ai})/\Gamma_0^{ai}$, where $\Gamma^{ai}$
is the corresponding one-loop corrected width.
From the previous formulae, $\delta^{ai}$ can be worked out
as follows:
\begin{eqnarray}
  \delta^{ai}=\frac{2 \left(m_{\tilde{b}_a}^2-\mis-\mts\right)
\left( \Ambai\,
\Lambda_L^{ai}+\Apbai\,\Lambda_R^{ai} \right)-4\,\mt\,\epsilon_i\,\mi\,(
    \Ambai\Lambda_R^{ai}+\Apbai\,\Lambda_L^{ai})}
  {\left[(\Apbai)^2+(\Ambai)^2\right]
\left(m_{\tilde{b}_a}^2-\mis-\mts\right)
-4\,\Apbai\,\Ambai\,\mt\,\epsilon_i\,\mi}\,.
\label{eq:sbtcharcorrection}
\end{eqnarray}

As mentioned in chapter~\ref{cap:Renorm} we use an on-shell renormalization
procedure, so the input parameters for the sbottom sector will be the masses as
well as the mixing angle
\begin{equation}
(m_{\tilde{b}_1}, m_{\tilde{b}_2}, \theta_{\tilde{b}})\,,
\label{eq:inputb}
\end{equation}
whereas for the stop sector we just have in addition 
\begin{equation}
(m_{\tilde{t}_1}, \theta_{\tilde{t}})\,\,,
\label{eq:inputt}
\end{equation}
since by $SU(2)_L$ gauge invariance 
the value of the other
stop mass $m_{\tilde{t}_2}$ is already determined. Of course, $\tan\beta$ and
the \SUSY\   Higgs mixing 
parameter $\mu$ are also additional independent inputs for our calculation.
Similarly, the sbottom and stop trilinear terms $A_b$ and $A_t$ are fixed by
the previous parameters as follows:
\begin{equation}
A_{b}=\mu\,\tan\beta+
{m_{\tilde{b}_2}^2-m_{\tilde{b}_1}^2\over 2\,m_b}\,\sin{2\,\theta_{\tilde{b}}}\,;
\ \ \ \
A_{t}=\mu\,\cot\beta+
{m_{\tilde{t}_2}^2-m_{\tilde{t}_1}^2\over 2\,\mt}\,\sin{2\,\theta_{\tilde{t}}}\,,
\label{eq:Abt}
\end{equation}
of course the $\stopp_2$ mass, as well as the $A$ parameters form
eq.~(\ref{eq:Abt}) will receive radiative corrections, but these would be second
order corrections to our process. We must be careful with the value of these $A$
parameters, as explained in section~\ref{sec:sfmas} they are bounded by the
condition of colour-breaking vacua~(\ref{eq:necessary}), as well as the
condition of perturbativity of the Higgs-squark-squark couplings. These bounds
will translate into a 
forbidden region in the parameter space defined by~(\ref{eq:inputb})
and~(\ref{eq:inputt}). As the conditions mentioned above are fairly qualitative
we will present our results in all of the parameter space, but at the same time
we will single out the regions where the conditions~(\ref{eq:necessary})
are fulfilled.

\section{Tree-level results}
\label{sec:sbtchartree}
In this section we make a simple tree-level analysis of the \MSSM\  parameter
space, trying to single out the regions where the process under study should be
interesting. In order to achieve this we focus on the sbottom decay
itself, but also on the sbottom production mechanism and on the top quark decay.

As we are interested in the effects that this decay could have at
the Tevatron we should use  relatively light sbottom masses (a
few hundred GeV). On the other hand,  as this processes would imply a growing of
the top quark production cross-section~(\ref{eq:productionMSSM}), 
it is necessary a mechanism that would keep this cross-section at its
measured value, so the top quark should have available 
non-standard decay channels, e.g.\  $t\rightarrow H^+\,b$, thus the charged Higgs
mass should comply with
\begin{equation}
  \label{eq:mhpltmt}
  \mHp < \mt-\mb\,\,,
\end{equation}
another possibility could be the supersymmetric channel
$t\rightarrow\stopp_1\,\neut_1$, however this latter channel cannot be under
control, for example, in the Yukawa approximation performed in
section~\ref{sec:sbtcharYukawa} it is necessarily blocked up.

It is clear that the radiative corrections to the process 
$\sbottom_a\rightarrow t\,\cmin_i$ will only be interesting in the region where it
also has a large tree-level branching ratio. Apart from the already stated
gluino decay channel there are also other channels ($\sbottom_a\rightarrow
b\,\neut_\alpha$, $\sbottom_2\rightarrow \sbottom_1 h^0$,~\ldots) that will
contribute to this decay width. 
To have an appreciable branching ratio $\sbottom_a \rightarrow t\,\cmin_i$ we
start out supposing that the gluino is much heavier than the squarks
\begin{equation}
  \label{eq:mggtmsb}
  \mg > \msba\,\,\, (a=1,\,2)\,\,,
\end{equation}
neutralino masses, on the other hand, are related to chargino ones, so no
additional conditions can be put on this side. Let us define the branching ratio
in which we are interested:
\begin{eqnarray}
BR_0(\tilde{b}_a\rightarrow t\,\cmin_1)&=&
{\Gamma_0(\tilde{b}_a\rightarrow
  \cmin_1\,t)\over\Gamma_0^T(\sbottom_a)}\,\,,\nonumber\\
\Gamma_0^T(\sbottom_a)&=&
\sum_\alpha \Gamma_0(\sbottom_a\rightarrow b\,\neut_\alpha) +
\Gamma_0(\tilde{b}_a\rightarrow t\,\cmin_1)+
\Gamma_0(\tilde{b}_a\rightarrow \stopp_1\,H^-)\nonumber\\
&&+\sum_i\,\Gamma_0(\tilde{b}_a\rightarrow \sbottom_b\,\Phi^0_i)\,,
\label{eq:sbdecayBR}
\end{eqnarray}
(where $\Phi_i^0=h^0,H^0,A^0$). To maximize this branching ratio we should work
in an scenario  
where the lightest chargino is higgsino-dominated, and \tb\  should have a
low-moderate value, if \tb\  is large ($\gsim 40$) then $\Gamma_0^T$ is dominated
by the neutral higgsino contribution (first summand of expr.~(\ref{eq:sbdecayBR})). 

In figure~\ref{fig:sbdecaybr} we have plot the
value of the branching ratio~(\ref{eq:sbdecayBR}) as a function of \tb,
$m_{\sbottom_1}$ and $\theta_{\sbottom}$ for given values of the other
parameters. From the figure it is 
clear that low \tb\  enhances this branching ratio. From now on we will
concentrate in the region of $\tb \simeq 20$, with this typical value the
branching ratio still has an appreciable branching ratio, whereas the electroweak
corrections can be enhanced by means of the bottom Yukawa
coupling~(\ref{eq:Yukawasgeneric}). In Fig.\,\ref{fig:sbdecaybr}(b) we can see the
opening of the Higgs channels, namely $\sbottom_2\rightarrow\sbottom_1\,\Phi_1^0$ (at
the left end of the figure) and $\sbottom_1\rightarrow\stopp_1\,H^-$ (at
its right end), it is clear
that when this channels are open they tend to take the branching
ratio~(\ref{eq:sbdecayBR}) to undetectable values. 
This large value of the Higgs decay
width is due to the fact that in this regions the $A$ parameters~(\ref{eq:Abt})
acquire large values. Of course one could fix the input
parameters~(\ref{eq:inputb}) in such a way that $A_{\{t,b\}}$ are small in one of
this regions (say at $m_{\sbottom_1}$ light), but at the price of making them
large at its central value and even larger at the other end. This effect is also
seen in Fig.\,\ref{fig:sbdecaybr}(c), as the $A$ parameters are related to the
angle trough~(\ref{eq:Abt}). Another possibility would be to push the Higgs
masses to a high value, which would be in contradiction with the
requirement~(\ref{eq:mhpltmt}). Note that the allowed range of
$\theta_{\sbottom}$ is rather narrow, so that the physical sbottom
masses basically coincide with
the LH and RH electroweak eigenstates.

\figsbdecaybr

\section{\QCD\  corrections}
The first step to the computation of the quantum effects on
$\sbottom_a\rightarrow t\,\cmin_i$ is to compute the \QCD\  corrections, as they
are expected to be larger than the \EW\  ones.

The evaluation of the one-loop \QCD\  corrections to the decay
$\sbottom_a\rightarrow t\,\cmin_i$ 
comprise the computation of the gluon and the gluino mediated diagrams. 
These corrections were
computed in~\cite{Djouadi:1997wt,Kraml:1996kz}, however we use slightly
different renormalization conditions, and we 
present our numerical results in a completely different way.
We have checked analytically our results with that
of~\cite{Djouadi:1997wt,Kraml:1996kz}, and also numerically in the case
of~\cite{Djouadi:1997wt}.

In this process it is not possible to separate between the gluon-mediated and
the gluino-mediated contributions, this is so because there are supersymmetric
(i.e.\  $R$-odd) particles in the external lines, and thus the supersymmetric
theory must be taken as a whole to be renormalizable.

\diagsbtcharQCDvertex

The one-loop vertex Feynman diagrams contributing to the one-loop form
factors~(\ref{eq:sbdecayLambdes}) are depicted in
figure~\ref{diag:sbtcharQCDvertex}. The gluon 
contribution~(Fig.\,\ref{diag:sbtcharQCDvertex}(a)) can be written as
\begin{eqnarray}
  \label{eq:sbtcharQCDFormgluon}
  F_L&=&i\,4\,\pi\,\alpha_s\,C_F(\Ambai\,\delta_{g1}+\Apbai\,\delta_{g2})\,\,, \nonumber\\
  F_R&=&i\,4\,\pi\,\alpha_s\,C_F(\Apbai\,\delta_{g1}+\Ambai\,\delta_{g2})\,\,,
\end{eqnarray}
where  $C_F=(N_C^2-1)/2\,N_C=4/3$  is a colour factor and
\begin{eqnarray}
  \label{eq:sbthcharQCDformgluonfactor}
  \delta_{g1}&=&C_0\,(\mts-\mis)+(\mis-\mts-\msbas) (C_{11}-C_{12})+
  \tilde{C_0}\,\,,\nonumber\\
  \delta_{g2}&=&2\,\mi\,\mt\,C_{12}\,\,,\nonumber\\
  C_{*}&=&C_{*}(k,-p,\msba,\lambda,\mt)\,\,,
\end{eqnarray}
where we have introduced a small gluon mass $\lambda$ to regularize the infrared
divergences. The gluino contribution from Fig.\,\ref{diag:sbtcharQCDvertex}(b) is
far more complicated
\begin{eqnarray}
  \label{eq:sbtcharQCDformgluinofactor}
  F_L&=&-i\,8\,\pi\,\alpha_s\,C_F \left[
    -(\msbas\,C_{11}+\tilde{C_0}-(\msbas-\mis)\,C_{12}) 
    \Aptbi{}^*\,R_{1a}^{(b)*}\, R_{2b}^{(t)*}\right. \nonumber\\
  &+&\mt\,\mg (C_{11}-C_{12})  
\Aptbi{}^*\,  R_{1a}^{(b)*}\, R_{1b}^{(t)*} \nonumber\\
  &-&\mg\,\mi\,C_{11}\, \Amtbi{}^*\,  R_{2a}^{(b)*}\, R_{2b}^{(t)*} \nonumber\\
  &+&\mi\,\mt\,C_{12}\, \Amtbi{}^{*}\,  R_{2a}^{(b)*}\, R_{1b}^{(t)*} \nonumber\\
  &+&\mg\,\mb\,C_0\, \Aptbi{}^{*}\,  R_{2a}^{(b)*}\, R_{2b}^{(t)*} \nonumber\\
  &-&\mb\,\mt (C_0+C_{11}-C_{12}) \Aptbi{}^{*}\,  R_{2a}^{(b)*}\, R_{1b}^{(t)*}
  \nonumber\\
  &+&\left.\mb\,\mi\,(C_0+C_{11}) \,\Amtbi{}^{*}\,  R_{1a}^{(b)*}\, R_{2b}^{(t)*}\right]
\,\,,  \nonumber\\
  F_R&=&F_L \left( A_{+} \leftrightarrow A_{-} , R_{1*}^{(*)} \leftrightarrow R_{2*}^{(*)}
  \right)\,\,,\nonumber\\
  C_{*}&=&C_{*}(k,-p,\mb,\mg,\mstc)\,\,\,.
\end{eqnarray}

\diagsbtcharbrems

The infrared divergences from~(\ref{eq:sbthcharQCDformgluonfactor}) and
from~(\ref{eq:QCDselfb}), (\ref{eq:wavesbottomQCDgluonsimple}) cancel with the
real corrections from the diagrams~\ref{diag:bremssbtcharQCD}. The gluon
bremsstrahlung contribution to the $\sbottom_a$ decay width is
\begin{eqnarray}
  \label{eq:sbtcharQCDbrems}
       \Gamma_{\rm Brems}^{ai}&=& -\frac{g^2\,\alpha_s}{6\,\pi^2\,\msba}\times\nonumber\\
     && \left\{(\Apbai)^2+(\Ambai)^2\right)\left[
        2\,\msbas(\msbas-\mts-\mis) I_{00}+2\,(\msbas-\mis-\mts)\,I_0 \right. \nonumber\\
      &&\left.      
        -2\,\mts\,I_1-2\,\mts(\mts+\mis-\msbas)I_{11}+I_{1}^0
      \right.\nonumber\\
      &&\left.
        +2\,(\msbas(\msbas-\mis)+\mis(\mis-\msbas)-\mt^4) I_{01}
        +2\,(\msbas-\mis)I_{1}\right] \nonumber \\
      &-&\left. 8\,\Apbai\,\Ambai\,\mt\,\mi\,\left[
        I_{00}+I_1+\mts I_{11}
        +(\mts-\mis+\msbas)I_{01}+I_0
        \right] \right\}\,\,,\nonumber\\
        I_{*}&=&I_{*}(\msba,\mt,\mi)\,\,,
\end{eqnarray}
where we have used the bremsstrahlung functions defined
in~\cite{Denner:1993kt}\footnote{We have corrected a typo present in expressions
  D.11 and D.12 of Ref.\cite{Denner:1993kt}.}. We have checked explicitly
(analytically and numerically) that after adding
up the one-loop~(\ref{eq:sbtcharcorrection}) and the
real~(\ref{eq:sbtcharQCDbrems}) corrections the final result
\begin{equation}
  \label{eq:sbtcharQCDdelta}
  \delta_{QCD}^{ai}=\delta^{ai}+\Gamma_{\rm Brems}^{ai}/\Gamma_0^{ai}\,\,,
\end{equation}
is free of 
ultraviolet and infrared divergences.

\figsbdecayQCDtbmu

In Figs.\,\ref{fig:sbdecayQCDtbmu}-\ref{fig:sbdecayQCDmgM} we present the evolution
of the corrections~(\ref{eq:sbtcharQCDdelta}) with some of the parameters. For
the numerical evaluation we use $\alpha_s(\msba)$, using the one-loop \MSSM\
$\beta$-function, but, for the $\msba$ we use, it is basically the 4-flavour \SM\
$\beta$-function, as the scale is almost always below the threshold of coloured
\SUSY\  particles (and top quark).
In
Fig.\,\ref{fig:sbdecayQCDtbmu} we can see the evolution with \tb\  and $\mu$,
which are the most interesting ones. The corrections
are large ($>10\%$) and present a weak evolution for large values of \tb\  
($\gsim 20$).
We remark that for 
$\mu<-120\,\GeV$ and $\tan\beta>20$ the corrections can be very large
near the phase space limit of the 
lightest sbottom decay. However, this effect
has nothing to do with the phase space exhaustion, which is described by
the kinematic function $\lambda (a, i, t)$ on the RHS of the tree-level
expression~(\ref{eq:sbtchartree}), but rather with the presence of the dynamical
factor 
in brackets on that equation which also goes to the denominator
of $\delta$ in eq.(\ref{eq:sbtcharcorrection}). That factor is fixed by the 
structure of the interaction Lagrangian
of the sbottom decay into charginos and top; and, for the parameters
in Fig.\,\ref{fig:sbdecayQCDtbmu}, it turns out to vanish near (actually past)
the phase space 
limit in the case of the lightest sbottom ($\tilde{b}_1$) decay.
However, this is not so either for the heaviest sbottom ($\tilde{b}_2$) or for
$\mu>120\,\GeV$ as it is patent in the same figure. The different evolution that
present the corrections of the two sbottoms
has more relation with the electroweak nature of the process than with the
purely \QCD\  loops, it is due to the fact that, being
$\theta_{\sbottom}$ and $\theta_{\stopp}$ so
small, the squarks are mostly chiral, namely
\begin{equation}
  \sbottom_1 \simeq \sbottom_R \,\,\,\,,\,\,\, \sbottom_2\simeq \sbottom_L  
  \,\,\,\,,\,\,\,
  \stopp_1 \simeq \stopp_R \,\,\,\,,\,\,\, \stopp_2\simeq \stopp_L\,\,\,, 
\label{eq:sbdecaySQchiral} 
\end{equation}
so its very different couplings to charginos~(\ref{V1Apm}) provides
a very different evolution of~(\ref{eq:sbtcharcorrection}), even if
$\Lambda_{L,R}$ were constant. In fact the sbottom mixing angle plays a crucial
role in this corrections as seen in Fig.\,\ref{fig:sbdecayQCDangle}, however we
also see that its value is highly constrained by 
the condition~(\ref{eq:necessary}). Finally we would like to comment on the
effect of the gaugino mass parameter $M$ and the gluino mass in
Fig.\,\ref{fig:sbdecayQCDmgM}. The gluino evolution is rather flat once the
pseudo-thresholds of $\sbottom_a\rightarrow b\,\sg$ are passed, so even though the
gluino could not be produced at the Tevatron it would have an effect on the
sbottom decay\footnote{In~\cite{Djouadi:1997wt} it is shown that there exist a
  non-decoupling effect at large gluino masses, however this effect is
  numerically small and is not the one reflected in Fig.\,\protect{\ref{fig:sbdecayQCDmgM}}(a).}. As for the gaugino mass parameter the correction is saturated
for $M\gsim 200\GeV$, so the corrections computed in this section can be compared
with the ones obtained in the higgsino approximation discussed in the next
section.

The other parameters of the model present a rather mild effect on the
corrections for squark masses in the ballpark of several hundreds of GeV. In
summary the \QCD\  corrections on the decay $\sbottom_a \rightarrow t\,\cmin_i$
are large ($\simeq -20\%$ for $\sbottom_2$, $\simeq -60\%$ for $\sbottom_1$) and
negative for values of the parameter space relevant to the Tevatron energies,
with a higgsino-like chargino and moderate or large \tb.

\figsbdecayQCDangle

\figsbdecayQCDmgM

\section{Yukawa corrections}
\label{sec:sbtcharYukawa}
At large ($\geq 20$) or small ($<1$) $\tan\beta$ 
these effects could be
competitive with the \QCD\  corrections of the previous section.
Since
in these conditions the full \MSSM\  
quantum effects can be rather large, their calculation is indispensable 
to account for the observed 
top quark production cross-section (\ref{eq:productionMSSM}) in the \MSSM\  
or, alternatively, to better assess how much the determination of the \SM\  
branching ratio $BR(t\rightarrow W^+\,b)$ is affected in the \MSSM\  context 
after plugging in the experimental number on 
the LHS of eq.(\ref{eq:productionMSSM}).

The analytical formulation developed so far in chapter~\ref{cap:Renorm} and in
section~\ref{sec:sbdecayrenorm} is well
suited to tackle the general problem of the \SUSY-\EW\  
corrections to squark decays. 
Since the dominant part is from the Yukawa sector we wish to pursue
our calculation in the following  within the Yukawa coupling approximation.
This means that we are going to compute the leading 
electroweak effects of ${\cal O}(\lambda^2_t)$ 
{\it and} ${\cal O}(\lambda^2_b)$ that emerge 
for large values of the Yukawa couplings (\ref{eq:Yukawas}) when the remaining
gauge contributions -- of ${\cal O}(g^2)$-- are subdominant.
In practice we shall only explore the large $\tan\beta$
 regime, typically $\tan\beta\geq 20$;
the possibility  $\tan\beta<1$ is not so appealing from the theoretical
point of view.  Thus within our approximation we will include the correction 
$\Delta_{\tau}$ in leading 
order ${\cal O}(\lambda_{\tau}^2)$ of the $\tau$ Yukawa-coupling, $\lambda_{\tau}$.
Notice furthermore that for $\lambda_b\gg 1$ the
tree-level decay rate, eq.\,(\ref{eq:sbtchartree}), is maximized. Therefore,
the large $\tan\beta$ range is expected to be the most relevant
one for the decay under consideration. 

In our approach, we set the $SU(2)_L$ gaugino mass parameter $M\gg |\mu|, M_W$ in
the chargino mass matrix (see section~\ref{sec:cnmas}),
and therefore the chargino $\chi^\pm_1$
is mainly higgsino, whereas the chargino $\chi^\pm_2$ is mainly gaugino and does
not contribute to our decays.
It is only in this case that the Yukawa-coupling approximation makes
sense. 
Thus, since $m_{\tilde{t}_1}>80-90\,\GeV$, in this approach the decay into stop
and neutralino
$t\rightarrow\tilde{t}_a\,\neut_{\alpha}$ is kinematically
forbidden.
In this approximation the relevant counterterms
$\delta A_{\pm ai}^{(b)}$ in eq.(\ref{eq:sbtchardcounter}) boil down to 
\begin{eqnarray}
\label{eq:sbtchardeltaAsEW}
\delta \Apbao&=&-\delta R_{2a}^{(b)}\,\lambda_b -R_{2a}^{(b)}\, \delta \lambda_b\nonumber\\
\delta \Ambao&=&-\delta R_{1a}^{(b)}\,\lambda_t -R_{1a}^{(b)}\, \delta \lambda_t\,.
\end{eqnarray}

\diagautoenerchar

In the higgsino approximation only two Feynman graphs contribute to the
$\cmin_1$ self-energy, and they are depicted in Fig.\,\ref{diag:autoenerchar}. We
can write the chargino $\cmin_1$ self-energies, defined like the fermion
self-energies in 
chapter~\ref{cap:Renorm}, as
\begin{eqnarray}
  \label{eq:charself}
  \Sigma_L^1(k^2) &=& - i\, g^2\, N_C (\Ambao)^2 B_1(k,\mt,\msba) \nonumber \\
  \Sigma_R^1(k^2) &=& - i\,g^2\,N_C (\Apbao)^2 B_1(k,\mt,\msba) \nonumber \\
  \Sigma_S^1(k^2) &=&+ i\,g^2\,N_C\frac{\mt}{\mi} \ei\,\Apbao\,\Ambao\,B_0(k,\mt,\msba) \,\,,
\end{eqnarray}
for diagram~\ref{diag:autoenerchar}(a), note that in this approximation the
chargino $\cmin_1$ coupling matrices are simply \diagautoenersbew
$$
\Ambao = -\lambda_b\,R_{2a}^{(b)}\,\,,\,\,
\Apbao = -\lambda_t\,R_{1a}^{(b)}\,\,,
$$
we prefer however to maintain a more general notation. The bottom-stop contribution
(diagram~\ref{diag:autoenerchar}(b)) can be obtained  performing the following
substitutions to~(\ref{eq:charself}): 
$\mt\rightarrow\mb$, $\msba\rightarrow\msta$, $A_{\pm ai}^{(b)}\rightarrow
A_{\mp ai}^{(t)}$. From these self-energies we compute the wave function
renormalization 
constants with the help 
of expression~(\ref{eq:DSRC}). A brief comment is mandatory respecting
the relation between the renormalization constants~(\ref{eq:charself}) and that
of the charged Higgs~(\ref{dZHSusy}). The chargino
definition~(\ref{eq:cinos}), in the case of the higgsino approximation, is
\begin{equation}
  \label{eq:defhiggsi}
  \Psi_1^{-}=\left(\begin{array}{c}
    \tilde{H}_1^- \\
    \bar{\tilde H}_2^{+} 
    \end{array}
    \right)\,\,, 
\end{equation}
thus in a supersymmetric renormalization we should obtain that the wave function
renormalization constants obtained from~(\ref{eq:charself}) are
\begin{equation}
  \label{eq:relateHiggsHiggsi}
  \delta Z_L^1 = \delta Z_{H_1} \,\,\,,\,\,\, \delta Z_R^1 = \delta Z_{H_2} \,\,,
\end{equation}
as we are not dealing with a supersymmetric renormalization procedure this is
not the case, however we have checked that the divergent part of this different
renormalization constants is the same. This fact is crucial in the cancellation
of the divergences of $\delta Z_{\{L,R\}}^1$ with the term $\delta Z_{H^\pm}$
appearing in the definition of $\delta \tb$~(\ref{eq:deltabeta}).

From   diagrams of figure~\ref{diag:autoenersbew} we can
obtain the sbottom self-energies. The chargino top contribution from
Fig.\,\ref{diag:autoenersbew}(a) is
\begin{eqnarray}
  \label{eq:sbottomselfEWchar}
  \Sigma_{ab}^{\cmin}(k^2)&=&-i\,2\,g^2 \left[
    \left(\Apbbi\,\Apbai+\Ambbi\,\Ambbi\right) (\tilde{B}_0+k^2\,B_1)
    \right.\nonumber\\
    &&\left.+
    \ei\,\mi\,\mt(\Apbbi\,\Ambai+\Ambbi\,\Apbai) B_0 \right](k,\mi,\mt)\,\,,
\end{eqnarray}
and the Higgs contributions from diagrams~\ref{diag:autoenersbew}(c) and (d) is
simply
\begin{equation}
  \label{eq:sbottomselfEWHiggs}
  \Sigma_{ab}^{H}(k^2)=i\,g^2 \sum_c R_{ia}^{(b)}\,R_{jc}^{(q)*}\,
  G^{H}_{ij}\,R_{kc}^{(q)}\,R_{lb}^{(b)*}\, G^H_{lk} B_{0}(k,\msqc,M_H)\,\,,
\end{equation}
with $H=(H^+,G^+;h^0,H^0,A^0,G^0)$, $\sq=(\stopp;\sbottom)$, and $G^H_{ij}$ is
the Feynman rule for $H\rightarrow\sq_i^*\,\sq_j'$, with $\sq_{1,2}$ the weak
eigenstates squarks, divided by $-ig$\,\cite{Hunter,Tesina}\footnote{Note that
  with this convention $G^A_{ij}=-G^A_{ji}$.}. 
On the other hand diagram\,\ref{diag:autoenersbew}(e) only contributes to the
mass counterterm and mixing wave function renormalization constants,
\begin{equation}
  \label{eq:sbottomselfEWHiggstadpole}
  \Sigma_{ab}^{HH}=i\,g^2\sum_{H^+,G^+} R_{ia}^{(b)}\,R^{(b)*}_{jb}\,E^H_{ij}
  \Az(M_H)+ \frac{i\,g^2}{2}  \sum_{h^0,H^0,A^0,G^0}  R_{ia}^{(b)}\,R^{(b)*}_{jb}\,E^H_{ij}\Az(M_H)
\,\,,
\end{equation}
with $E^H_{ij}$ the corresponding Feynman rule of $H\,H^* \rightarrow
\sbottom^{\prime *}_i\,\sbottom^{\prime}_j$, $\sbottom_{1,2}$ the weak
eigenstate squarks,  divided by $-ig^2$\cite{Hunter}. 

Then the diagonal wave function renormalization constant is
\begin{eqnarray}
  \label{eq:sbottomwaveEWchar}
  \delta Z_{aa}^{\cmin}&=&-i\,2\,g^2\left[
    \frac{1}{2}\left((\Apbai)^2+(\Ambai)^2\right) \left(
      (\mts+3\,\mis+\msbas) B_{0}^\prime-B_0
    \right) \right.\nonumber\\
    &+&\left.2\,\ei\,\mi\,\mt\,\Apbai\,\Ambai\,B_0^\prime
  \right] (\msba,\mi,\mt)\,\,, \\
  \label{eq:sbottomwaveEWHiggs}
  \delta Z_{aa}^{H}&=&i\,g^2  \sum_c R_{ia}^{(b)}\,R_{jc}^{(q)*}\,
  G^{H}_{ij}\,R_{kc}^{(q)}\,R_{lb}^{(b)*}\, G^H_{lk} B_{0}^\prime(\msba,\msqc,M_H)\,\,.
\end{eqnarray}
The neutralino-bottom contribution form Fig.\,\ref{diag:autoenersbew}(b) can be
easily found by performing the following transformations in
expressions~(\ref{eq:sbottomselfEWchar}) and~(\ref{eq:sbottomwaveEWchar}):
substitute  $\mt\rightarrow\mb$, the chargino indices by neutralino ones
$i\rightarrow\alpha$, and divide the expressions by $2$.

The rest of the renormalization constants are computed using the very same
expressions of chapter~\ref{cap:Renorm} by taking the Yukawa approximation,
i.e.\  by removing the interactions with the gauge bosons (but maintaining that of
the Goldstone bosons) and with the gauginos.

\diagsbdecayEWvertex

The one-loop Feynman diagrams contributing to the vertex form factors can be
seen in Fig.\,\ref{diag:sbdecayEWvertex}. There exists other possible diagrams,
but they do not contribute in the Yukawa approximation. From this diagrams we
can compute the corresponding $F_{\{L,R\}}$ form factors,
\begin{eqnarray}
  \label{eq:sbdecayEWformHfactor}
  F_L^{\{h^0,H^0\}}&=&N_{\{h^0,H^0\}}\left[ \mt\,\ei\,\Ambbi (C_{12}-C_{11})
  -\mi\,\Apbbi C_{12}
  + \mt\,\ei\,\Ambbi\,C_0\right]\,\,,\nonumber\\
  F_R^{\{h^0,H^0\}}&=&N_{\{h^0,H^0\}}\left[ \mt\, \Apbbi (C_{12}-C_{11})
    -\mi\,\ei\,\Ambbi C_{12}+
    \mt\,\Apbbi\,C_0 \right]\,\,,\nonumber\\
  N_{\{h^0,H^0\}}&=&-i\,g^2\frac{\lambda_t}{2} \{\sa,\ca\}
  G^{\{h^0,H^0\}}_{ij}R^{(b)}_{ia} R^{(b)*}_{jb} \,\,,\nonumber\\
  F_L^{\{A^0,G^0\}}&=&N_{\{A^0,G^0\}}\left[
    -\mt \,\ei\,\Ambbi (C_{12}-C_{11})
    -\mi\,\Apbbi\,C_{12}
    +\mt\,\ei\,\Ambbi\, C_0
    \right]\,\,,\nonumber\\
  F_R^{\{A^0,G^0\}}&=&N_{\{A^0,G^0\}}\left[
    \mt\,\Apbbi\, (C_{12}-C_{11})
    +\mi\,\ei\Ambbi\,C_{12}
    -\mt\,\Apbbi\, C_0
    \right]\,\,, \nonumber\\
    N_{\{A^0,G^0\}}&=&-i \frac{g^2}{2\,\sqrt{2}\,\mw} \{-\frac{\mt\,\ctb}{\sqrt{2}}
    ,-\mt\} G^{\{A^0,G^0\}}_{ij}R^{(b)}_{ia} R^{(b)*}_{jb}\,\,,\nonumber\\
    C_{*}&=&C_{*}(-p,k,\mt,m_{\Phi^0},\msbb) \,\,,
\end{eqnarray}
for diagram~\ref{diag:sbdecayEWvertex}(a), and with the help of the following combinations 
\begin{equation}
  \label{eq:sbdecayEWshortneut}
\ba{rclrcl}
  \LLL&=&\Aptba\,\Amtbi\,\Ambaa\,, &  \RLL&=&\epa\,\Amtba\,\Amtbi\,\Ambaa\,, \\
  \LLR&=&\Aptba\,\Amtbi\,\epa\,\Apbaa\,,&  \RLR&=&\Amtba\,\Amtbi\,\Apbaa\,, \\
  \LRL&=&\Aptba\,\ei\Aptbi\,\Ambaa\,,&\RRL&=&\epa\,\Amtba\,\ei\,\Aptbi\,\Ambaa\,, \\
  \LRR&=&\Aptba\,\Aptbi\,\epa\,\Apbaa\,,&\RRR&=&\Amtba\,\ei\,\Aptbi\,\Apbaa\,,
  \\
\ea
\end{equation}
the contribution from diagram diagram~\ref{diag:sbdecayEWvertex}(b) is
\begin{eqnarray}
  \label{eq:sbdecayEWformneutfactors}
  F_L^{\chi}&=&-i\frac{g^2}{4} \left[
    C_0 (\mi\,\mt\, \LRL + \mt\,\mb\, \LRR + \ma\,\mi\,\RLL+\ma\,\mb\,\RRR)
    \right.\nonumber\\
    &-&\mt\,(C_{11}-C_{12})(\mi\,\LRL+\mb\,\LRR+\mt\,\RLR+\ma\,\LLR) \nonumber\\
    &+&\left.\mi\,C_{12}(\mi\,\RLR+\mb\,\RLL+\mt\,\RLR+\ma\,\RRL)+\tilde{C}_0
    \right]\,\,,\nonumber\\
  F_R^{\chi}&=&F_L^{\chi} \left\{ A^{(i)}\leftrightarrow A^{(9-i)} \right\} \,\,,\nonumber\\
   C_{*}&=&C_{*}(-p,k,\mstb,\ma,\mb) \,\,.
\end{eqnarray}

For the numerical analysis, we follow the directions given in
section~\ref{sec:sbtchartree}.
In the relevant large $\tan\beta$ segment under consideration,
namely 
\begin{equation}
20\lsim\tan\beta\lsim 40\,,
\label{eq:tansegment}
\end{equation}
the bottom quark Yukawa coupling $\lambda_b$ 
is comparable to the top quark Yukawa coupling, $\lambda_t$. 
Even though the extreme interval $40<\tan\beta<60$ can be tolerated
by perturbation theory, we shall confine ourselves to the moderate range
(\ref{eq:tansegment}). This is 
necessary to preserve the condition (\ref{eq:necessary}) for the
typical set of sparticle masses used in our analysis. 
We point out that the colour stability requirement (\ref{eq:necessary})
could be satisfied independently of $\tan\beta$ if the $A$-parameters would be
chosen directly as a part of the set of inputs and then taken
sufficiently small. Nevertheless this possibility is not so
convenient in our analysis where the sparticle masses are the natural inputs
that we wish to control in order to make sure that
sparticles can be produced and decay at the Tevatron as explained
in connection to eq.(\ref{eq:productionMSSM}).

\figsbdecaytres

\figsbdecayquatre

\figsbdecaycinc

The corresponding corrections $\delta^{ai}$~(\ref{eq:sbtcharcorrection}) are shown in 
Figs.\,\ref{fig:sbdecaytres}(a) and \ref{fig:sbdecaytres}(b) as a function of the
lightest stop and sbottom masses, 
respectively.
The allowed range for the sbottom and stop mixing angles is conditioned by
the upper bound on the trilinear couplings and is obtained
from eqs.(\ref{eq:Abt}) and (\ref{eq:necessary}). 
In the physical
$\theta_{\sbottom}$ range, the variation
of the correction (\ref{eq:sbtcharcorrection}) is shown in Fig.\,\ref{fig:sbdecayquatre}(a).
On the other hand the permitted range for the stop mixing
angle, $\theta_{\tilde{t}}$, is much larger and we have plotted the corrections
within the allowed region in Fig.\,\ref{fig:sbdecayquatre}(b). Notice that
the sign of the quantum effects changes
within the domain of variation of $\theta_{\tilde{t}}$.
Finally, we display the evolution of the \SUSY-\EW\  
effects as a function of $\tan\beta$ (Fig.\,\ref{fig:sbdecaycinc}(a)) and of $\mu$ (Fig.\,\ref{fig:sbdecaycinc}(b))
within the region of compatibility with the constraint (\ref{eq:necessary}).

A few more words are in order to explain the origin of the leading
electroweak effects. One could expect that they come from 
the well-known large $\tan\beta$ enhancement stemming
from the chargino-stop corrections to the bottom mass  
(Cf.\ Fig.\,\ref{diag:deltambEW}(b)). 
Nonetheless this is only partially true, for in the present
case the remaining contributions (Cf.\ Figs.\,\ref{diag:autoenerchar},
\ref{diag:autoenersbew} and~\ref{diag:sbdecayEWvertex}) 
can be sizeable enough. One can also think on the \SUSY\  counterpart of
Fig.\,\ref{diag:deltambEW}(b), 
which we have depicted in
Fig.\,\ref{diag:leadingautosbottm}, as an additional leading contribution, as,
in addition to the \tb\  enhancement, has an $A_q$ enhancement. However the
addition of these two kind of contributions does not account for the total
behaviour in all of the parameter space.
To be more precise, in the region of the
parameter space that we have dwelled upon
the bottom mass contribution is seen to be
dominant only for the lightest sbottom decay
and for the lowest values of $\tan\beta$ in the
range (\ref{eq:tansegment}). This is indeed the case in Fig.\,\ref{fig:sbdecayquatre}(b) where
$\tan\beta=20$ and therefore the bottom mass effect modulates the 
electroweak correction in this process and 
$\delta^{11}$ becomes essentially an odd function of the stop mixing angle. 
This fact is easily understood since, as noted above, sbottoms are mostly chiral 
--eq.\,(\ref{eq:sbdecaySQchiral})-- and the $\sbottom_R$ is the only one with
couples with $\lambda_b$ --eq.\,(\ref{eq:vyukawa}).
On the other hand, from Fig.\,\ref{fig:sbdecaycinc}(a) it is obvious that the 
(approximate) linear behaviour on $\tan\beta$ expected from bottom 
mass renormalization becomes
completely distorted by the rest of the contributions,
especially in the high $\tan\beta$ end.
In short, the final electroweak correction cannot be simply ascribed 
to a single renormalization source but to the full Yukawa-coupling combined yield.

In general the \SUSY-\EW\   corrections to 
$\Gamma(\sbottom_a\rightarrow t\,\cmin_i)$ are smaller than 
the \QCD\   corrections. 
The reason why the electroweak corrections are smaller
is in part due to
the condition (\ref{eq:necessary}) restricting our analysis within the
$\tan\beta$ interval (\ref{eq:tansegment}). From  Figs.\,\ref{fig:sbdecayquatre} and~\ref{fig:sbdecaycinc}(a) it is
clear that outside this interval the \SUSY-\EW\   contributions could be much higher
and with the same or opposite
sign as the \QCD\   effects, depending on the choice of the sign of the 
mixing angles. 
Moreover, since we have focused our analysis to sbottom masses
accessible to Tevatron, again the theoretical bound (\ref{eq:necessary})
severely restricts the maximum value of the trilinear couplings and this
prevents the electroweak corrections from being larger. 
This cannot be cured by assuming larger values of $M_H$ and/or of $\mu$
due to our assumption that $t\rightarrow H^+\,b$ is operative and because
$\mu$ directly controls the value of the (higgsino-like) chargino final
state in our decay, so that basically we have  $|\mu|<m_{\sbottom_a}-M_H$.
The restriction cannot be circumvented either if we
assume larger values of $m_{\tilde{t}_a}$, for it has been shown
that too heavy stops are incompatible with the CLEO data on
 $b\rightarrow s\,\gamma$
both at low and high $\tan\beta$\,\cite{Coarasa:1997ky,Rattazzi:1996gk,Barbieri:1993av,Garisto:1993jc,Diaz:1994fc,Borzumati:1994zg,Bertolini:1995cv,Carena:1995ax,Bertolini:1991if}. 
We point out that the \MSSM\  analysis of $b\rightarrow s\,\gamma$ 
also motivated the sign choice 
$A\,\mu<0$ in our numerical calculation\,\cite{Coarasa:1997ky}. 
Admittedly, the situation with radiative $B$-decays is still under 
study and there are many sources of 
uncertainties
that deserve
further experimental consideration. Still, we have used this information
to focus on a limited domain of the \MSSM\  parameter space. 

\diagsbdecayleadingEW

\section{Conclusions}

In summary, the \MSSM\  corrections to squark decays into charginos and
neutralinos can be significant and therefore must be included in any reliable
analysis of top quark physics at the Tevatron within the \MSSM. The
main corrections stem from 
the strongly interacting sector of the theory
(i.e. the one involving gluons and gluinos), but also
non-negligible effects may appear from the electroweak sector 
(characterized by chargino-neutralino exchange) at large (or very small)
values of $\tan\beta$. 
Failure of including these corrections in future studies
of top quark physics at the Tevatron, both in the production and decay
mechanisms, might seriously hamper the possibility of
discovering clear-cut traces of \SUSY\   physics from the identification of 
large non-SM quantum corrections in these processes. 
As already stated, we have mainly concentrated on the impact of
these quantum signatures in the physics of
the Tevatron, but important effects are also expected  for 
experiments aiming at the production
and decay of ``obese'' squarks  
at the LHC. The latter type of squarks could be free of some of the restrictions that
have been considered for the present calculation. 

The present study has also an impact on the determination of squark parameters
at the LC. The squark masses used in it are available already for a LC running
at a center of mass energy of $800\GeV$. The large corrections of both, the
\QCD\  and the \EW\  sector (in the Yukawa approximation), makes them necessary,
not only for prospects of precision measurements in the
sbottom-chargino-neutralino sectors, but also for a reliable first determination
of its parameters.


%% file: conclu.tex
\chapter{Conclusions}
\label{conclusions}

In this Thesis we have performed a study of some of the possible
phenomenological consequences deriving from  the interactions between the
third generation matter supermultiplet and the
the super-Higgs boson sector of  the \MSSM\  at the one-level,  with
especial emphasis on the implications for the top quark and Higgs-boson
physic  at the Tevatron collider. 
We have done so in the on-shell renormalization scheme,
using a physically motivated definition of \tb. Our definition of \tb\  has the
virtue of automatically incorporating the one-loop radiative corrections to the
most plausible signature of the charged Higgs 
(if $\tb>2$), namely the $\tau\,\nu_\tau$ channel. 
Remarkably
enough, this definition of \tb\  can also be extended to the situation
when $\mHp > \mt$ for a wide range of heavy charged Higgs masses to be explored
at the LHC rather than at the Tevatron\,\cite{Coarasa:1997ky,TesiToni}. For it turns out that the
branching ratio of the charged Higgs into $\tau$ never becomes negligible in
that range.
\vspace{2cm}

\begin{itemize}
\item 
The effects of one-loop \EW\  radiative corrections to the unconventional top quark
decay mode $t\rightarrow H^+ b$ are large in the moderate and specially in the high regime
of \tb, where they can easily reach values of
\begin{equation}
  \label{eq:conclustbhdeltamult0}
  \delta_{EW} (t\rightarrow H^+\,b) \simeq +30\%
\end{equation}
for negative $\mu$ (and positive $A_t$) and a ``light'' sparticle spectrum
(Fig.\,\ref{fig:tbhdeltatbmult0}), and  
\begin{equation}
  \label{eq:conclustbhdeltamugt0}
  \delta_{EW} (t\rightarrow H^+\,b) \simeq +20\%
\end{equation}
for positive $\mu$ (and negative $A_t$) and heavy sparticle spectrum
(Fig.\,\ref{fig:tbhdeltamsbmugt0}). In both cases we have singled out
the domain $\mu A_t<0$ of the parameter space, which is the one  preferred by
the experimental data on  radiative B-meson decays ($b\rightarrow s \gamma$). 

The previous results should be compared 
with the \QCD\ and \SUSY-\QCD\  corrections which in previous
studies\,\cite{TesiRicci,Guasch:1995rn} were shown to reach values of
$\delta_{QCD}\simeq-60\%$,  $\delta_{SUSY-QCD} \simeq +80\%$ and  $\delta_{SUSY-QCD} \simeq -40\%$ 
for the same scenarios. 
In the
perturbative regime of the calculation, the positive \EW\  corrections
attained for $\mu>0$ ($\mu A_t<0$) can be of the same order as they are in the
$\mu<0$ ($\mu A_t<0$)
case. Unfortunately, the \EW\  effects never become  huge enough so as to
prevent the total \MSSM\  correction from being highly negative for $\mu>0$
($\mu A_t <0$)
-- an unlucky fact which unavoidably leads to a severe suppression of the
corresponding branching ratio in this case. Quite in contrast, in the
$\mu<0$ ($\mu A_t<0$) situation there are indeed regions of the parameter space where
the positive \EW\  corrections could be perfectly visible; namely, in those
places where the total \QCD\  correction in the \MSSM\  largely cancels out,
e.g. around $\tb=30$ in Fig.\,\ref{fig:tbhdeltatbmult0}(a). Negative corrections 
of the same order can be obtained provided $\mu A_t>0$
(Cf. Fig\,\ref{fig:tbhdeltaAmult0}(b)). 

From these considerations
it follows that, if there exists 
supersymmetric partners of the
standard particles at a scale $100-500\GeV$,  the unconventional top quark decay
mode $t\rightarrow H^+\,b$  
has a partial width that differs significantly from the conventional \QCD\
expectations. In 
this case the analyses of the Tevatron data could exclude a region in the
$\tb-\mHp$ plane which is substantially modified as compared to
recent analyses from the Tevatron collaborations. Thus e.g. for a charged Higgs mass of
$110\GeV$, and using the conventional \QCD\  corrected value for the decay width
$\Gamma(t\rightarrow H^+\,b)$, present Tevatron data implies that the excluded
region is
\begin{equation}
  \label{eq:tbhexclusiondeltaQCD}
  \tb \geq 50 \,\,.
\end{equation}
If, instead, we assume that the charged Higgs belongs to the
Higgs sector of the \MSSM, then we find that the excluded values of
\tb\  are
\begin{equation}
  \label{eq:tbhexclusiondeltaEW}
  \tb\geq 35 \,\, , \,\, \tb\geq 75\,\,,
\end{equation}
for the two scenarios presented above respectively (see Figs\,\ref{fig:tanbtop3}
and~\ref{fig:tanbtop4}). Remarkably there is a range in the parameter space
(characterized by $\mu>0$, $\mu A_t<0$) where no value of the $\tb-\mHp$ plane
is excluded at 
all by present data on the charged Higgs decay of the top quark.
Although in mSUGRA models in the literature one usually claims $A_t<0$, we
already emphasized in chapter~\ref{cap:tbh} that this is not necessary the case, 
especially at large \tb\  and for sufficiently large values of the
trilinear Soft-\SUSY-breaking parameter, $A_0$, at the unification scale. 
Thus,  also in specific  minimal SUGRA models, one can have $A_t>0$ and  so
$\mu<0$, which is the most attractive possibility since one achieves large
positive corrections to $t\rightarrow H^+\,b$ compatible with $BR(b\rightarrow s \gamma)$.

The bulk of the \EW\  corrections is
given by the finite corrections to the bottom quark mass (or the bottom Yukawa coupling),
which are proportional to $-\mu A_t$ -see eq.~(\ref{eq:dmbEW}). 
However, it should be mentioned that there are
cancellations among other sources of significant corrections. This fact
implies that the rest of the \SUSY\  corrections was not obviously
negligible from the very beginning.
Thus, present
CLEO data on the partial 
decay branching ratio $BR(b\rightarrow s \gamma)$\cite{Alam:1995aw} favours
positive values of the 
\EW\  corrections. However, the leading component of the quantum effects is the
\SUSY-\QCD\  contribution which depends on the sign of $\mu$ (rather than that of
$A_t\mu$). Therefore, in the end the sign that really matters for this
process is that of $\mu$ alone. The best possible situation for the charged
Higgs decay of the top quark would occur for negative $\mu$, since then the
\SUSY-\QCD\  corrections are positive, and the \EW\  corrections are also
positive 
due to the $b \rightarrow s\, \gamma$ constraint. On the other hand, if $\mu$ is positive,
then  the total \MSSM\  correction is negative (in spite of the \EW\  component
which must stay positive).  
In this case the bounds on
the $(\tb, \mHp)$ space could disappear, as we said above. And in these
circumstances, as  explained
in chapter~\ref{cap:tbh}, there is an alternative scenario with relatively light
neutral \MSSM\ Higgs boson which in combination with the negative 
$t\rightarrow H^+\,b$ 
searches could strongly  point towards the \SUSY\  nature of  these Higgs
bosons.
As for the one-loop Higgs corrections sector of the \MSSM, the
over-all correction to the decay under consideration is very small due to
huge cancellations triggered by the SUSY structure of the Higgs
potential to this decay except for very low values of \tb\ (Fig.\,\ref{fig:tbhdeltatbmult0}). 
Work is currently in progress to determine
the effects of these large corrections to the top quark and charged Higgs
associated production at the Tevatron and at the LHC.
\vspace{2cm}

\item \FCNC\  top quark decays into neutral Higgs particles have been reviewed. The
correct inclusion of the left-right mixing between squarks implies an
enhancement of the partial branching ratio $BR(t \rightarrow c h)$ of two orders
of magnitude with respect previous estimates. We have performed a separate
analysis of the \SUSY-\EW\  and \SUSY-\QCD\  effects, with different
approximations. For the \SUSY-\EW\  sector we have used the super-\ckm\
basis, whereby \FCNC\  are produced through the charged sector, as in the \SM. For
the \SUSY-\QCD\  estimate we have supposed a non-flavour-diagonal mass elements in
the left-chiral squark matrix. We have applied present bounds from \EW\
precision data to these elements. The theoretical upper limits of these contributions
to the decay width of the top quark are found to be (Figs.\,\ref{fig:resultsew} and~\ref{fig:resultsqcd})
\begin{eqnarray}
  \label{eq:conclusFCNClimits}
  BR^{\rm SUSY-EW}(\tch) &\lsim& \mbox{ several }\times10^{-6}\nonumber\\
  BR^{\rm SUSY-QCD}(\tch)&\lsim& \mbox{ several }\times10^{-4}\,\,\,,
\end{eqnarray}
and the typical values for this ratio are $10^{-8}$ and $10^{-5}-10^{-4}$ for
the \SUSY-\EW\ and \SUSY-\QCD\  induced \FCNC\  decays respectively.
We have found that the \SUSY-\QCD\  induced \FCNC\
decay widths are at least two orders of magnitude larger than the \SUSY-\EW\  ones in most
of the parameter space, thus
making unnecessary the computation of both the interference terms and \FCNC\
\EW\  induced effects through non-flavour-diagonal mass 
terms. If data improves the bounds on these mass terms they will become more strict,
bringing the \SUSY-\QCD\  induced branching ratios down to the values of the
\SUSY-\EW\  ones, and then a complete computation would be needed, if there would
be any hope at all to see these effects at the small branching ratio predicted by
\EW\  corrections! 
The value found for  $BR(t\rightarrow c h)$  is not
sufficiently large to yield measurable effects at the Tevatron or at the
LC\footnote{\textbf{Note Added}: See however note on section~\ref{sec:con} (pg.\,\pageref{pagenoteLC}).}. There
is, however, a possibility that this decay mode could be measured at the LHC.
\vspace{2cm}

\item If bottom-like squarks are heavy enough they could decay into a top quark and a
chargino. This could serve as an unexpected source of top quarks at the
Tevatron, at the LHC, or at the LC. The radiative corrections to the partial decay width 
$\sbottom\rightarrow t\,\cmin$ are large, both the \QCD\   and the \EW-like in the
Yukawa approximation. In the case of the \QCD\  corrections, they are
negative in most of the \MSSM\  parameter space accessible to 
Tevatron. These corrections are of the order
\begin{eqnarray}
  \label{eq:conclustbtcharQCD}
  \delta_{QCD} (\sbottom_1\rightarrow t\,\cmin_1) &\simeq& -60 \% \nonumber\\
  \delta_{QCD} (\sbottom_2\rightarrow t\,\cmin_1) &\simeq& -20 \% 
\end{eqnarray}
for a wide range of the parameter space (Fig.\,\ref{fig:sbdecayQCDtbmu}). In certain corners of this space, though,
they vary in a wide range of values.
\EW\  corrections can be of both signs. Our renormalization
prescription uses the mixing angle between squarks as an input parameter.
This prescription forces the physical region to a narrow range when we  require
that colour 
breaking vacua is not generated. Within this restricted region the typical corrections
vary in the range (Figs.\,\ref{fig:sbdecayquatre}, \ref{fig:sbdecaycinc})
\begin{eqnarray}
  \label{eq:conclusstbcharEW}
  \delta_{EW} (\sbottom_1\rightarrow t\,\cmin_1) &\simeq& +25 \% \mbox{ to } -15 \% \nonumber\\
  \delta_{EW} (\sbottom_2\rightarrow t\,\cmin_1) &\simeq& +5 \% \mbox{ to } -5 \% \,\,,
\end{eqnarray}
However we must recall  that these limits are
qualitative. In the edge of such regions  we find the largest \EW\
contributions. 
We stress that in this case
it is not possible to narrow down the bulk of the corrections to just the
renormalization of the bottom quark Yukawa coupling.
Although it is true that for moderate values of the parameters
(and for 
the lightest sbottom decay) the finite threshold corrections to the bottom
Yukawa coupling yields 
most of the contribution, as soon as we take the parameters away of this
central values the total corrections deviate significantly from the ones
obtained using this only term.  More work is presently in progress to generalize
this results to the full \SUSY-\EW\  sector, by incorporating the
neutralino-like decays, and also the inclusion of gaugino-higgsino mixing.
\end{itemize}
\vspace{3cm}

Our general conclusion is that the supersymmetric strong and electroweak radiative
corrections can be very important in the top/bottom-Higgs super-sector of the
\MSSM. Therefore, it is necessary
to account for these corrections in the theoretical computation of the high energy physics
observables, otherwise highly significant information on the potentially
underlying \SUSY\  dynamics could be missed. 
This is true, not only for the future experiments at the LHC and
the LC, but also for the present Run I data (and the Run II data around
the corner) at the Fermilab Tevatron collider\,\cite{OurTevatron}.


%% file: biblio.tex


%% file: tbh_c.tex
\chapter{Vertex functions} 
\label{ap:pointfun} 
 
In this appendix we briefly collect, for notational  
convenience, the basic vertex functions frequently referred  
to in the text.  
The given  
formulas are exact for arbitrary internal masses and external  
on-shell momenta. Most of them are an adaptation to the  
$g_{\mu\nu}=\{+\,-\,-\,-\}$ metric of the standard formulae  
of Refs.\cite{'tHooft:1979xw,Passarino:1979jh,Axelrod:1982yc}.  
The basic one-, two- and  
three-point scalar functions are: 
\begin{eqnarray} 
    A_{0}(m)&=& \int d^{n}\tilde{q}\, \frac{1}{[q^2-m^2]},\\ 
    B_{0}(p,m_1,m_2)&=&\int d^{n}\tilde{q}\, \frac{1}{[q^2-m_1^2]\, 
    [(q+p)^2-m_2^2]}\,, 
\end{eqnarray} 
\begin{equation} 
    C_{0}(p,k,m_1,m_2,m_3)=\int d^{n}\tilde{q}\,  
    \frac{1}{[q^2-m_1^2]\,[(q+p)^2-m_2^2]\,[(q+p+k)^2-m_3^2]}\,; 
\end{equation} 
using the integration measure 
\begin{equation} 
\label{eq:muUV} 
d^n\tilde{q}\equiv {\mu }^{(4-n)}\frac{d^nq}{(2\pi )^n}\,. 
\end{equation} 
The two and three-point tensor functions needed for our calculation  
are the following 
\begin{equation} 
    [\tilde{B}_0,B_{\mu },B_{\mu \nu }](p,m_1,m_2)=\int d^{n}\tilde{q}\,  
    \frac{[q^2,q_{\mu },q_{\mu }q_{\nu}]}{[q^2-m_1^2]\,[(q+p)^2-m_2^2]}\,, 
\end{equation} 
\begin{eqnarray} 
     \lefteqn{[\tilde{C}_0,C_{\mu },C_{\mu \nu }](p,k,m_1,m_2,m_3)=}  
     \nonumber \\ 
               & &\int d^{n}\tilde{q}\, \frac{[q^2,q_{\mu},q_{\mu}  
               q_{\nu}]}{[q^2-m_1^2]\,[(q+p)^2-m_2^2]\,[(q+p+k)^2-m_3^2]}\,. 
\end{eqnarray} 
By Lorentz covariance, they can be decomposed in terms of the  
above basic scalar functions and the external momenta: 
\begin{eqnarray} 
  \tilde{B}_0 (p,m_1,m_2) &=& A_0(m_2)+m_1^2 B_0(p,m_1,m_2)\,\,,\nonumber\\ 
    B_{\mu }(p,m_1,m_2)    &=& p_{\mu } B_1(p,m_1,m_2)\,\,,\nonumber \\ 
    B_{\mu \nu }(p,m_1,m_2)&=& p_{\mu } p_{\nu }B_{21}(p,m_1,m_2)+ 
                            g_{\mu \nu }B_{22}(p,m_1,m_2)\,\,, \nonumber \\ 
    \tilde{C}_0(p,k,m_1,m_2,m_3) &=& B_0(k,m_2,m_3)+ 
                                  m_1^2 C_{0}(p,k,m_1,m_2,m_3)\,\,, \nonumber\\ 
    C_{\mu}(p,k,m_1,m_2,m_3)&=&p_{\mu} C_{11}+k_{\mu} C_{12}\,\,,\nonumber\\ 
    C_{\mu \nu }(p,k,m_1,m_2,m_3)&=& p_{\mu }p_{\nu}C_{21}+ 
                                  k_{\mu }k_{\nu}C_{22}+ 
                                  (p_{\mu}k_{\nu}+k_{\mu}p_{\nu})C_{23}+ 
                                  g_{\mu \nu }C_{24}\,\,, 
\end{eqnarray} 
where we have defined the Lorentz invariant functions: 
\begin{eqnarray} 
B_1(p,m_1,m_2)&=&\frac{1}{2p^2}[A_0(m_1)-A_0(m_2)- 
    f_1B_0(p,m_1,m_2)], \\ 
B_{21}(p,m_1,m_2)&=&\frac{1}{2p^2(n-1)}[(n-2)A_0(m_2) 
            -2m_1^2B_0(p,m_1,m_2) \nonumber \\ 
            &-& nf_1B_1(p,m_1,m_2)],\\ 
B_{22}(p,m_1,m_2)&=&\frac{1}{2(n-1)}[A_0(m_2)\!+\!2m_1^2B_0(p,m_1,m_2) 
            \!+\!f_1B_1(p,m_1,m_2)], 
\end{eqnarray} 
\begin{equation} 
     \left(\begin{array}{c} 
       C_{11}\\C_{12} 
     \end{array}\right)=Y 
     \left(\begin{array}{c} 
       B_0(p+k,m_1,m_3)-B_0(k,m_2,m_3)-f_1C_0 \\  
       B_0(p,m_1,m_2)-B_0(p+k,m_1,m_3)-f_2C_0 
     \end{array}\right)\,, 
\end{equation} 
\begin{equation} 
     \left(\begin{array}{c} 
       C_{21}\\C_{23} 
     \end{array}\right)=Y  
     \left(\begin{array}{c} 
       B_1(p+k,m_1,m_3)+B_0(k,m_2,m_3)-f_1C_{11}-2C_{24} \\  
       B_1(p,m_1,m_2)-B_1(p+k,m_1,m_3)-f_2C_{11} 
     \end{array}\right)\,, 
\end{equation} 
\begin{eqnarray} 
        C_{22}&=&\displaystyle{\frac{1}{2[p^2k^2-(pk)^2]}}  
              \{-pk[B_1(p+k,m_1,m_3)-  
              B_1(k,m_2,m_3)-f_1C_{12}]  \nonumber\\ 
              &&+p^2[-B_1(p+k,m_1,m_3)-f_2C_{12}-2C_{24}]\}\,,             
\end{eqnarray} 
\begin{equation} 
    C_{24}=\frac{1}{2(n-2)} [B_0(k,m_2,m_3)+ 
           2m_1^2C_0+f_1C_{11}+f_2C_{12}]\,,\\ 
\end{equation} 
the factors $f_{1,2}$ and the matrix $Y$, 
\begin{eqnarray*} 
   f_1 &=& p^2+m_1^2-m_2^2, \\ 
   f_2 &=& k^2+2pk+m_2^2-m_3^2\,, 
\end{eqnarray*} 
\begin{equation} 
   Y=\frac{1}{2[p^2k^2-(pk)^2]}  
   \left(\begin{array}{cc} 
        k^2 & -pk \\-pk & p^2 
   \end{array}\right)\,. 
\end{equation} 
The UV divergences for $n\rightarrow 4$ can be parametrized as 
\begin{eqnarray} 
\label{eq:DeltaUV} 
    \epsilon &=& n-4,\nonumber \\ 
      \Delta &=& \frac{2}{\epsilon }+ 
                 {\gamma }_{\scriptscriptstyle E} -\ln (4\pi)\,, 
\end{eqnarray} 
being ${\gamma }_{\scriptscriptstyle E}$ the Euler constant. 
In the end one is left with the evaluation of the scalar one-loop 
functions: 
\begin{eqnarray} 
    A_0(m) &=& \left(\frac{-i}{16{\pi }^2} \right)  
           m^2(\Delta-1+\ln \frac{m^2}{\mu ^2})\,, \\ 
    B_0(p,m_1,m_2) &=& \left(\frac{-i}{16{\pi }^2} \right)  
           \left[\Delta+\ln \frac{p^2}{{\mu }^2}-2  
           +\ln [(x_1-1)(x_2-1)] \right.\nonumber \\ 
           & & \left . +x_1\ln \frac{x_1}{x_1-1}+ 
                        x_2\ln \frac{x_2}{x_2-1}\right], \\ 
    C_0(p,k,m_1,m_2,m_3) &=& \left(\frac{-i}{16{\pi }^2} \right)  
           \frac{1}{2}\;\frac{1}{pk+p^2\xi }\Sigma\,  
\end{eqnarray} 
with 
\begin{eqnarray} 
    x_{1,2}=x_{1,2}(p,m_1,m_2) &=& \frac{1}{2}+\frac{m_1^2-m_2^2}{2p^2} 
           \pm \frac{1}{2p^2}\lambda^{1/2}(p^2,m^2_1,m^2_2), \\ 
           \lambda (x,y,z) &=& [x-(\sqrt{y}-\sqrt{z})^2] 
                               [x-(\sqrt{y}+\sqrt{z})^2]\,, \nonumber 
\end{eqnarray} 
and where $\Sigma$ is a bookkeeping device for the following alternate  
sum of twelve (complex) Spence functions: 
\begin{eqnarray} 
    \Sigma &=& 
      Sp\left(\frac{y_1}{y_1-z_{1}^{i}}\right)-Sp\left(\frac{y_1-1} 
      {y_1-z_{1}^{i}}\right)+ 
      Sp\left(\frac{y_1}{y_1-z_{2}^{i}}\right)-Sp\left(\frac{y_1-1} 
      {y_1-z_{2}^{i}}\right)  
\nonumber \\ 
  &  -&Sp\left(\frac{y_2}{y_2-z_{1}^{ii}}\right)+Sp\left(\frac{y_2-1} 
      {y_2-z_{1}^{ii}}\right)- 
      Sp\left(\frac{y_2}{y_2-z_{2}^{ii}}\right)+Sp\left(\frac{y_2-1} 
      {y_2-z_{2}^{ii}}\right) 
\nonumber \\ 
  &  +&Sp\left(\frac{y_3}{y_3-z_{1}^{iii}}\right)-Sp\left(\frac{y_3-1} 
      {y_3-z_{1}^{iii}}\right)+ 
      Sp\left(\frac{y_3}{y_3-z_{2}^{iii}}\right)-Sp\left(\frac{y_3-1} 
      {y_3-z_{2}^{iii}}\right). 
\end{eqnarray} 
The Spence function is defined as 
\begin{equation} 
     Sp(z)=-\int_{0}^{1}\frac{\ln (1-zt)}{t}\,dt\,, 
\end{equation} 
and we have set, on one hand: 
\begin{eqnarray} 
     z_{1,2}^{i}   & = & x_{1,2}(p,m_2,m_1)\,,   \nonumber \\ 
     z_{1,2}^{ii}  & = & x_{1,2}(p+k,m_3,m_1)\,, \nonumber \\ 
     z_{1,2}^{iii} & = & x_{1,2}(k,m_3,m_2)\,; 
\end{eqnarray} 
and on the other: 
\begin{equation} 
     y_1=y_0+\xi\,,\;\;\;  
     y_2=\frac{y_0}{1-\xi }\,, \;\;\;\;  
     y_3=-\frac{y_0}{\xi }\,, \;\;\;\;  
     y_0=-\frac{1}{2}\;\frac{g+h\xi }{pk+p^2\xi }\,, \;\;\;\; 
\end{equation} 
where 
\begin{equation} 
\begin{array}{ll} 
     g=-k^2+m_2^2-m_3^2\,,\;\;\; & h=-p^2-2pk-m_2^2+m_1^2\,,  
\end{array} 
\end{equation} 
and $\xi $ is a root (always real for external on-shell momenta) of  
\begin{equation} 
     p^2\xi^2+2pk\xi +k^2=0\,. 
\end{equation} 
 
Derivatives of some 2-point functions are also needed in the  
calculation of self-energies, and we use the 
following notation: 
\begin{equation} 
     \frac{\partial}{\partial p^2}B_{*}(p,m_1,m_2)\equiv  
                             B^{\prime}_{*}(p,m_1,m_2).  
\end{equation} 
We can obtain all the derivatives from the basic $B^{\prime}_0$: 
\begin{eqnarray} 
  B^{\prime}_0(p,m_1,m_2)  
   &=&      \left( \frac{-i}{16\pi^2} \right) 
            \left\{    \frac{1}{p^2} + 
            \frac{1}{\lambda^{1/2}(p^2,m^2_1,m^2_2)} 
          \right. \nonumber \\    
   & &  \times  \left.  
            \left[   x_1(x_1-1)\ln \left( \frac{x_1-1}{x_1} \right) 
                         -x_2(x_2-1)\ln \left( \frac{x_2-1}{x_2} 
                         \right)  
\right]\right\}\,, 
\end{eqnarray} 
which has a threshold for $|p|=m_1+m_2$ and a pseudo-threshold 
for $|p|=|m_1-m_2|$.
